\documentclass[11pt,a4paper]{article}

\usepackage{graphicx}
\usepackage{amssymb}
\usepackage{textcomp}
\usepackage{amsmath}
\usepackage{bm}
\usepackage{times}
\usepackage{epsfig}
\usepackage{color}
\usepackage{array,multirow}
\usepackage{jheppub}

\newcommand{\Tr}{\mbox{Tr$\;$}}

\newcommand{\pmns}{{\mbox{\tiny PMNS}}}

\newcommand{\trix}[1]{\left(\begin{array}{#1}}
\newcommand{\notrix}{\end{array}\right)}
\newcommand{\comment}[1]{}
\newcommand{\rulesep}{\unskip\ \vrule\ }

\def\beq{\begin{equation}}
\def\eeq{\end{equation}}
\def\bea{\begin{eqnarray}}
\def\eea{\end{eqnarray}}

\usepackage[T1]{fontenc}

\usepackage[utf8]{inputenc}

\usepackage{mathtools} 

\usepackage{amsfonts}

\usepackage{ amssymb }

\usepackage[toc,page,title]{appendix}

\usepackage{xcolor}

\usepackage[compat=1.1.0]{tikz-feynman}

\usepackage{graphicx}

\usepackage{subcaption}

\usepackage{float}

\usepackage{caption}

\usepackage{breqn}

\usepackage{amsmath}

\usepackage{etoolbox}
\makeatletter
\patchcmd{\maketitle}{\@fpheader}{}{}{}
\makeatother

\usepackage[compat=1.1.0]{tikz-feynman}
\numberwithin{equation}{section}
\title{\Large  {{\bf{The $R$-parity Violating Decays of \\
Charginos and Neutralinos \\
in the B-L MSSM}}}}
\author{{Sebastian Dumitru$^{1}$, Burt A.~Ovrut$^{1}$ and Austin Purves$^{2}$} \\[2mm
]
    {\it $^{1}$ Department of Physics and Astronomy, University of Pennsylvania} \\
   {\it Philadelphia, PA 19104--6396}\\
   {\it $^{2}$ Department of Physics, Manhattanville College}\\
   {\it Purchase, NY 10577} \\[4mm]
}
\date{\today}

{\footnotetext{\mbox{sdumitru@sas.upenn.edu, ~ovrut@elcapitan.hep.upenn.edu, ~austin.purves@mville.edu} }}
\abstract{
The $B-L$ MSSM is the MSSM with three right-handed neutrino chiral multiplets and gauged $B-L$ symmetry. The $B-L$ symmetry is broken by the third family right-handed sneutrino acquiring a VEV, thus spontaneously breaking $R$-parity. Within a natural range of soft supersymmetry breaking parameters, it is shown that a large and uncorrelated number of initial values satisfy all present phenomenological constraints; including the correct masses for the $W^{\pm}$, $Z^0$ bosons, having all sparticles exceeding their present lower bounds and giving the experimentally measured  value for the Higgs boson. For this ``valid'' set of initial values, there are a number of different LSPs, each occurring a calculable number of times. We plot this statistically and determine that among the most prevalent LSPs are chargino and neutralino mass eigenstates. In this paper, the $R$-parity violating decay channels of charginos and neutralinos to standard model particles are determined, and the interaction vertices and decay rates computed analytically. These results are valid for any chargino and neutralino, regardless of whether or not they are the LSP. For chargino and neutralino LSPs, we will-- in a subsequent series of papers --present a numerical study of their RPV decays evaluated statistically over the range of associated valid initial points.} 
\begin{document}

\maketitle

\newpage

\newpage
%
\section{Introduction}
\label{sec:1}
%
The discovery of the Higgs particle at the Large Hadron Collider (LHC) \cite{Aad:2012tfa,Chatrchyan:2012xdj} completed the experimental search for the spectrum of the ``standard model'' of particle physics. The three chiral families of quarks and leptons, as well as the gluons, the $W^{\pm}$, $Z^{0}$ and photon vector bosons corresponding to the standard model gauge group $SU(3)_{C} \times SU(2)_{L} \times U(1)_{Y}$, had long since been known. However, the discovery of the Higgs scalar boson had special significance, since its vacuum expectation value (VEV) is required to spontaneously break the $SU(2)_{L} \times U(1)_{Y}$ electroweak symmetry to the $U(1)_{EM}$ of electromagnetism and to give masses to the matter fields and the electroweak vector bosons. This completion of the standard model confirms that the ``low energy'' world we observe is made up of precisely this spectrum, the particles of which interact with each other via strong, weak and electromagnetic gauge interactions.

However, as it presently stands, the standard model has several very significant shortcomings. To begin with, there is no theoretical explanation for either the particle content or the explicit gauge group of the interactions. 
Furthermore, the standard model contains a large number of undetermined parameters. These include 1) the Yukawa couplings that give rise, after spontaneous electroweak breaking, to the masses of the quarks/leptons, 2) the gauge coupling parameters of the strong, weak and electromagnetic interactions as well as 3) the mass and coupling parameters of the pure Higgs boson Lagrangian. At the present state of our knowledge, these parameters are simply input at a fixed scale so as to lead to the experimentally determined values of the particle masses and the observed strength of the gauge interactions. Finally, the standard model makes no attempt to couple its spectrum to gravitation, let alone to explain the origin of this fundamental force.
It seems clear, therefore, that despite the remarkable successes of the standard model, there must exist ``beyond the standard model'' physics which addresses and solves all of the shortcomings discussed above. 

At the present time, the only fundamental theory that would appear to have the potential to do this is superstring/M-theory--although at the price of introducing spontaneously broken $N=1$ supersymmetry (SUSY) as a major component of ``beyond the standard model'' physics. In this paper, we will consider a specific subset of such theories; namely, the $E_{8} \times E_{8}$ heterotic string \cite{Gross:1984dd} and its possible origin as a vacuum state of M-theory, called heterotic M-theory \cite{Horava:1996ma,Lukas:1997fg,Lukas:1998ew,Lukas:1998yy,Lukas:1998tt}. We do this for several important reasons. First of all, it is known using a series of papers \cite{Evans:1985vb,Donagi:2000fw,Donagi:2004ia,Braun:2005zv} that 
heterotic M-theory, when compactified on a specific Calabi-Yau threefold \cite{Braun:2005nv}, exhibits an observable sector with {\it exactly} the quark/lepton and Higgs spectrum of the so-called Minimal Supersymmetric Standard Model (MSSM); that is, three families of quark and lepton chiral superfields, each family with a right-handed neutrino chiral multiplet, and two doublets of Higgs chiral supermultiplets. Furthermore, the observable sector of this heterotic M-theory vacuum ``contains'' the gauge group $SU(3)_{C} \times SU(2)_{L} \times U(1)_{Y}$; significantly, multiplied by an additional gauged Abelian group $U(1)_{B-L}$, where $B$ and $L$ are baryon number and lepton number respectively. Below the scale of both spontaneous $B-L$ and SUSY breaking, the observable sector of this theory contains {\it precisely} the particle spectrum and gauge group of the standard model. It has also been demonstrated that the potential energy functions of the geometric, vector bundle and five-brane moduli can, in principle, be calculated and the vacuum for these moduli stabilized; see, for example \cite{Anderson:2011cza,Buchbinder:2002pr,Donagi:1999jp,Lima:2001jc}. 

Secondly, it has been shown in a number of contexts \cite{Anderson:2009nt,Anderson:2009ge,Blesneag:2015pvz,Blesneag:2016yag}  that the Yukawa couplings are, in principle, directly computable as integrals over the products of the harmonic representatives of the associated sheaf cohomology classes \cite{Braun:2006me}. 
Similarly, gauge couplings are potentially calculable from the threshold corrections at the string unification scale \cite{Deen:2016vyh,Kaplunovsky:1992vs,Kaplunovsky:1995jw,Mayr:1993kn,Dienes:1995sq,Dienes:1996du,Nilles:1997vk,Ghilencea:2001qq,deAlwis:2012bm,Bailin:2014nna}. 
Finally, making $N=1$ SUSY a local symmetry produces gravitation as the associated gauge field, although in the form of a gravity supermultiplet containing the gravitino as well as the graviton. That is, the existence of $N=1$ supersymmetry puts gravity on par with the strong, weak and electromagnetic gauge interactions. 
This property enables fundamental theories of early universe cosmology to occur; both within the context of inflation \cite{Deen:2016zfr,Ferrara:2015cwa,Linde:2016bcz} or ``bouncing universe'' \cite{Koehn:2012ar,Battarra:2014tga} scenarios. For all of these reasons, the $E_{8} \times E_{8}$ heterotic string and its possible origin as a vacuum state of heterotic M-theory appears to be a strong candidate for the theory of ``beyond the standard model'' physics.

With this in mind, we will confine ourselves in this paper to ``beyond the standard model'' physics arising in the observable sector of the $E_{8} \times E_{8}$ heterotic M-theory vacuum discussed above. At energies below the Calabi-Yau scale, the observable sector is precisely the MSSM with three right-handed neutrino chiral multiplets and gauge group $SU(3)_{C} \times SU(2)_{L} \times U(1)_{Y} \times U(1)_{B-L}$. Because of the existence of the additional $U(1)_{B-L}$ gauge factor, we call this theory the $B-L$ MSSM. The additional gauged $B-L$ symmetry plays a fundamental role in our analysis. Recall that to prevent unacceptably rapid nucleon decay in the conventional MSSM, it is necessary to postulate the existence of a finite symmetry group called $R$-parity. This acts on component fields as $(-1)^{3(B-L)+2s}$, where $s$ is the associated spin. While this finite symmetry indeed accomplishes its purpose, $R$-parity is, from  a theoretical viewpoint, completely {\it ad hoc}--without any fundamental justification.  However, continuous $U(1)_{B-L}$ symmetry arises naturally as a consequence of the compactification of heterotic M-theory and, indeed, has long been known in a non-supersymmetric context to be the minimal extra gauging of the standard model that remains quantum mechanically anomaly free. That is, the gauged $U(1)_{B-L}$ that arises in our context gives a ``natural way'' to suppress unwanted baryon and lepton number violating decays. Of course, the symmetry must be spontaneously broken at a scale sufficiently high to account for the fact that its associated massive vector boson $Z_{B-L}$ has, so far, not been observed. As discussed in the text, $U(1)_{B-L}$ symmetry is spontaneously broken by the right-handed sneutrino acquiring an non-vanishing VEV. This breaks lepton number $L$ and, hence, $B-L$ symmetry. However, baryon number $B$ remains unbroken and, therefore, proton decay continues to be suppressed below its present experimental bounds. However, the parameters of the $B-L$ MSSM must be chosen so as to adequately suppress lepton number violating processes. This will indeed be the case, as originally discussed in \cite{Marshall:2014cwa} and elaborated on in the text below. 

We conclude that {\it the $B-L$ MSSM is the simplest possible phenomenologically realistic low energy theory of heterotic superstring/M-theory; being the exact MSSM with right-handed neutrinos and spontaneously broken $R$-parity} This theory was originally presented in the series of papers \cite{Braun:2005nv,Ambroso:2009jd,Ambroso:2009sc,Ambroso:2010pe,Ovrut:2012wg,Ovrut:2014rba,
Ovrut:2015uea}. 
It is interesting to point out that the $B-L$ MSSM was also constructed from a ``bottom-up'' point of view, completely unrelated to superstring theory \cite{FileviezPerez:2008sx,Barger:2008wn,FP:2009gr,Everett:2009vy,FileviezPerez:2012mj,Perez:2013kla}. This simply postulated that the standard model should be extended to $N=1$ supersymmetry with three right-handed neutrino chiral multiplets--that is, the MSSM--with the problem of motivating $R$-parity solved by postulating spontaneously broken gauged $U(1)_{B-L}$ symmetry. That is, the $B-L$ MSSM as the simplest ``beyond the standard model'' theory is obtained from both a ``top-down'' superstring analysis as well as a ``bottom-up'' phenomenological approach.
It follows that the $B-L$ MSSM represents a strongly motivated ``beyond the standard model'' paradigm. Decay channels and decay rates for various sparticles of arbitrary mass have been identified and computed within the context of the $R$-parity conserving MSSM. Some of these have been searched for experimentally; so far without success. In the $B-L$ MSSM, these same decay channels remain. In addition, there are new $R$-parity violating (RPV) decay modes that now occur. These are, however, generically much weaker and, hence, harder to search for experimentally. There is, however, an important exception to this!

As is well-known, see for example \cite{Martin:1997ns}, the original $R$-parity invariant MSSM
must contain a lightest supersymmetric particle (LSP) that is completely stable and cannot further decay to standard model particles. The mass of this LSP depends on the scale of spontaneous SUSY breaking introduced into the model. For different choices of input parameters, the LSP sparticle will vary but, in all cases, is stable and cannot further decay. However, {\it this fundamentally changes in the $B-L$ MSSM}. In the $B-L$ MSSM, prior to the spontaneous breaking of $U(1)_{B-L}$, there will, as in the MSSM, be an LSP whose species again depends on the input initial conditions. As in the MSSM, this LSP cannot decay via $R$-parity preserving processes. However, the spontaneous breaking of $U(1)_{B-L}$ in this theory now leads to specific, and completely calculable, $R$-parity violating decays of this LSP into standard model particles. Not only are the decay modes of the LSP explicitly determined, but the associated vertex coefficients and, hence, the decay rates and branching ratios are exactly calculable. It follows that these $R$-parity violating decays should be amenable to direct detection at the ATLAS and CMS detectors at the LHC. Detection of these processes would not only be an explicit indication of ``beyond the standard model'' physics, but would also strongly hint at the existence of $N=1$ SUSY with spontaneously broken $R$-parity.

The first calculation of such an $R$-parity decay was presented in \cite{Marshall:2014cwa,Marshall:2014kea}. In this example,
the initial conditions of the $B-L$ MSSM were chosen so that the LSP was the lightest real scalar superpartner of the top quark; the so-called admixture stop. Its $R$-parity violating decay modes and branching ratios were determined theoretically. We refer the reader to \cite{Marshall:2014cwa,Marshall:2014kea} for details. In recent work, an ATLAS group analyzed the relevant LHC data looking for these experimental signatures \cite{Aaboud:2017opj}. None were found, but a new experimental lower bound on the stop mass was determined. We emphasize that the lightest stop LSP was chosen as a ``test case'' since it is exotic, carrying both color and electric charge, and also has a large production cross section. 
Therefore, the stop could never have been chosen as the LSP in the ordinary MSSM, since it would be stable and contribute to 
at least a portion of dark matter, which must be neutral in all interactions. As shown in \cite{Ovrut:2014rba,Ovrut:2015uea} however, the stop sparticle occurs as the LSP for only a relatively small fraction of possible initial conditions in the $B-L$ MSSM. As discussed in \cite{Ovrut:2014rba,Ovrut:2015uea}, and analyzed in more detail in this paper, almost all other sparticles are much more likely to be the LSP in the $B-L$ MSSM. In particular, due both to their frequent appearance as the LSP and because their experimental signatures are easily detected by the LHC experimental groups, the $R$-parity violating decays of both charginos and neutralinos are very interesting to analyze. Therefore, in this paper, we determine the $R$-parity violating decay modes for both charginos and neutralinos, compute the explicit interaction coefficients for each such mode and calculate the explicit decay rates. We emphasize that the results of this paper are applicable to RPV decays of {\it any} chargino or neutralino. However, they are most easily experimentally observed when applied to the LSP of the $B-L$ MSSM. Specifically, we do the following.

In Section \ref{sec:2}, we give a brief summary of the $B-L$ MSSM. In particular, the spontaneous breaking of gauged $U(1)_{B-L}$ symmetry by a non-vanishing VEV of the third family right-handed sneutrino is discussed. The associated $R$-parity violating interactions induced in the Lagrangian are presented in detail. The VEV of the right-handed sneutrino produces a mixing of the third family right-hand neutrino and the three left-handed neutrinos with all fermionic superpartners of the neutral gauge bosons and the up and down neutral Higgsinos. This is presented in Section \ref{sec:3}. The general form of this $9 \times 9$ mass matrix is given, as well as the explicit form of the unitary matrix required to diagonalize it. However, in that section, we focus on the diagonal $3 \times 3$ left-handed neutrino Majorana submatix $m^{D}_{\nu} $ only. The mass eigenvalues of this matrix can be determined from the explicit form of the PMNS mixing matrix as well as the off-diagonal left-handed neutrino matrix $m_{\nu}$. This latter matrix is a function of the $R$-parity violating parameters as well as three additional quantities. The result is compared with the experimentally determined mass eigenvalues of both the ``normal'' and the ``inverted'' neutrino mass hierarchies. In Section \ref{sec:4}, we list all of the presently known experimental data that must be satisfied in any phenomenologically acceptable vacuum. These include the masses of the $W^{\pm}$,$Z^{0}$ electroweak vector bosons, the Higgs mass, the present lower bounds on the SUSY sparticles and so on. Having done this, we present the mass interval in which we will statistically throw all dimensionful parameters of the soft SUSY breaking terms. The reason for choosing the specific median value and width of this interval is discussed in detail. Having presented this interval, we do a statistical analysis involving 100 million independent throws. A plot of the the ``valid'' points, that is, all parameters solving all required phenomenological bounds, is given and analyzed. A histogram of the LSPs associated with these valid points is presented. Section \ref{sec:5} is devoted to analyzing the mass matrices, including the terms induced by both spontaneous electroweak and $R$-parity violation, for both the charginos and, independently, for the neutralinos of the $B-L$ MSSM. Important technical details of these calculations are discussed in the Appendix. The mass eigenvalues and eigenstates are explicitly calculated for both charginos and neutralinos. In Section \ref{sec:6} we present the relevant portions of the complete $B-L$ MSSM Lagrangian, including the effect of both electroweak symmetry breaking and $R$-parity violation. Rewriting the original fields in terms of the chargino and neutralino mass eigenstates calculated in the previous section, we determine the three-point interaction vertices involving either a chargino or a neutralino eigenstate decaying into two standard model particles. The Feynman diagrams associated with these decays are presented pictorially and, for each decay process, the exact expression for the vertex coefficient is given. Finally, in Section \ref{sec:7} we summarize the possible $R$-parity violating decays with their associated vertex parameters. These are then used to compute the decay rate for each of these processes. We emphasize that the results we present are valid for any chargino and neutralino, regardless of whether or not they are the LSP.

Before continuing to our analysis of the $R$-parity violating $B-L$ MSSM chargino and neutralino LSP decays, it is useful to point out that the subject of RPV in the MSSM and related $N=1$ supersymmetric particle physics models has a long history in the literature. Papers in this context, up to and including most of 2005, are cited and discussed in the comprehensive review in \cite{Barbier:2004ez}. Relevant to the content of our present paper, this review discussed both explicit and, more briefly, spontaneous RPV due to both left- and right-chiral sneutrinos developing VEVs, the associated massless ``Majoron'', and possibly gauging lepton number global symmetry to make the Majoron heavy. The review also discussed the RGEs in some RPV theories, the RPV decays of some of the LSPs in various models and their impact on theories of dark matter. More recently, the subject was reviewed in 2015 \cite{Mohapatra:2015fua}. This discussed explicit RPV in the MSSM but, in particular, focussed on spontaneous breaking of $R$-parity in theories where the standard model symmetry is extended by a gauged $U(1)_{B-L}$. 
This review post-dates the papers in \cite{Braun:2005nv,Ambroso:2009jd,Ambroso:2009sc,Ambroso:2010pe,Ovrut:2012wg,Ovrut:2014rba,Ovrut:2015uea} and \cite{FileviezPerez:2008sx, Barger:2008wn, FP:2009gr, Everett:2009vy, FileviezPerez:2012mj, Perez:2013kla} and cites some of them. In particular, this review highlights what it refers to as ``Minimal models with automatic $R$-parity breaking''; that is, the $B-L$ MSSM. It then introduces, and devotes the remainder of the work, to models with two pairs of Higgs doublets. More recently, there was a comprehensive paper \cite{Dercks:2017lfq} on these subjects within the context of the RPV-CMSSM; that is, the MSSM with an additional RPV trilinear coupling at the unification scale. Within this context, that paper discussed the RGEs, taking into account the then recently discovered Higgs mass, and the associated LSPs. It goes on to discuss the RPV decay of some of the LSPs; specifically the Bino neutralino and the stau sparticle. It is not the intention of our paper to review this vast RPV literature. We refer the readers to the references in the mentioned papers. We do wish to point out that, although some of the broad topics
that occur in our present paper are mentioned and discussed in previous RPV literature, such as RG evolution, the associated LSP calculations and their RPV decays, relationship to neutrino masses and so on, these previous discussions all occur in contexts considerably different than the heterotic M-theory $B-L$ MSSM. The present paper works strictly within this context and presents specific results for both chargino and neutralino decays not previously discussed. Finally, we note that there has been significant experimental studies of chargino and neutralino decays in $R$-parity {\it conserving} theories--see M. Tanabashi et al. (Particle Data Group), Phys. Rev. D 98, 030001 (2018). The main purpose of the present paper is to analyze RPV chargino and neutralino decays in the spontaneously broken RPV MSSM. Some of the present authors have already applied this formalism to RPV decays of admixture stop LSPs in \cite{Marshall:2014cwa,Marshall:2014kea} and to Wino chargino and Wino neutralino decays in \cite{Dumitru:2018nct}--analyzing potential LHC signatures. Other LSP RPV decays, and the predictions for LHC signatures in this context, will be presented in future publications.

\section{The B-L MSSM}
\label{sec:2}

In this section, we briefly review the contents of the $B-L$ MSSM theory relevant to a phenomenological discussion of its $R$-parity violating decay processes and their potential signatures at the LHC.

The low energy manifestation of the ``heterotic standard model'', that is, the $B-L$ MSSM, arises from the breaking of an $SO(10)$ GUT theory via two independent Wilson lines, denoted by $\chi_{3R}$ and $\chi_{B-L}$, associated with the diagonal $T_{3R}$ generator of $SU(2)_{R}$ and the generator $T_{B-L}$ of $U(1)_{B-L}$ respectively. These specific generators  are chosen since it can be shown that there is no kinetic mixing of their respective Abelian gauge kinetic terms at any energy scale-- thus simplifying the RG calculations \cite{Ovrut:2012wg}. However, identical physical results will be obtained for any linear combination of these generators. Associated with these Wilson lines are two mass scales, $M_{\chi_{3R}}$ and $M_{\chi_{B-L}}$, with three possible relations between them; 1) $M_{\chi_{B-L}} > M_{\chi_{3R}}$, 2) $M_{\chi_{3R}} > M_{\chi_{B-L}}$ and 3) $M_{\chi_{3R}} = M_{\chi_{B-L}}$. As discussed in \cite{Ovrut:2012wg}, the masses in the first two relations can be adjusted so as to enforce exact unification at one loop of all gauge couplings at the SO(10) unification scale $M_{U}$, whereas gauge unification cannot occur for the third mass relationship without accounting for threshold effects at the unification scale or the SUSY scale \cite{Deen:2016vyh,Ovrut:2012wg,Fundira:2017vip}. For this reason, we will not consider the third option in this paper. The gauge coupling RG equations associated with each of the first two mass relations were discussed in detail in \cite{Ovrut:2012wg} and, as far as low energy LHC phenomenology is concerned, give almost identical results. For specificity, therefore, in this paper we will focus on the first relationship and, without loss of accuracy, choose $M_{\chi_{B-L}}=M_{U}$. The lower scale $M_{\chi_{3R}}$, which we henceforth denote by $M_{I}$, is adjusted so as to obtain exact gauge coupling unification. We emphasize, however, that the low energy results predicted for the LHC are almost unchanged even if $M_{I}$ is chosen to yield only ``approximate'' gauge unification --- with moderate sized gauge ``thresholds''. Conventionally, the scale of supersymmetry breaking is defined to be
\begin{equation}
M_{SUSY} = \sqrt{m_{{\tilde{t}}_{1}}m_{{\tilde{t}}_{2}}} \ ,
\label{cham1}
\end{equation}
where $m_{{\tilde{t}}_{1}}$ and $m_{{\tilde{t}}_{2}}$ are the lightest and heaviest stop masses respectively; see ,for example, \cite{Ovrut:2015uea}.
Suffice it here to say that for supersymmetry breaking to occur between the electroweak scale and 10~TeV, which will be the case in this paper, the unification scale $M_{U}$ is found to be $\cal{O}$$(3 \times 10^{16}~{\rm GeV})$. Over the same range of supersymmetry breaking, however, the  intermediate scale $M_{I}$ changes from $\cal{O}$$(2 \times 10^{16}~{\rm GeV})$ to $\cal{O}$$(3 \times 10^{15}~{\rm GeV})$ respectively \cite{Deen:2016vyh}.

The details of the symmetry breaking and the respective mass spectra for this choice of Wilson line hierarchy were given in \cite{Ovrut:2012wg}. Here, we simply note that in the mass regime between $M_{U}$ and $M_{I}$, the gauge group is broken from $SO(10)$ to $SU(3)_{C} \times SU(2)_{L} \times SU(2)_{R} \times U(1)_{B-L}$ with the spectrum shown in Figure 1. This theory is referred to as the ``left-right'' model \cite{Senjanovic:1975rk,Grimus:1993fx}. As discussed above, for the supersymmetry breaking scales of interest in this paper, this mass regime will on average be considerably smaller than one order of magnitude in GeV. 
\setlength{\unitlength}{.9cm}
\begin{figure}[!ht]
\begin{center}
\scriptsize
\begin{picture}(5,10)(0,1)

\put(0.8,10){\color{blue}$\underline{SO(10)}$}
\put(-2,9.5){\color{red}\line(1,0){6}}
\put(4.3,9.3){\color{red} $M_{U}=M_{\chi_{B-L}}$}
\put(1.,9.4){\vector(0,-1){1.3}}
\put(1.2,9){$\chi_{B-L}$}

\put(-1.0,7.5){\color{blue}$\underline{SU(3)_C\otimes SU(2)_L\otimes SU(2)_R\otimes U(1)_{B-L}}$}
\put(-0.5,6.5){$L=(\textbf{1},\textbf{2},\textbf{1},-1)$}
\put(-0.5,5.9){$L^c=(\textbf{1},\textbf{1},\textbf{2},1)$}
\put(-0.5,5.3){$Q=(\textbf{3},\textbf{2},\textbf{1},1/3)$}
\put(-0.5,4.7){$Q^c=(\bar{\textbf{3}},\textbf{1},\textbf{2},-1/3)$}
\put(-0.5,3.9){$\mathcal{H}=(\textbf{1},\textbf{2},\textbf{2},0)$}
\put(-0.5,3.3){$H_C=(\textbf{3},\textbf{1},\textbf{1},2/3)$}
\put(-0.5,2.7){${\bar{H}}_C=(\bar{\textbf{3}},\textbf{1},\textbf{1},-2/3)$}

\put(-0.5,6.8){\line(-1,0){.2}}
\put(-0.5,4.6){\line(-1,0){.2}}
\put(-0.7,6.8){\line(0,-1){2.2}}
\put(-0.7,5.7){\line(-1,0){.2}}
\put(-1.4,5.6){\textbf{16}}
\put(3.,6.6){\oval(.4,.4)[tr]}
\put(3.2,6.6){\line(0,-1){.7}}
\put(3.4,5.9){\oval(.4,.4)[bl]}
\put(3.4,5.5){\oval(.4,.4)[tl]}
\put(3.2,5.5){\line(0,-1){.7}}
\put(3.,4.8){\oval(.4,.4)[br]}
\put(3.6,5.6){$\times 9$}

\put(-0.5,4.2){\line(-1,0){.2}}
\put(-0.5,2.6){\line(-1,0){.2}}
\put(-0.7,4.2){\line(0,-1){1.6}}
\put(-0.7,3.4){\line(-1,0){.2}}
\put(-1.4,3.3){\textbf{10}}
\put(3.1,4.2){\oval(.2,.2)[tr]}
\put(3.2,4.2){\line(0,-1){.1}}
\put(3.3,4.1){\oval(.2,.2)[bl]}
\put(3.3,3.9){\oval(.2,.2)[tl]}
\put(3.2,3.9){\line(0,-1){.1}}
\put(3.1,3.8){\oval(.2,.2)[br]}
\put(3.6,3.85){$\times 2$}

\put(-2,2.2){\color{red}\line(1,0){6}}
\put(4.3,2.1){\color{red}$M_{I}=M_{\chi_{3R}}$}
\put(1.,2.1){\vector(0,-1){1.}}
\put(1.2,1.7){$\chi_{3R}$}

\put(-1,.5){\color{blue}$\underline{SU(3)_C\otimes SU(2)_L\otimes U(1)_{3R}\otimes U(1)_{B-L}}$}
\put(-.5,-.5){$L=(\textbf{1},\textbf{2},0,-1)$}
\put(-.5,-1){$e^c=(\textbf{1},\textbf{1},1/2,1)$}
\put(-.5,-1.5){$\nu^c=(\textbf{1},\textbf{1},-1/2,1)$}
\put(-.5,-2){$Q=(\bar{\textbf{3}},\textbf{2},0,1/3)$}
\put(-.5,-2.5){$u^c=(\textbf{3},\textbf{1},-1/2,-1/3)$}
\put(-.5,-3.0){$d^c=(\textbf{3},\textbf{1},1/2,-1/3)$}
\put(-.5,-3.5){$H_u=(\textbf{1},\textbf{2},1/2,0)$}
\put(-.5,-4.0){$H_d=(\textbf{1},\textbf{2},-1/2,0)$}

\put(-.5,-.2){\line(-1,0){.2}}
\put(-.5,-3.1){\line(-1,0){.2}}
\put(-.7,-.2){\line(0,-1){2.9}}
\put(-.7,-1.7){\line(-1,0){.2}}
\put(-1.4,-1.8){\textbf{16}}
\put(3,-.4){\oval(.4,.4)[tr]}
\put(3.2,-.4){\line(0,-1){1.1}}
\put(3.4,-1.5){\oval(.4,.4)[bl]}
\put(3.4,-1.9){\oval(.4,.4)[tl]}
\put(3.2,-1.9){\line(0,-1){1}}
\put(3,-2.9){\oval(.4,.4)[br]}
\put(3.6,-1.8){$\times 3$}

\put(-.5,-3.2){\line(-1,0){.2}}
\put(-.5,-4.1){\line(-1,0){.2}}
\put(-.7,-3.2){\line(0,-1){.9}}
\put(-.7,-3.7){\line(-1,0){.2}}
\put(-1.4,-3.8){\textbf{10}}

\put(6.6,-1.5){MSSM}
\put(7,-2){+}
\put(5.3,-2.5){3 right-handed neutrino}
\put(6.0,-2.9){supermultiplets}

\end{picture}
\end{center}
 \vspace{5cm}
\caption{The particle spectra in the scaling regimes of the sequential Wilson line breaking pattern of $SO(10)$ in which 
$M_{\chi_{B-L}}=M_{U}~>~M_{\chi_{3R}}=M_{I}$.}
\label{fig:matterContent}
\end{figure}
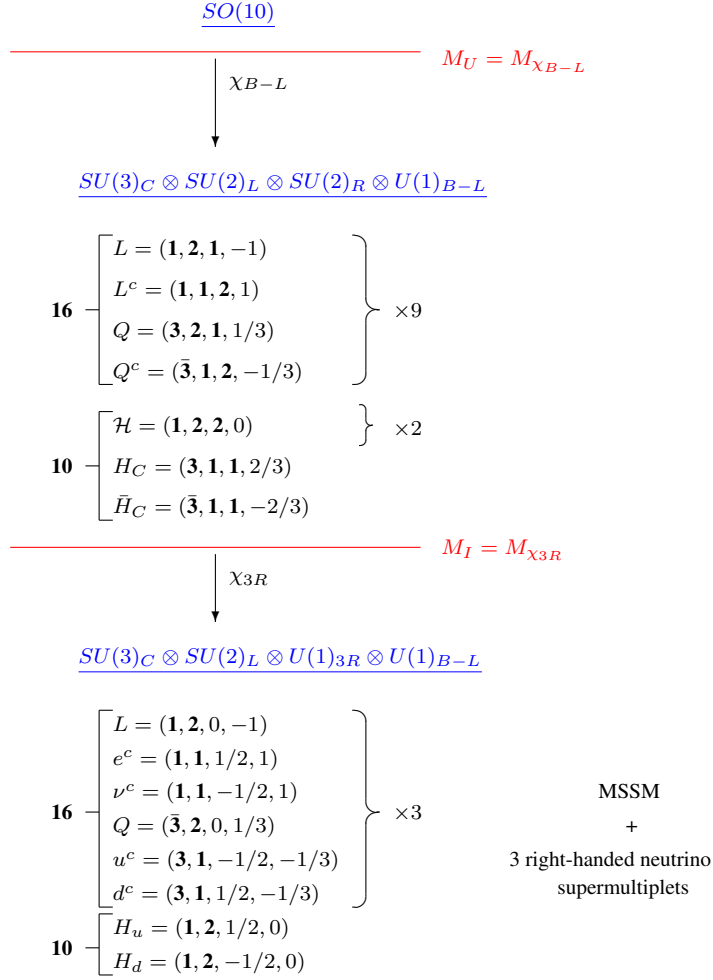
At the ``intermediate'' scale $M_{I}$, the second Wilson line breaks this ``left-right'' model down to the exact $B-L$ MSSM. This theory has the $SU(3)_{C} \times SU(2)_{L} \times U_{Y}(1)$ gauge group of the standard model augmented by an additional $U(1)_{B-L}$ Abelian symmetry. As mentioned above, it is convenient-- and equivalent --to use the Abelian group $U(1)_{3R}$ with the generator 
\begin{equation}
T_{3R}=Y-\frac{B-L}{2}
\label{eq:1}
\end{equation}
in the RGE's since the associated gauge kinetic term cannot mix  with the  gauge kinetic energy of $U(1)_{B-L}$. That is, the $B-L$ MSSM gauge group is chosen, for computational convenience, to be
\begin{equation}
	SU(3)_C\otimes SU(2)_L\otimes U(1)_{3R}\otimes U(1)_{B-L} \ .
	\label{eq:2}
\end{equation}
The associated gauge couplings will be denoted by $g_3$, $g_2$, $g_{R}$ and $g_{BL}$.
The spectrum, as shown in Figure 1, is exactly that of the MSSM with three right-handed neutrino chiral multiplets, one per family; that is, three generations of matter superfields
\begin{eqnarray}
	Q=\trix{c}u\\d\notrix\sim({\bf 3}, {\bf 2}, 0, \frac{1}{3}) & \begin{array}{rl}u^c\sim&(\bar{\bf 3}, {\bf 1}, -1/2, -\frac{1}{3}) \\
	d^c\sim&(\bar{\bf 3}, {\bf 1}, 1/2, -\frac{1}{3})\end{array} \ , \nonumber\\
	L=\trix{c}\nu\\e\notrix\sim({\bf 1}, {\bf 2}, 0, -1)&\begin{array}{rl}\nu^c\sim&({\bf 1}, {\bf 1}, -1/2, 1)\\
	e^c\sim&({\bf 1}, {\bf 1}, 1/2, 1)\end{array} \ ,
	\label{eq:3}
\end{eqnarray}
along with two Higgs supermultiplets
\begin{equation}
H_u=\trix{c}H_u^+\\H_u^0\notrix\sim({\bf 1}, {\bf 2}, 1/2, 0) ~~, ~ H_d=\trix{c}H_d^0\\H_d^-\notrix\sim({\bf 1}, {\bf 2}, -1/2, 0) \ .
\label{eq:4}
\end{equation}
%
%

The superpotential of the $B-L$ MSSM is given by
\begin{eqnarray}
	W=Y_u Q H_u u^c - Y_d Q H_d d^c -Y_e L H_d e^c +Y_\nu L H_u \nu^c+\mu H_u H_d \ ,
	\label{eq:5}
\end{eqnarray}
where flavor and gauge indices have been suppressed and the Yukawa couplings are three-by-three matrices in flavor space. In principle, the Yukawa matrices are arbitrary complex matrices. However, the observed smallness of the three CKM mixing angles and the CP-violating phase dictate that the quark Yukawa matrices be taken to be nearly diagonal and real. The charged lepton Yukawa coupling matrix can also be chosen to be diagonal and real. This is accomplished  by moving the rotation angles and phases into the neutrino Yukawa couplings which, henceforth, must be complex matrices. Furthermore, the smallness of the first and second family fermion masses implies that all components of the up, down quark and charged lepton Yukawa couplings-- with the exception of the (3,3) components --can be neglected for the purposes of the RG running. Similarly, the very light neutrino masses imply that the neutrino Yukawa couplings are sufficiently small so as to be neglected for the purposes of RG running. However, the $Y_{\nu i3}$, $i=1,2,3$ neutrino Yukawa couplings cannot be neglected for the calculations of the neutralino, neutrino and chargino mass matrices, as well as in decay rates/branching ratios.
The $\mu$-parameter can be chosen to be real, but not necessarily positive, without loss of generality. We implement these constraints in the remainder of our analysis.

Spontaneous supersymmetry breaking is assumed to occur in a hidden sector-- a natural feature of both strongly and weakly coupled $E_{8} \times E_{8}$ heterotic string theory --and be transmitted through gravitational mediation to the observable sector and, hence, to the $B-L$ MSSM. Since the $B-L$ MSSM first manifests itself at the scale $M_{I}$, we will begin our analysis by presenting the most general soft supersymmetry breaking interactions at that scale. That is, at scale $M_{I}$,
the soft supersymmetry breaking Lagrangian is given by
\begin{align}
 \label{eq:6}
\begin{split}
	-\mathcal L_{\mbox{\scriptsize soft}}  = &
	\left(
		\frac{1}{2} M_3 \tilde g^2+ \frac{1}{2} M_2 \tilde W^2+ \frac{1}{2} M_R \tilde W_R^2+\frac{1}{2} M_{BL} \tilde {B^\prime}^2
	\right.
		\\
	& \left.
		\hspace{0.4cm} +a_u \tilde Q H_u \tilde u^c - a_d \tilde Q H_d \tilde d^c - a_e \tilde L H_d \tilde e^c
		+ a_\nu \tilde L H_u \tilde \nu^c + b H_u H_d + h.c.
	\right)
	\\
	& + m_{\tilde Q}^2|\tilde Q|^2+m_{\tilde u^c}^2|\tilde u^c|^2+m_{\tilde d^c}^2|\tilde d^c|^2+m_{\tilde L}^2|\tilde L|^2
	+m_{\tilde \nu^c}^2|\tilde \nu^c|^2+m_{\tilde e^c}^2|\tilde e^c|^2  \\  
	&+m_{H_u}^2|H_u|^2+m_{H_d}^2|H_d|^2 \ .
\end{split}
\end{align}
The $b$ parameter can be chosen to be real and positive without loss of generality. The gaugino soft masses can, in principle, be complex. This, however, could lead to CP-violating effects that are not observed. Therefore, we proceed by assuming they all are real. The $a$-parameters and scalar soft mass can, in general, be Hermitian matrices in family space. Again, however, this could lead to unobserved flavor and CP violation. Therefore, we will assume they all are diagonal and real. Furthermore, we assume that only the (3,3) components of the up, down quark and charged lepton $a$-parameters are significant and that the neutrino $a$ parameters are negligible for the RG running and all other purposes. For more explanation of these assumptions, see \cite{Ovrut:2015uea}.

As discussed in \cite{Ovrut:2015uea}, without loss of generality one can assume that the third generation right-handed sneutrino, since it carries the appropriate $T_{3R}$ and $B-L$ charges, spontaneous breaks the $B-L$ symmetry by developing a non-vanishing VEV
\begin{equation}
\left< \tilde \nu^c_3 \right> \equiv \frac{1}{\sqrt 2} v_R \ .
\label{eq:7}
\end{equation}
This VEV spontaneously breaks $U(1)_{3R}\otimes U(1)_{B-L}$ down to the hypercharge gauge group $U(1)_Y$. We denote the associated gauge parameter by $g^{\prime}$. However, since sneutrinos are singlets under the $SU(3)_C\otimes SU(2)_L\otimes U(1)_{Y}$ gauge group, it does not break any of the SM symmetries.  
At a lower mass scale, electroweak symmetry is spontaneously broken by the neutral components of both the up and down Higgs multiplets acquiring non-zero VEV's.  In combination with the right-handed sneutrino VEV, this also induces a VEV in each of the three generations of left-handed sneutrinos. The notation for the relevant VEVs is
\beq
        \left<\tilde \nu_{i}\right> \equiv \frac{1}{\sqrt 2} {v_L}_i, \ \
	\left< H_u^0\right> \equiv \frac{1}{\sqrt 2}v_u, \ \ \left< H_d^0\right> \equiv \frac{1}{\sqrt 2}v_d,
	\label{eq:8}
\eeq
where $i=1,2,3$ is the generation index. The neutral gauge boson that becomes massive due to $B-L$ symmetry breaking, $Z_R$, has a mass at leading order, in the relevant limit that $v_R \gg v$, of
\beq
	M_{Z_R}^2 = \frac{1}{4}\left(g_R^2+g_{BL}^2 \right) v_R^2
				\left(1+\frac{g_R^4}{g_R^2+g_{BL}^2} \frac{v^2}{v_R^2}\right) \ ,
	\label{eq:9}
\eeq
where
\beq
	v^2 \equiv v_d^2 + v_u^2 \ .
\eeq
The second term in the parenthesis is a small effect due to mixing in the neutral gauge boson sector. 

A discussion of the neutrino masses is presented in the next section, where they are shown to be roughly proportional to the ${Y_\nu}_{ij}$ and ${v_L}_i$ parameters. It follows that ${Y_\nu}_{ij} \ll 1$ and ${v_L}_i\ll v_{u,d}, v_R$. In this phenomenologically relevant limit, the minimization conditions of the potential are simple, leading to the VEV's
\begin{align}
	\label{eq:10}
	v_R^2=&\frac{-8m^2_{\tilde \nu_{3}^c}  + g_R^2\left(v_u^2 - v_d^2 \right)}{g_R^2+g_{BL}^2} \ ,
	\\
	{v_L}_i=&\frac{\frac{v_R}{\sqrt 2}(Y_{\nu_{i3}}^* \mu v_d-a_{\nu_{i3}}^* v_u)}
			{m_{\tilde L_{i}}^2-\frac{g_2^2}{8}(v_u^2-v_d^2)-\frac{g_{BL}^2}{8}v_R^2} \ , \label{eq:11}
	\\
	\label{eq:12}
	\frac{1}{2} M_{Z^0}^2 =&-\mu^2+\frac{m_{H_u}^2\tan^2\beta-m_{H_d}^2}{1-\tan^2\beta} \ ,
	\\
	\label{eq:13}
	\frac{2b}{\sin2\beta}=&2\mu^2+m_{H_u}^2+m_{H_d}^2 
\end{align}
where 
\begin{equation}
{\rm tan} \beta= \frac{v_{u}}{v_{d}} \ .
\label{eq:14}
\end{equation}
Here, the first two equations correspond to the sneutrino VEVs. The third and fourth equations are of the same form as in the MSSM, but new $B-L$ scale contributions to $m_{H_u}$ and $m_{H_d}$ shift their values significantly compared to the MSSM. Eq.~(\ref{eq:10}) can be used to re-express the $Z_R$ mass as
\beq
	\label{eq:15}
	M_{Z_R}^2 = -2 m_{\tilde \nu^c_3}^2
				\left(1+\frac{g_R^4}{g_R^2+g_{BL}^2} \frac{v^2}{v_R^2}\right) \ .
\eeq
This makes it clear that, to leading order, the $Z_R$ mass is determined by the soft SUSY breaking mass of the third family right-handed sneutrino. The term proportional to $v^2/v_R^2$ is insignificant in comparison and, henceforth, neglected  in our calculations.

Recall that $R$-parity is defined as 
\begin{equation}
R=(-1)^{3(B-L)+2s}.
\label{16}
\end{equation}
It follows that a direct consequence of generating a VEV for the third family sneutrino is the spontaneous breaking of $B-L$ symmetry and, hence, $R$-parity. The $R$-parity violating operators induced in the superpotential are given by
\begin{equation}
	\label{eq:17}
	W \supset \epsilon_i \,  e_i \,  H^{+}_u - \frac{1}{\sqrt 2 }{Y_e}_i \, {v_L}_i \,  H_d^- \,   e^c_i \ ,
\end{equation}
where
\begin{equation}
	\epsilon_i  \equiv \frac{1}{\sqrt 2} {Y_\nu}_{i3} v_R \ 
	\label{eq:18}
\end{equation}
and $Y_{ei}$ is the $i$th component of the diagonal lepton Yukawa coupling.
This general pattern of $R$-parity violation is referred to as bilinear $R$-parity breaking and has been discussed in many different contexts \cite{Mukhopadhyaya:1998xj,Chun:1998ub,Chun:1999bq,Hirsch:2000ef}. In addition, the Lagrangian contains bilinear terms generated by ${v_L}_i$ and $v_R$ in the super-covariant derivatives. These are
\begin{align}
\begin{split}
	\mathcal{L} \supset &
	- \frac{1}{2}{v_L}_i^* \left[ g_2 \left(\sqrt 2 \, e_i \tilde W^+ 
	+  \nu_{L_i} \tilde W^0\right) - g_{BL} \nu_{L_i} \tilde B' \right] \label{eq:19}
	\\
	&
	-\frac{1}{2} v_R \left[-g_R \nu_3^c \tilde W_R + g_{BL} \nu_3^c \tilde B' \right]+ \text{h.c.}
\end{split}
\end{align}

The consequences of spontaneous $R$-parity violation are quite interesting, and have been discussed in a number of papers \cite{FileviezPerez:2008sx,Barger:2008wn,FP:2009gr,Everett:2009vy,FileviezPerez:2012mj,Perez:2013kla,Ghosh:2010hy,Barger:2010iv,Mohapatra:1986aw}. 
In this paper, we will present the decay channels for arbitrary mass charginos and neutralinos, and analytically determine their decay rates. However, in a series of following works, we will explore the phenomenological consequences of the $R$-parity violating (RPV) decays of the lightest, and next-to-lightest, supersymmetric particles; referred to as the LSP and NLSP respectively. These decays are potentially observable at the ATLAS detector of the LHC. Hence, if detected, these explicit decays could verify the existence of low energy $N=1$ supersymmetry, shed light on the structure of the precise supersymmetric model-- such as the $B-L$ MSSM --and, as will become apparent, even constrain whether the neutrino mass hierarchy is ``normal'' or ``inverted''. 
However, as is clear from expressions \eqref{eq:17} and \eqref{eq:19}, these results will depend explicitly on the values of the parameters $\epsilon_{i}$, $i=1,2,3$ and $v_{L_{i}}$, $i=1,2,3$ defined in \eqref{eq:18} and \eqref{eq:11} respectively. In turn, these parameters are dependent on the present experimental values of the neutrino masses. These reduce the number of independent RPV parameters from six to one and potentially restrict the value of the remaining independent coefficient. For that reason, we will discuss the neutrino masses and their direct  relationship to the $\epsilon_{i}$ and $v_{L_{i}}$ parameters in the next section.

\section{Neutrino Masses and the RPV Parameters}
\label{sec:3}
As discussed in \cite{Marshall:2014cwa,Marshall:2014kea,Ovrut:2014rba,Ovrut:2015uea,Barger:2010iv}, it follows from the above Lagrangian that the third family right-handed neutrino and the three left-chiral neutrinos $\nu_{i}$, $i=1,2,3$ mix with the fermionic superpartners of the neutral gauge bosons and with the up- and down- neutral Higgsinos. In other words, the neutralinos now mix with the neutral fermions of the standard model. The mixing with the third-family right-handed sneutrino, through terms proportional to $\epsilon_i = Y_{\nu_{i3}}v_R/\sqrt 2 $ and $v_{L_i}$, allows the third-family right-handed sneutrino to act as a seesaw field giving rise to Majorana neutrino masses. This is reviewed in this section.

First, we note that this paper is focused on the consequences of RPV decays at the LHC. There is the possibility that the RPV parameters are so small that the LSP decay length is too long for it to decay within the detector. Then the LSP would be effectively stable within the detector. For certain cases, such effectively stable sparticles have been searched for in \cite{ATLAS:2014fka}. If the LSP decay length is small enough to decay within the detector, but greater than about 1 mm, this would lead to ``displaced'' vertices, such as those searched for in, for example, \cite{Aad:2015rba}. In the present paper, we will choose parameters so that the decay length of the LSP, whatever sparticle that may be, is less than about 1 mm. We refer to such decays as ``prompt'' decays. Therefore, even though the analysis in this work is valid for any mass chargino and neutralino, should we choose the initial conditions so that they are the LSP, then their RPV decays will be prompt.

As was shown in the case of stops and sbottoms in \cite{Marshall:2014cwa,Marshall:2014kea}, prompt decays require the RPV parameters to be large enough to allow for significant Majorana neutrino masses. We expect the same to hold true for a variety of LSPs.  Therefore, in this paper we focus on the case of significant Majorana neutrino masses.
Note that, in addition to these Majorana neutrino masses, there can be pure Dirac mass contributions coming from the neutrino Yukawa coupling. The components $Y_{\nu_{i3}}$, which couple the left-handed neutrinos to the third-family right-handed neutrino, allow the third-family right-handed neutrino to act as a seesaw field and give rise to Majorana neutrino masses. The other components, $Y_{\nu_{i1}}$ and $Y_{\nu_{i2}}$, couple the left-handed neutrinos to the first- and second-family right-handed neutrinos. Note that in this model, the heavy third-family right-handed neutrino acts as a seesaw field, while the first- and second-family right-handed neutrinos remain as light sterile neutrinos. This means that the Dirac mass terms related to $Y_{\nu_{i1}}$ and $Y_{\nu_{i2}}$ can give rise to active-sterile oscillations in the neutrino sector. There have been some experimental hints of such oscillations, see \cite{PDG} for review. However, it is not yet clear that these results are due to true active-sterile oscillations. Hence, we proceed under the assumption that no such oscillations exist and that the $Y_{\nu_i1}$ and $Y_{\nu_i2}$ components of the neutrino Yukawa coupling must, therefore, be negligible, so they do not appear in the neutralino mass matrix below. It may be interesting to revisit the question of active-sterile neutrino oscillations in the $B-L$ MSSM in the future, perhaps after there is more experimental data.

In the basis $\left(\tilde W_R, \ \tilde W^0, \ \tilde H_d^0, \ \tilde H_u^0, \ \tilde B', \ \nu_3^c, \ \nu_{i}\right)$ with $i=1,2,3$, the neutralino mass matrix is of the form
\begin{equation}
	\mathcal{M}_{\tilde\chi^0} =
	\begin{pmatrix}
		M_{\tilde\chi^0}
		&
		m_D
		\\
		m_D^T
		&
		0_{3 \times 3}
	\end{pmatrix},
	\label{eq:20}
\end{equation}
where $M_{\tilde\chi^0}$ is a six-by-six matrix of order a TeV given by
\begin{equation}
{\tiny
\label{eq:21}
	M_{\tilde\chi^0} =
	\begin{pmatrix}
			M_R
		&
			0
		&
			-\frac{1}{2} \, g_{R} \, v_d
		&
			\frac{1}{2} \, g_R \, v_u
		&
			0
		&
			-\frac{1}{2} g_R v_R
	\\
			0
		&
			M_2
		&
			\frac{1}{2} \, g_2 \, v_d
		&
			-\frac{1}{2} \, g_2 \, v_u
		&
			0
		&
			0
	\\
			-\frac{1}{2} \, g_{R} \, v_d
		&
			\frac{1}{2} \, g_{2} \, v_d
		&
			0
		&
			-\mu
		&
			0
		&
			0
	\\
			\frac{1}{2} \, g_R \, v_u
		&
			-\frac{1}{2} \, g_2 \, v_u
		&
			-\mu
		&
			0
		&
			0
		&
			0
	\\
			0
		&
			0
		&
			0
		&
			0
		&
			M_{BL}
		&
			\frac{1}{2} \, g_{BL} \, v_{R}
	\\
			-\frac{1}{2} g_R v_R
		&
			0
		&
			0
		&
			0
		&
			\frac{1}{2} \, g_{BL} \, v_{R}
		&
			0
	\\
	\end{pmatrix}, 
}
\end{equation}
and $m_D$ is a six-by-three matrix 
\begin{equation}
{\tiny
\label{eq:22}
        m_{D} =
	\begin{pmatrix}
	&
  	                0_{1 \times 3}
	\\
         &
			\frac{1}{2} \, g_2 \, {v_L}_i^*
	\\
        &
	         	0_{1 \times 3}
	\\
        &
			\epsilon_i
	\\
	&
			-\frac{1}{2} \, g_{BL} \, {v_L}_i^*
	\\
         &
			\frac{1}{\sqrt 2} \, {Y_\nu}_{i3} \, v_u
			
	\end{pmatrix}
}
\end{equation}
\noindent of order an MeV. This allows the mass matrix to be diagonalized perturbatively. Note that we have suppressed all terms of the form $v_{L_{i}}Y_{\nu ij}$ in $\mathcal{M}_{\tilde\chi^0}$ since both $v_{L_{i}}$ and the neutrino Yukawa parameters are small. In addition, we emphasize that since only the third family right-handed sneutrino gets a non-vanishing VEV, only $\nu_3^c$ couples to the gauginos/Higgsinos. It follows that only the Dirac mass of the third-family neutrino enters the above mass matrix, whereas the the first and second family Dirac neutrino masses are excluded.

The entire mass matrix $\mathcal{M}_{\tilde\chi^0}$ in \eqref{eq:20} can be diagonalized to
\begin{equation}
	\label{eq:23}
	\mathcal{M}_{\tilde\chi^0}^D = \mathcal{N}^* \mathcal{M}_{\tilde\chi^0} \mathcal{N}^\dagger
	\end{equation}
with
\begin{equation}
	\mathcal{N} =
	\begin{pmatrix}
		N & 0_{3 \times3}
		\\
		 0_{3\times3} & V_\pmns^\dagger
	\end{pmatrix}
	\begin{pmatrix}
		1_{6\times6} & -\xi_0
		\\
		\xi_0^\dagger & 1_{3 \times 3}     \label{eq:24}
	\end{pmatrix},
	\end{equation}
where $N$ is the matrix that diagonalizes $M_{\tilde\chi^0}$ given in eq. \eqref{eq:21}. Requiring that $\mathcal{M}^D_{{\tilde \chi}^0}$ be diagonal yields
\begin{equation}
\xi_0 = M_{\tilde\chi^0}^{-1} m_D.
\label{eq:25} 
\end{equation}
The second matrix on the right-hand side of $\mathcal{N}$ rotates away the neutralino/left-handed neutrino mixing, whereas the first matrix diagonalizes the six neutralino/third family right-handed neutrino states as well as the three left-chiral neutrino states. In this section, we will consider the diagonal $3 \times 3$ left-handed neutrino Majorana mass matrix only, returning to the diagonal neutralino mass matrix later in the paper. 

The diagonal left-chiral neutrino Majorana mass matrix is found to be
\begin{align}
\begin{split}
	\label{eq:26}
	{m_\nu^D}_{ij} & = \left(V_\pmns^T \, m_\nu \, V_\pmns\right)_{ij} \ .
\end{split}
\end{align}
The $3 \times 3$ matrix $m_{\nu}$ is given by \cite{Marshall:2014cwa}
\begin{equation}
	\label{eq:27}
	{m_\nu}_{ij} = A {v_L}_i^* {v_L}_j^* + B \left({v_L}_i^* \epsilon_j + \epsilon_i {v_L}_j^* \right) + C \epsilon_i \epsilon_j \ ,
\end{equation}
where
\begin{align}
\label{eq:28}
	A & = \frac
		{
			\mu \, M_{\tilde \gamma}
		}
		{
			2 \, M_{\tilde \gamma} v_u v_d - 4 M_2 M_{\tilde Y} \mu
		}
	\\
\label{eq:29}	
	B & = \frac
		{
			M_{\tilde \gamma} v_d \left( 2 M_{Z_R}^2+ g_{Z_R}^2 v_u^2\right) - 2 g_{Z_R}^2 g_{BL}^2 M_2 M_R \, \mu \, v_u
		}
		{
			4 M_{Z_R}^2 (M_{\tilde \gamma} v_u v_d - 2 M_{\tilde Y} M_2 \mu)
		}
	\\
	\begin{split}
	\label{eq:30}
	C & = 
		\Big(
				 2 g_{Z_R}^4 M_2 M_{BL} M_R \, \mu^2 v_u^2\\
&\quad\quad
				- g_{Z_R}^2 M_{BL} \mu
				\left(
					g_2^2 \, g_{Z_R}^2 M_R v_u^2
					+ g_R^2 M_2 \left(4 M_{Z_R}^2 + g_{Z_R}^2 v_u^2\right)
				\right) v_d v_u
		\Big)\\
		&\quad\quad/\Big(
			4 M_{Z_R}^4 \mu \left(2 M_{\tilde Y} M_2 \, \mu - M_{\tilde \gamma} v_d v_u\right)
		\Big)
		\\
		& \quad \quad 
		-\frac
		{
			M_{\tilde \gamma}  v_d^2
		}
		{
			2 \mu \left(2 M_{\tilde Y} M_2 \, \mu - M_{\tilde \gamma} v_d v_u\right)
		} 
	\end{split}
\end{align}
and 
\begin{equation}
	g_{Z_R}^2 \equiv g_{BL}^2+g_{R}^2 \ . \label{eq:31}
\end{equation}
As will be discussed in detail below, the soft mass parameters are all initialized statistically at the scale $M_{I}$, whereas the measured values of the gauge couplings are introduced at the electroweak scale. All of these parameters are then run to the appropriate energy scale using the RGEs discussed in detail in \cite{Ovrut:2015uea}. Additionally, the value of tan$\beta$ will be chosen statistically within a physically relevant interval and, for a given value of tan$\beta$, the parameters $v_{u}$ and $v_{d}$ are the measured Higgs VEVs. Finally, for any given set of statistical initial data, we fine-tune the value of the parameter $\mu$ using equation \eqref{eq:12}, so as to obtain the experimental value of the electroweak gauge boson $Z^{0}$ and, hence, the measured values for $W^{\pm}$ as well. 
 The $3 \times 3$ Pontecorvo-Maki-Nakagawa-Sakata matrix is
\begin{eqnarray}
	V_\pmns &=& 
	\begin{pmatrix}
		c_{12} c_{13}
		&
		s_{12} c_{13}
		&
		s_{13} e^{-i \delta}
		\\
		-s_{12} c_{23} - c_{12} s_{23} s_{13} e^{i \delta}
		&
		c_{12} c_{23} - s_{12} s_{23} s_{13} e^{i \delta}
		&
		c_{13} s_{23}
		\\
		s_{12} s_{23} - c_{12}  c_{23} s_{13} e^{i \delta}
		&
		-c_{12} s_{23} - s_{12}  c_{23} s_{13} e^{i \delta}
		&
		c_{13} c_{23}
	\end{pmatrix}\nonumber\\ &&\times \text{diag}(1, e^{i \mathcal{A}/2}, 1) \ , \label{eq:32}
\end{eqnarray}
with $c_{ab} (s_{ab}) = \cos \theta_{ab} (\sin \theta_{ab})$. The mixing angles and phases are determined by neutrino experiments. For the mixing angles, we use the values and uncertainties from \cite{Capozzi:2018ubv}. They are
\begin{equation}
\sin^2\theta_{12}=0.307\pm0.013~, \qquad  \sin^2\theta_{13}=(2.12\pm0.08)\times 10^{-2}
\label{eq:33}
\end{equation}
for both the normal and inverted neutrino mass hierarchies. For $\theta_{23}$, however, the best-fit values depend on the hierarchy, and the data admits multiple best-fit values. In the normal hierarchy, one finds
\begin{equation}
\sin^2\theta_{23}=0.417^{~+0.025}_{~-0.028}\qquad\mbox{or}\qquad0.597^{~+0.024}_{~-0.030}\ ,
\label{eq:normalHierarchy}
\end{equation}
while in the inverted hierarchy
\begin{equation}
\sin^2\theta_{23}=0.421^{~+0.033}_{~-0.025}\qquad\mbox{or}\qquad0.529^{~+0.023}_{~-0.030}\ .
\label{eq:invertedHierarchy}
\end{equation}
In this paper, we will do a complete study of  all four of the cases in equations \eqref{eq:normalHierarchy} and \eqref{eq:invertedHierarchy}.
Regarding the CP-violating phase, $\delta$, we use the recent results in \cite{Capozzi:2018ubv} that in the normal hierarchy
\begin{equation}
\delta = {231.6^{\circ}}^{~+41.4^{\circ}}_{~-30.6^{\circ}},
\end{equation}
while in the inverted hierarchy
\begin{equation}
\delta = {273.6^{\circ}}^{~+18.7^{\circ}}_{~-27.0^{\circ}}.
\end{equation}
In addition, note that there is only one ``Majorana'' phase, that is, parameter $\mathcal{A}$, since in both the normal and the inverted hierarchy one of the neutrinos is massless and, therefore, does not have a Majorana mass. The value of $\mathcal{A}$ is unknown and, hence, in this paper we will simply throw it statistically in the interval $[0^{\circ}, 360^{\circ}]$.

The mathematical expressions for the mass eigenvalues of the Majorana neutrino mass matrix $m^{D}_{\nu ij}$ can be constructed from the A,B,C components of $m_{\nu ij}$ given in \eqref{eq:28}, \eqref{eq:29} and \eqref{eq:30} respectively, as well as from the PMNS matrix given in \eqref{eq:32}. This has been done in detail in \cite{Marshall:2014cwa}, to which we refer the reader for details. Given values for all the relevant parameters discussed above, and the measured values for the neutrino mass eigenvalues for the normal and inverted hierarchies, this allows one to solve for the RPV parameters $\epsilon_{i}$, $v_{L_{i}}$ $i=1,2,3$. Respectively, the experimental values of the mass eigenvalues of the normal and inverted hierarchies are \cite{PDG}\\

\begin{itemize}
\item Normal Hierarchy:
\begin{equation}
	m_1 = 0, \quad m_2 = (8.68 \pm 0.10) \times 10^{-3} ~{\rm eV},\quad m_3 = (50.84 \pm 0.50) \times 10^{-3}~{\rm eV}
\label{b2}
\end{equation}
\item Inverted Hierarchy:
\begin{equation}
	m_1 = (49.84 \pm 0.40) \times 10^{-3} ~{\rm eV}, \quad m_2 = (50.01 \pm 0.40) \times 10^{-3} ~{\rm eV},\quad m_3 = 0.
\label{b3}
\end{equation}
\end{itemize}

In each case, all three $v_{L}$ parameters as well as two of the $\epsilon$ parameters can be determined in terms of a third $\epsilon$ parameter. The explicit expressions, of course, differ in the normal and inverted hierarchy cases, and are presented in detail in \cite{Marshall:2014cwa}. These are encoded into the computer program by which we determine all decay rates and branching ratios and won't be presented here. Suffice it to say that, in each case, which parameter $\epsilon_i$ is inputted is undetermined. Thus, we will statistically decide which of the three dimension one $\epsilon_{i}$ parameters is selected. Furthermore, we choose its value by randomly throwing it to be in the interval $[10^{-4}, 1.0]$ GeV with a log-uniform distribution. We limit the upper bound to $1.0$ GeV to avoid excessive fine-tuning in the neutrino masses. Furthermore, we cut off the lower bound at $10^{-4}$ GeV-- although this could be taken to $0$ --to enhance the readability of our branching ratio plots. It is important to note that, having statistically chosen one of the $\epsilon$ parameters in the above range, the other two $\epsilon$ parameters are determined by the computer code and are not necessarily bounded by this interval. For example, at least one of the remaining two $\epsilon$ parameters could, in principle, be considerably larger than $1.0$ GeV. 
If so, this could have important consequences for for the suppression of lepton number violating interactions. The reason is the following.

It was shown in \cite{Barbier:2004ez}, and discussed in \cite{Marshall:2014cwa}, that the experimental bound on the decay of $\mu \longrightarrow e\gamma$ leads to the constraint on $\epsilon_{1}$ and $\epsilon_{2}$ that
\begin{equation}
\big|\frac{\epsilon_{1}\epsilon_{2}}{\mu^{2}} \big| \lesssim 2.5 \times 10^{-3}\big(\frac{m_{{\tilde{\nu}}^{c}_{3}}}{100~{\rm GeV}} \big)^{-2} \ .
\label{burta}
\end{equation}
Scanning the initial parameters in the $B-L$ MSSM, using \eqref{eq:9}, \eqref{eq:15} and the values for the gauge parameters discussed in \cite{Ovrut:2012wg}, we find that this becomes
\begin{equation}
\epsilon_{1}\epsilon_{2} \lesssim 68~ {\rm GeV}^{2} \ .
\label{burtb}
\end{equation}
Therefore, to adequately suppress lepton number violating decays, it is essential to show that this bound is satisfied for any physically interesting set of initial data in this analysis. At the end of Section 4, we will demonstrate that for the LSPs of interest in this and in the follow-up papers, that is, for charginos and neutralinos, constraint \eqref{burtb} is easily satisfied.


\section{Physically Acceptable Vacua}
\label{sec:4}

Having stated the structure of the $B-L$ MSSM, and defined all of the associated parameters, we now use the computer code specified in detail in \cite{Ovrut:2015uea} to find the initial values of the parameters leading to completely acceptable physical vacua. The choice of initial parameters will be subject to all the constraints listed previously in Sections 2 and 3. For example, as discussed in Section \ref{sec:2}, we will assume that the ratio of the Wilson line masses $M_{\chi_{B-L}}$ and $M_{\chi_{3R}}$ is such that all gauge parameters unify at $M_{U}$, that all quark and charged lepton Yukawa couplings can be taken to be nearly diagonal and real and, as discussed in Section \ref{sec:3}, only the $Y_{\nu i3}$ components of the neutrino Yukawa couplings are non-negligible. In addition, we solve for the physically acceptable initial conditions subject to several new, and important, constraints. These are the following:

\begin{itemize}

\item Scattering Range of the Dimensionful Soft SUSY Breaking Parameters:\\
\begin{equation}
\big[~\frac{M}{f},Mf~\big] \quad {\rm where}~~~M=1.5~{\rm TeV}~, f=6.7  \ .
\label{eq:36}
\end{equation}
\end{itemize}
The median supersymmetry breaking mass $M$ and the parameter $f$ are chosen so that the range of dimensionful soft masses can have random values from just above the electroweak scale to a scale approaching the upper bound of what will be observable at the LHC. That is, range (\ref{eq:36}) is approximately $[200$ GeV$\>, 10$ TeV$]$.

\begin{itemize}
\item Random Sign of the Soft SUSY Breaking Parameters $\mu$, $M$ and $a$:
\begin{equation}
\big[-,+\big ] \ .
\label{eq:37}
\end{equation}
\end{itemize}
The sign of $\mu$ and the various soft parameters of the form $M$ and $a$ are chosen randomly to have either a + or - sign. \\

\begin{itemize}
\item Randomly Scattered Choice of ${\rm tan}\beta$:
\begin{equation}
{\rm tan}\beta \in [1.2~,~65] \ .
\label{eq:38}
\end{equation}
\end{itemize}
The upper and lower bounds for ${\rm tan} \beta$ are taken from \cite{Martin:1997ns} and are consistent with present bounds that ensure perturbative Yukawa couplings.\\

In addition to being subject to the above constraints, physically acceptable initial conditions are those which lead to the following phenomenological results. First, $B-L$ gauge symmetry must be spontaneously broken at a sufficiently high scale. Presently, the measured lower bound on the $Z_{R}$ mass is given by \cite{Aaboud:2017buh}\\
\begin{equation}
M_{Z_{R}} = 4.1 ~{\rm TeV} \ .  \label{eq:39}
\end{equation}

\noindent Secondly, electroweak (EW) symmetry must be spontaneously broken so that the $Z^{0}$ and $W^{\pm}$ masses have the measured values of \cite{PDG}

\begin{equation}
\quad M_{{Z^{0}}}= 91.1876 \pm 0.0021~{\rm GeV} , \quad M_{W^{\pm}}= 80.379 \pm0.012~{\rm GeV} \ .  \label{eq:40}
\end{equation}
\\
\noindent Third, the remaining sparticles must be above their measured lower bounds \cite{Ovrut:2015uea} given in Table \ref{tab:lower_bounds}.\\

\begin{table}[H]

\begin{center}

\begin{tabular}{ |c|c| }

\hline

SUSY Particle & Lower Bound \\

\hline

Left-handed sneutrinos & 45.6 GeV\\

Charginos, sleptons& 100 GeV \\

Squarks, except stop or bottom LSP& 1000 GeV \\

Stop LSP (admixture)& 550 GeV \\

Stop LSP (right-handed)& 400 GeV\\

Sbottom LSP& 500 GeV\\

Gluino& 1300 GeV\\

\hline

\end{tabular}

\caption{Current lower bounds on the SUSY particle masses.}

\label{tab:lower_bounds}

\end{center}

\end{table}
\noindent Finally, the Higgs mass must be within the $3\sigma$ allowed range from ATLAS combined run 1 and run 2 results \cite{Aaboud:2018wps}. This is found to be
\begin{equation}
M_{h^{0}}=124.97 \pm 0.72~ {\rm GeV} \ .
\label{eq:41}
\end{equation}
\\
\indent We now want to search for physically acceptable initial data, subject to all of the constraints and phenomenological conditions introduced above. Before applying any of these constraints, the number of parameters appearing in the $B-L$ MSSM greatly exceeds 100. However, subject to the constraints discussed above, this number is significantly reduced-- down to only 24 soft SUSY breaking parameters, as well as ${\rm tan} \beta$ and $\mu$. The RG code \cite{Ovrut:2015uea} that we use in this analysis involves all 24 SUSY breaking parameters.  It is, 
however, helpful to point out that many of the RGE's are dominated by two specific sums of these parameters given by
\begin{eqnarray}
	\label{eq:S.BL}
	&S_{BL'}=\Tr(2m_{\tilde Q}^2-m_{\tilde u^c}^2-m_{\tilde d^c}^2-2m_{\tilde L}^2+m_{\tilde \nu^c}^2+m_{\tilde e^c}^2) \label{eq:42} \\
	\label{eq:S.R}
	&\qquad \quad S_{3R}=m_{H_u}^2-m_{H_d}^2+\Tr\left(-\frac{3}{2}m_{\tilde u^c}^2+\frac{3}{2}m_{\tilde d^c}^2-\frac{1}{2} m_{\tilde \nu^c}^2+\frac{1}{2} m_{\tilde e^c}^2\right) \label{eq:43}\ ,
\end{eqnarray}
where the traces are over generational indices. This is helpful in that one can now reasonably plot initial data points in two-dimensional $S_{BL^{\prime}}- S_{3R}$ space, rather than in the full 24-dimensional space of all parameters. At the electroweak scale, we randomly set the value of ${\rm tan}\beta$. Furthermore, and importantly, we do not run the parameter $\mu$. Rather, after running all other parameters down to the electroweak scale, we fine-tune $\mu$ to give the measured values for the electroweak gauge bosons, as discussed above.

\begin{figure}[h]

\centering

\begin{subfigure}[b]{0.7\textwidth}

\includegraphics[width=1.\textwidth]{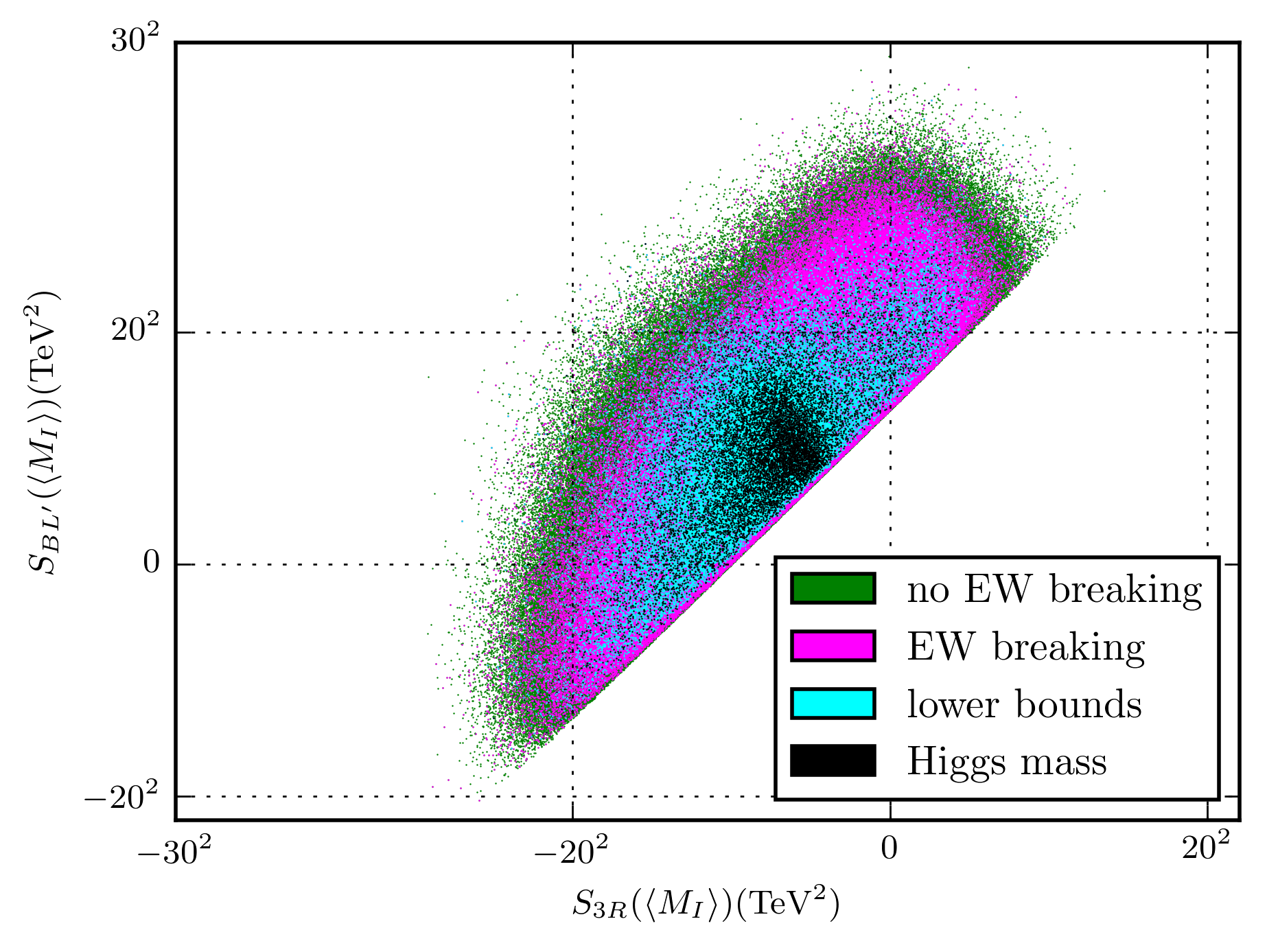}

\end{subfigure}

\caption{Plot of the 100 million initial data points for the RG analysis evaluated at $M_{I}$ . The 4,351,809 green points lead to appropriate breaking of the $B-L$ symmetry. Of these, the 3,142,657 purple points also break the EW symmetry with the correct vector boson masses. The cyan points correspond to 342,236 initial points that, in addition to appropriate $B-L$ and EW breaking, also satisfy all lower bounds on the sparticle masses. Finally, as a subset of these 342,236 initial points, there are 67,576 valid black points which lead to the experimentally measured value of the Higgs boson mass.}\label{fig:ScatterPlot}
\label{fig:eye}
\end{figure}

Searching for physically acceptable initial data, subject to all of the constraints and phenomenological conditions above, we find the following. For 100 million sets of randomly scattered initial conditions, it is found that 4,351,809 break $B-L$ symmetry with the $Z_{R}$ mass above the lower bound in equation \eqref{eq:39}. These are plotted as the green points in Figure \ref{fig:eye}. Running the RG down to the EW scale, one finds that of these 4,351,809 appropriate $B-L$ initial points, only 3,142,657 break electroweak symmetry with the experimentally measured values for $M_{Z^{0}}$ and $M_{W^{\pm}}$ given in equation \eqref{eq:40}. These are shown as the purple points in the Figure. Now applying the constraints that all sparticle masses be at or above their currently measured lower bounds presented in Table 1, we find that of these 3,142,657 initial points, only 342,236 are acceptable. These are indicated by cyan colored points in the Figure. Finally, it turns out that of these 342,236 points, only 67,576 also lead to the currently measured Higgs mass given in equation \eqref{eq:41}. That is, of the 100 million sets of randomly scattered initial conditions, 67,576 satisfy all present phenomenological requirements. In Figure \ref{fig:eye}, we represent these ``valid'' points in black. 
That is, of the 100 million randomly scattered initial points, approximately .067\% satisfy all present experimental conditions. Although this might-- at first sight --appear to be a small percentage, it is worth noting that these initial points not only break $B-L$ symmetry appropriately and have all sparticle masses above their present experimental lower bounds, but also give the measured experimental values for the mass of the EW gauge bosons and, remarkably, the Higgs boson mass as well! From this point of view, this percentage of valid black points seems remarkably high. The electroweak gauge boson masses were obtained, as discussed above, by fine-tuning the parameter $\mu$. For example, a typical value of the fine-tuning of $\mu$ is of the order of 1 in 1000 \cite{Ovrut:2014rba}. However, one might also be concerned that getting the Higgs mass correct might require some other fine-tuning of the 24 initial parameters that may not be apparent. However, in previous work \cite{Ovrut:2015uea} it was shown that the 24 parameters associated with any given black point are generically widely disparate with no apparent other fine-tuning.
%

We conclude that the $B-L$ MSSM, in addition to arising as a vacuum of heterotic M-theory and having exactly the mass spectrum of the MSSM, satisfies all present experimental low-energy physical bounds for a remarkably large number of disparate initial data points. Given this, it becomes of real interest to determine whether the RPV decays of the $B-L$ MSSM can be directly observed at the LHC at CERN. These decays are most easily observed in the lightest sparticles in the mass spectrum; that is, the LSP has the best prospects for RPV detection in general. There are, however, cases in which the next lightest supersymmetric particle (NLSP) is highly degenerate in mass with the LSP-- see examples presented in \cite{Ovrut:2015uea} --and, hence, their RPV decay channels become relevant as well. 
Hence, in the sections to follow, we compute the decays of charginos and neutralinos without making any assumptions regarding their masses.
As discussed in detail in \cite{Ovrut:2015uea}, the particle spectrum of each of the 67,576 valid black points is exactly determined by the computer code.  It follows that we can compute the LSP associated with each valid black point. It turns out that there are many possible different LSPs. Before enumerating these, however, we must be more specific about the definition and structure of any LSP. Although the original fields entering the $B-L$ MSSM Lagrangian are ``gauge'' eigenstates, the LSP associated with a given valid black point  is, by definition, a ``mass'' eigenstate-- generically a linear combination of the original fields. For example, as will be discussed in detail in subsection 5.1, the lightest mass eigenstate chargino of either charge, which we denote by $\tilde \chi_{1}^{\pm}$, is found to be an $R$-parity conserving linear combination of the charged Wino, ${\tilde{W}}^{\pm}$, and the charged Higgsino, ${\tilde{H}}^{\pm}$, added to  RPV terms proportional to the left and right chiral charged leptons. As discussed in subsection 5.1, the RPV coefficients are very small and, hence, can be ignored in the discussion of the masses of the charginos. Therefore, in this section, since we are analyzing the possible LSPs, we will consider the $R$-parity conserving part of the chargino states only. It then follows from the discussion in subsection 5.1 that when  $M_2<|\mu|$ the lightest chargino is given by
\begin{equation}
\tilde \chi^{\pm}_{1}=\cos \phi_{\pm}{\tilde{W}}^{\pm} + \sin \phi_{\pm}{\tilde{H}}^{\pm} \ ,
\label{cham2}
\end{equation}
whereas for $|\mu|<|M_2|$
\begin{equation}
\tilde \chi^{\pm}_{1}=-\sin \phi_{\pm}{\tilde{W}}^{\pm} + \cos \phi_{\pm}{\tilde{H}}^{\pm} \ .
\label{chamonix}
\end{equation}
The angles $\phi_{\pm}$ are exactly determined for any given black point. It follows that for some black points the mass eigenstate $\tilde \chi^{\pm}_{1}$ is predominantly a charged Wino, whereas for other black points it is mainly a charged Higgsino. We will, henceforth, denote the first and second type of mass eigenstates by $\tilde \chi^{\pm}_{W}$ and $\tilde \chi^{\pm}_{H}$, and refer to them as ``Wino charginos'' and ``Higgsino charginos'' respectively. That is, instead of labelling a chargino LSP simply as $\chi^{\pm}_{1}$, and counting the number of valid black points associated with it, we can be more specific-- breaking the chargino LSP into two different types of states, $\tilde \chi^{\pm}_{W}$ and $\tilde \chi^{\pm}_{H}$ respectively, and counting the number of black points associated with each type individually. This gives additional information about the structure of the LSPs.

\begin{figure}[H]

\centering

\begin{subfigure}[b]{0.7\textwidth}
\includegraphics[width=1.\textwidth]{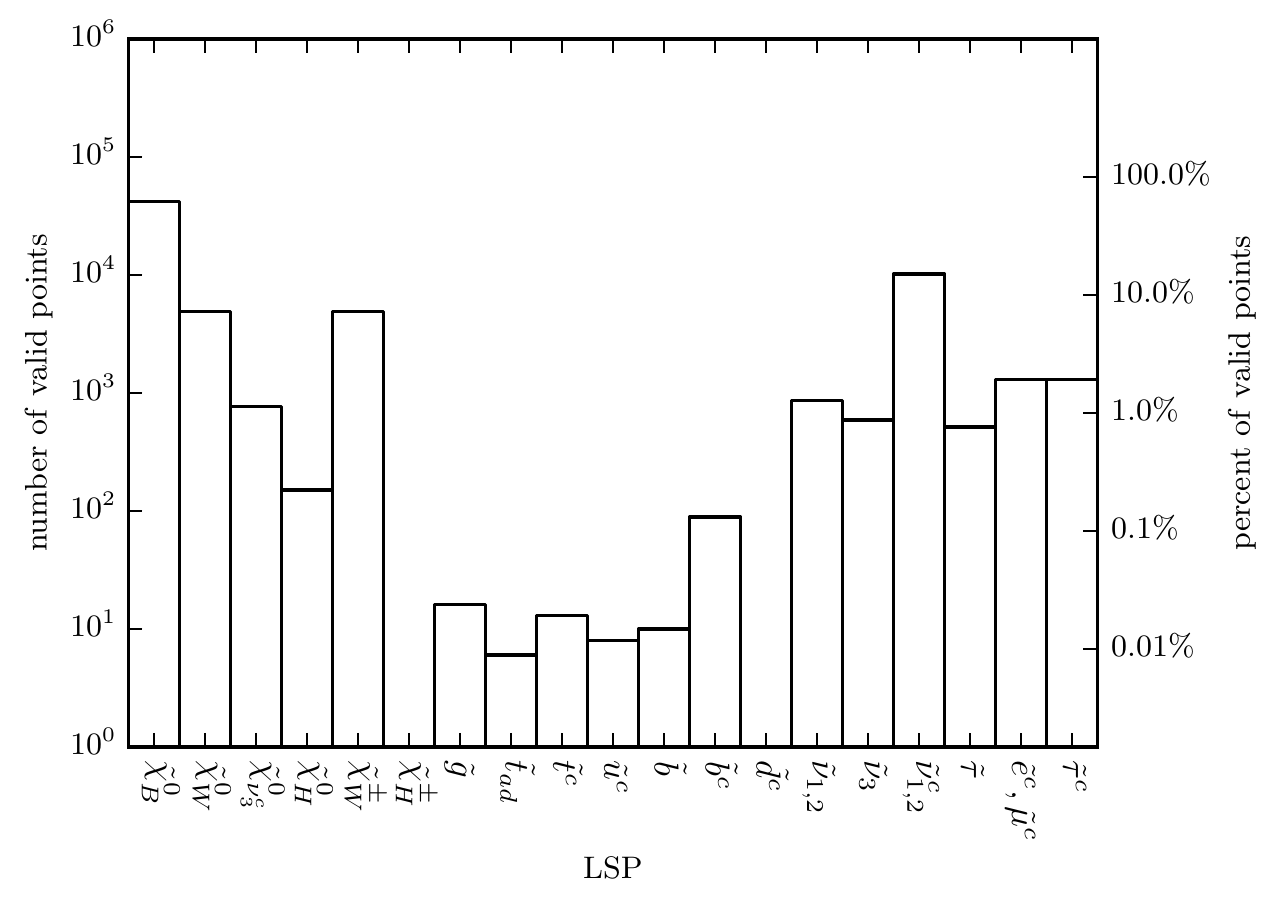}
\end{subfigure}
\caption{A histogram of the LSPs associated with a random scan of 100 million initial data points, showing the percentage of valid black points with a given LSP. Sparticles which did not appear as LSPs are omitted. The y-axis has a log scale. The notation and discussion of the sparticle symbols on the x-axis is presented in Table 2.}
\label{fig:lspHistogram}
\end{figure}

With this in mind, we have calculated the LSP associated with each of the 67,576 valid black points and plotted the results as a histogram in Figure \ref{fig:lspHistogram}. The notation for the various possible LSPs is specified in Table \ref{tab:notation}. 
For example, out of the 67,576 valid black points, there are 4,858 that have a ${\tilde{\chi}}_{W}^{\pm}$ Wino chargino as their LSP. Similarly, out of all the valid black point initial conditions, 4,869 have a  ${\tilde{\chi}}_{W}^{0}$ Wino neutralino as their LSP. And so on.
Notice that the cases in which the chargino LSP is dominantly a charged Higgsino-- that is, $\tilde \chi^\pm_H$ --are rare. In fact, in Figure \ref{fig:lspHistogram} there is precisely one such black point. As discussed above and shown in Section \ref{sec:5}, the lighter chargino state is dominantly Wino if $|M_2|<|\mu|$, and dominantly Higgsino if $|\mu|<|M_2|$. The little hierarchy problem tells us that $\mu$ is generally large, of the order of a few TeV. However, the $M_2$ parameter generally takes smaller values in our simulation. For this reason, the instances in which $|\mu|<|M_2|$-- required for the Higgsino chargino to be the LSP --are scarce. 

%

\begin{figure}[	h]
   \centering

   \begin{subfigure}[b]{0.39\textwidth}
 \centering
\includegraphics[width=1.0\textwidth]{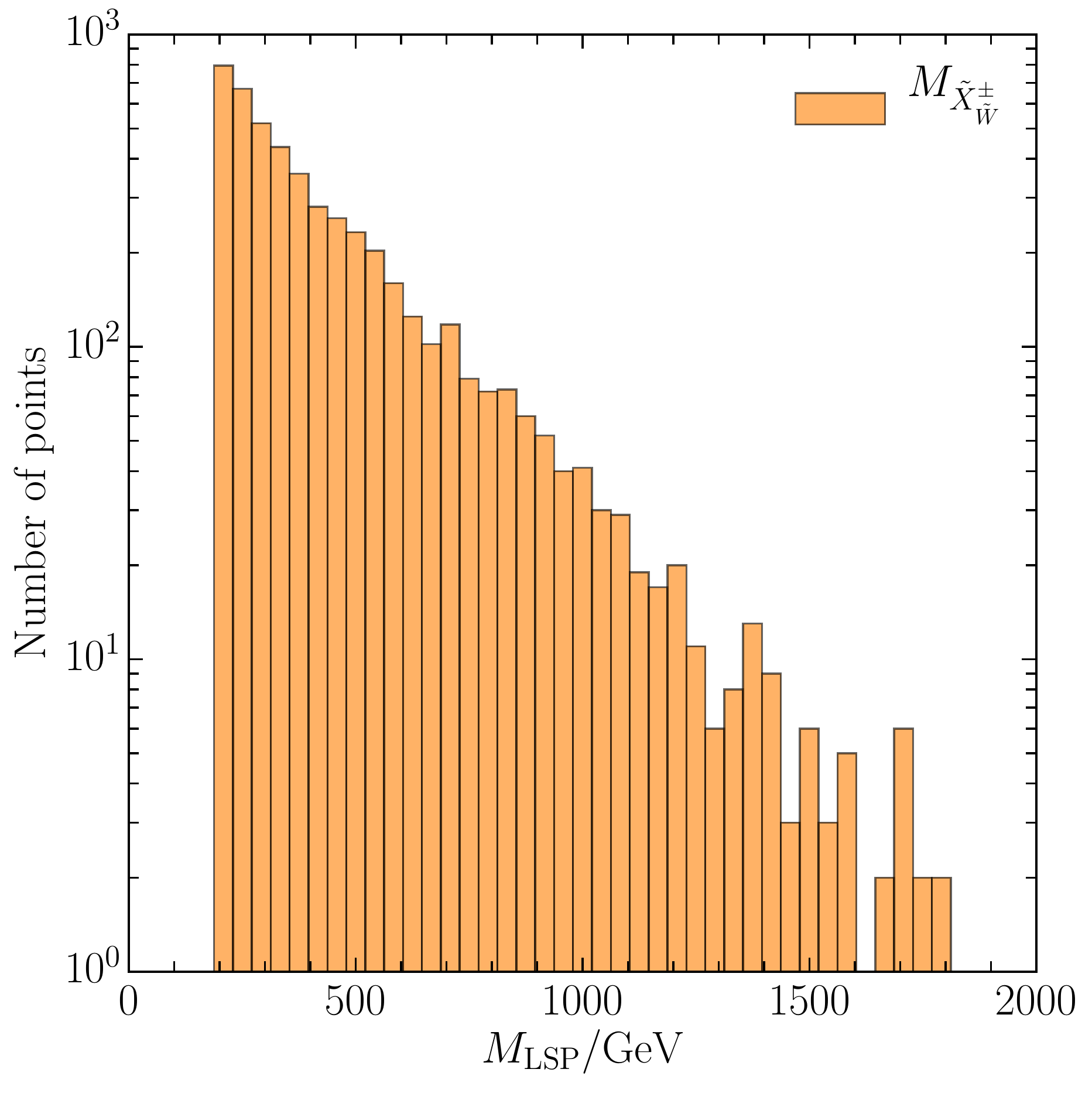}
\caption{}
\end{subfigure}
   \begin{subfigure}[b]{0.39\textwidth}
 \centering
\includegraphics[width=1.0\textwidth]{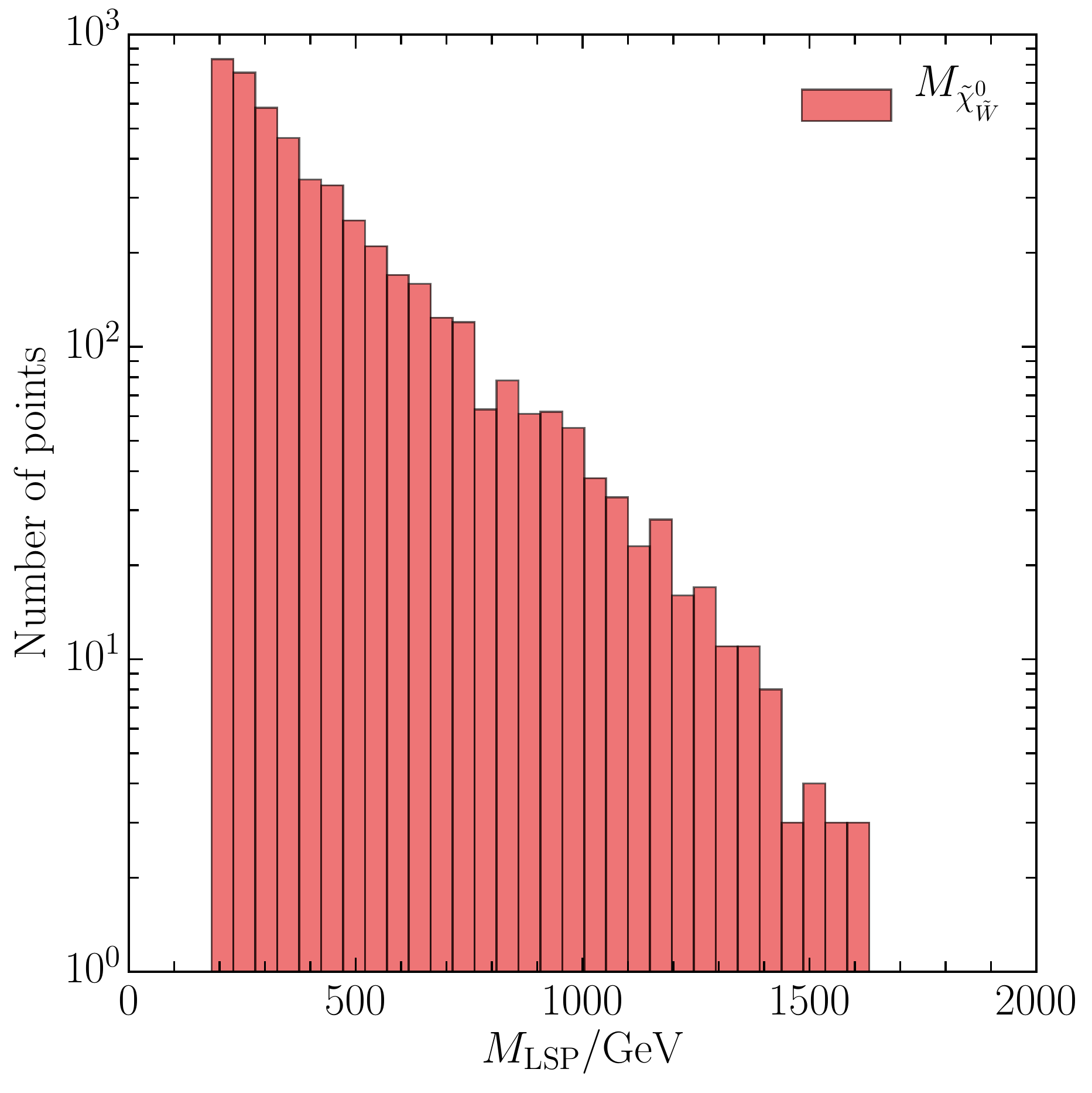}
\caption{}
\end{subfigure}
 \caption{(a) Mass distribution of the Wino chargino LSPs. The masses range is from 194 GeV to 1710 GeV, peaking towards the low mass end. (b) Mass distribution of the Wino neutralino LSPs. The masses range is from 192 GeV to 1630 GeV, peaking towards the low mass end.  }\label{fig:mass_hist}
\end{figure}

For any given choice of LSP, we can plot the number of such points as a function of their masses in GeV. As an example, Figures 4 (a) and (b) present such a mass distribution for Wino chargino and Wino neutralino LSPs respectively. We obtain viable supersymmetric spectra with Wino chargino and Wino neutralino LSP masses ranging from about 200 GeV to 1700 GeV. A striking feature of the Wino chargino and Wino neutralino LSP mass distributions in Figure \ref{fig:mass_hist} is the peak towards the low mass values. Higher LSP masses are exponentially less probable. The reason is that we sample all soft mass terms log-uniformly in the interval $[200$ GeV,  $10$ TeV]. This includes the $M_2$ gaugino mass term, which gives the dominant contribution for both the Wino chargino and Wino neutralino masses, see \eqref{eq:Wino_mass} and \eqref{eq:Wino_Neutralino_mass} respectively. If we would plot all the Wino chargino or Wino neutralino masses for all the viable points in our simulation, we would obtain an almost uniform mass distribution. However, for the Wino charginos or Wino neutralinos to be the LSPs, their masses must be lower than all the other random soft masses in our simulation. Conversely, it demands that all the other random soft mass terms be larger than a Wino chargino or Wino neutralino mass value for each viable point. This is exponentially less likely as this mass value increases, following a Boltzmann distribution. We point out that this discussion is a simplification of what actually happens, since it omits the running of the soft mass terms, as well as their mixing in the final mass eigenstates. These details, however, do not effect the essence of the above argument, since the mass runnings and the mass mixing couplings are generically very small.

\begin{figure}[h]
\centering
\begin{subfigure}[b]{0.4\textwidth}
\centering
\includegraphics[width=1.0\textwidth]{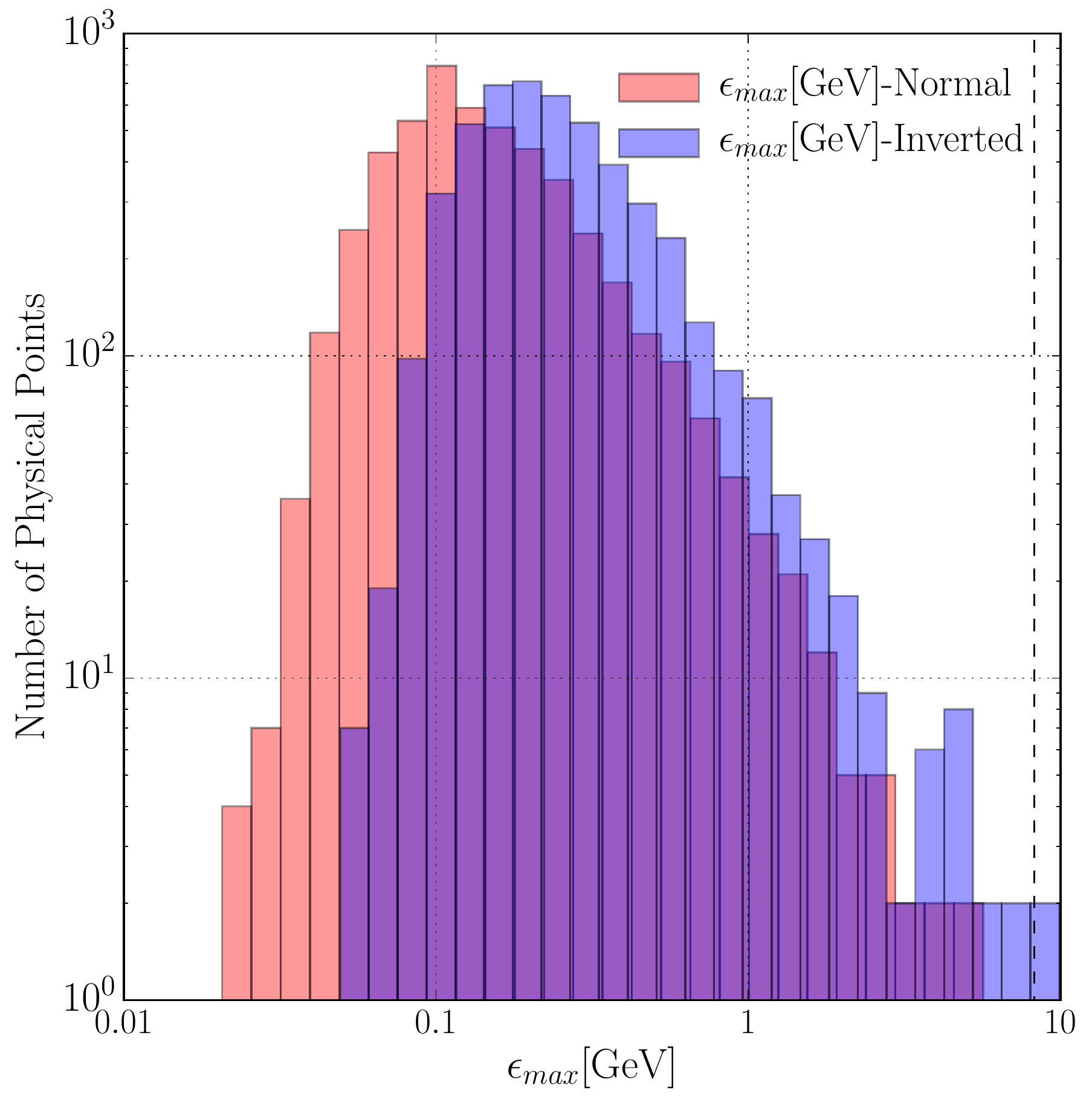}
\end{subfigure}
\caption{For each of the 4,858 black points with a Wino chargino LSP, we statistically sample one of the three $\epsilon_i$ parmeters in the range $[10^{-4}, 1.0]$ GeV and, for this graph, choose 1) a normal neutrino hierarchy and $\sin \theta_{23}=0.597$-- indicated in red --and 2) an inverted hierarchy with $\sin \theta_{23}=0.529$-- indicated in blue. We then solve for the other two $\epsilon$ values numerically. For each case, we plot a histogram of the number of valid black points associated with a Wino chargino LSP against the value of $\epsilon_{\text{max}}$, the largest of the three epsilon parameters. We find that most values of $\epsilon_{\text{max}}$ are smaller than 1 GeV. For larger values, the viable points become much less numerous, since such values would require more fine-tuning to match the existent neutrino data. The bound $\sqrt{68}$ GeV-- see the end of Section 3-- beyond which unphysical lepton number violation is possible, is indicated by the dashed line. We find only 1  and 4 points beyond this line for the normal and inverted hierarchy cases respectively. Hence, lepton number violation, if it occurs at all, is statistically insignificant in our simulation. }
\label{fig:hope}
\end{figure}

Associated with a given choice of LSP, there are a fixed number of valid initial points. For example, as mentioned above, a Wino chargino LSP arises from 4,858 black points. As discussed at the end of Section 3, for each such black point, we 1) statistically throw one of the parameters $\epsilon_{i}, i=1,2,3$ in the interval $[10^{-4},1.0]$ GeV, 2) choose the neutrino mass hierarchy to be either normal or inverted and, having done so, choose the associated value of $\theta_{23}$, 3) then, using \eqref{eq:27}, determine the remaining two epsilon parameters and the three $v_{L}$ parameters using the computer code. Let us denote the maximum one of the three $\epsilon$ parameters by $\epsilon_{{\rm max}}$. By running over all 4,858 black points subject to a fixed choice of the neutrino hierarchy and $\theta_{23}$, one can create a histogram of the 
number of valid points associated with a given value for $\epsilon_{{\rm max}}$. For example, the results of, first, choosing a normal neutrino hierarchy and $\theta_{23}$ such that $\sin \theta_{23}=0.597$ and, second, choosing an inverted neutrino hierarchy and $\theta_{23}$ with $\sin \theta_{23}=0.529$ are graphically depicted in Figure \ref{fig:hope}.
\noindent We find only a statistically insignificant number of points, in the case of the normal hierarchy 1 point and in the case of the inverted hierarchy 4 points, that exceed $\sqrt{68}$ GeV. If $\epsilon_{{\rm max}}=\epsilon_{3}$, then  constraint \eqref{burtb} is immediately satisfied. Even if this parameter is, say, $\epsilon_{2}$, it remains statistically extremely likely that constraint \eqref{burtb} remains satisfied. It follows that, for the choice of a normal neutrino hierarchy and  $\sin \theta_{23}=0.597$ and an inverted hierarchy with $\sin \theta_{23}=0.529$, lepton number violation via $\mu \rightarrow e\gamma$ is statistically highly  suppressed in our theory. We find  similar results for each of the other two choices of $\theta_{23}$. The identical conclusion can be drawn for  the Wino neutralino. We conclude that lepton number violation is highly suppressed in the $B-L$ MSSM-- at least when the LSP is a Wino chargino or a Wino neutralino.

In previous papers \cite{Marshall:2014kea,Marshall:2014cwa}, we have analyzed in detail the RPV decays of the ``admixture'' stop. Stops have a very high production cross section from proton-proton collisions. Furthermore, their decay products are relatively easy to observe at the LHC detectors. For these reasons, the ATLAS group at the LHC did a detailed study of the RPV decays of the admixture stop LSPs \cite{Aaboud:2018wps,ATLAS:2015jla,Jackson:2015lmj,ATLAS:2017hbw}. However, it is clear from Figure \ref{fig:lspHistogram} that neutralinos and charginos are much more prevalent as LSPs of the $B-L$ MSSM. Therefore, in the present paper, we begin a study of the RPV decays of neutralinos and charginos. Their RPV decay channels will be analyzed in detail.

\begin{table}
\begin{center}
\begin{tabular}{|c|c|}
\hline
\ Symbol \ & Description
\\
\hline
\hline
$\tilde \chi^0_{B}$ & Mostly a neutral Bino.
\\
$\tilde \chi^0_{W}$ & Mostly a neutral Wino.
\\
${\tilde{\chi}}^{0}_{\nu^{c}_{3a,b}}$ & Mostly third generation right-handed neutrino.
\\
$\tilde \chi^0_{H}$ & Mostly a neutral Higgsino.
\\
\hline
$\tilde \chi^{\pm}_{W}$ & Mostly a charged Wino.
\\
$\tilde \chi^{\pm}_{H}$ & Mostly a charged Higgsino. 
\\
\hline
$\tilde g$ & Gluino. 
\\
\hline
 $\tilde t_{ad}$ & Left- and right-handed stop admixture. 
\\
\hline
$\tilde t_{r}$ & Mostly right-handed stop (over 99\%). 
\\
\hline
$\tilde q^{c}$ & Right-handed 1st and 2nd generation squarks. 
\\
\hline
$\tilde b$ & Mostly left-handed sbottom. 
\\
\hline
$\tilde b^{c}$ & Mostly right-handed sbottom. 
\\
\hline
$\tilde \nu_{{1,2}}$ & 1st and 2nd generation left-handed sneutrinos.				
\\
&LSPs evenly split among two generations.
\\
\hline
$\tilde \nu_{3}$ & Third generation left-handed sneutrino.
\\
\hline
$\tilde \nu^c_{{1,2}}$ & 1st and 2nd generation right-handed sneutrinos.
\\
\hline
{$\tilde \tau$} &  {Third generation left-handed stau.}
\\
\hline
{$\tilde e^{c}, \tilde \mu^{c}$} &  1st and 2nd generation right-handed sleptons. 
\\
	& LSPs evenly split among two generations.	
\\
\hline
$\tilde \tau^{c}$ &  Third generation right-handed stau. 
\\
\hline
\end{tabular}
\end{center}
\caption{The notation used for the LSP states on the x-axis of Figure 3.}
\label{tab:notation}
\end{table}
\section{Chargino and Neutralino states}\label{Chargino and Neutralino states}
\label{sec:5}
\subsection{Chargino mass eigenstates}\label{Chargino_masses}

After EW breaking, the Higgs fields acquire a VEV which induces off-diagonal couplings 
between the charged gauginos of the theory. The terms that enter the chargino mass matrix, {\it in the
absence of RPV effects}, are
\begin{equation}
\mathcal{L}\supset \frac{-g_2}{\sqrt{2}}[v_u\tilde W^-\tilde H_u^+ +v_d\tilde W^+H_d^-]-
M_2|\tilde W|^2-\mu \tilde H_u^+\tilde H_d^-+h.c.
\end{equation}
The first terms come from the supercovariant derivative of the Higgs chiral fields, the Wino mass term originates in the soft SUSY breaking Lagrangian, while the last term is introduced in the superpotential $W$.
Combining the charged Higgsinos and the charged Winos into $\psi^+=(\tilde W^+,\> \tilde H_u^+)$ and 
$\psi^-=(\tilde W^-,\> H_d^-)$, we can write the previous terms in the form
\begin{equation}
\mathcal{L}\supset -\frac{1}{2}\left(\psi^+ \> \psi^-\right)
\left(\begin{matrix}
0&{M_{\tilde \chi^\pm}}^T\\
{M_{\tilde \chi^\pm}}&0
\end{matrix}
\right)
\left(\begin{matrix}\psi^+\\ \psi^-\end{matrix}\right)
+h.c.
\label{eq:generalCharginoMatrix}
\end{equation}
where $M_{\tilde \chi^\pm}$ is the $2 \times 2$ matrix given by
\begin{equation}
{M_{\tilde \chi^\pm}}=\left(\begin{matrix}
M_2&\frac{1}{\sqrt{2}}g_2 v_u\\
\frac{1}{\sqrt{2}}g_2v_d&\mu 
\end{matrix}
\right).
\end{equation}
The mass eigenstates ${\tilde \chi}^+=V\psi^+$ and ${\tilde \chi}^-=U\psi^-$ diagonalize $M_{\tilde \chi^\pm}$ to
\begin{equation}
U^*M_{\tilde \chi^\pm}V^{-1}=M^D=
\left(\begin{matrix}M_{{\tilde \chi}^\pm_1}&0\\0&M_{{\tilde \chi}^\pm_2}\end{matrix}\right)
\end{equation}
with $M_{{\tilde \chi}^\pm_1}$ and $M_{{\tilde \chi}^\pm_2}$ positive.
One can solve analytically for the eigenvalues and obtain
\begin{multline}
M^2_{{\tilde \chi}^\pm_1}, M^2_{{\tilde \chi}^\pm_2}=
\frac{1}{2} \Big[|M_2|^2+|\mu|^2+2M^2_{W^\pm} \\
\mp \sqrt{ \left(|M_2|^2+|\mu|^2+2M_{W^\pm}^2 \right)^2-4|\mu M_2-M^2_{W^\pm} \sin{2 \beta}|^2} \>\Big] \ ,
\end{multline}
where $M^2_{{\tilde \chi}^\pm_1}$ and $M^2_{{\tilde \chi}^\pm_2}$ correspond to the $-$ and $+$ sign in front of the square root respectively.
We will always choose the square root to be positive, so that the "minus" sign-- and, hence, ${\tilde{\chi}}^{\pm}_1$ --corresponds to the lighter mass eigenstate. 
That is, with this convention $M_{{\tilde \chi}^\pm_1}<M_{{\tilde \chi}^\pm_2}$. The expressions for the mass eigenvalues can be simplified by noting that the lower bounds on sparticle masses are well above $M_{W^\pm}$. It follows that $M^2_{W^\pm}\ll M^2_2, \mu^2$. Therefore, the mass eigenvalues depend primarily on the parameters $M_{2}$ and $\mu$.
When $|M_{2}| \lesssim |\mu|$, we find that
%
\begin{equation}\label{eq:Wino_mass}
M_{{\tilde \chi}^\pm_1} \simeq|M_2|-\frac{M^2_{W^\pm}(M_2+\mu \sin 2\beta)}{\mu^2-M_2^2},
\end{equation}
\begin{equation}
M_{{\tilde \chi}^\pm_2} \simeq|\mu|+\frac{\text{sgn}(\mu)M^2_{W^\pm}(\mu+M_2 \sin 2\beta)}{\mu^2-M_2^2}
\label{eq:Higgsino_mass}
\end{equation}
whereas for $|\mu| \lesssim |M_{2}|$, the expressions for the mass eigenvalues are simply exchanged; that is, 
\begin{equation}
M_{{\tilde \chi}^\pm_1} \simeq |\mu|+\frac{\text{sgn}(\mu)M^2_{W^\pm}(\mu+M_2 \sin 2\beta)}{\mu^2-M_2^2} \ ,
\label{again1}
\end{equation}
\begin{equation}
M_{{\tilde \chi}^\pm_2} \simeq |M_2|-\frac{M^2_{W^\pm}(M_2+\mu \sin 2\beta)}{\mu^2-M_2^2}.
\label{again2}
\end{equation}
%

The mixing matrices $U$ and $V$, defined by 
\begin{equation}
\left( \begin{matrix}{\tilde \chi}^-_1\\{\tilde \chi}^-_2\end{matrix}  \right)=
U\left( \begin{matrix}\tilde W^-\\\tilde H_d^-\end{matrix}  \right),\quad
\left( \begin{matrix}{\tilde \chi}^+_1\\{\tilde \chi}^+_2\end{matrix}  \right)=
V\left( \begin{matrix}\tilde W^+\\ \tilde H_u^+\end{matrix}  \right) \ ,
\end{equation}
also are dependent on the relative sizes of $M_{2}$ and $\mu$. For $|M_{2}| \lesssim |\mu|$ they are found to be
\begin{equation}\label{eq:U_matrix}
U=O_-~~ , ~~ V=\begin{cases}
    O_+, & \det M_{\tilde \chi^\pm}>0\\
    \sigma_3 O_+, & \det M_{\tilde \chi^\pm}<0 \ ,
  \end{cases}
  \end{equation}
 where
\begin{equation}
O_\pm=\left(\begin{matrix}\cos \phi_\pm&\sin \phi_\pm\\-\sin \phi_\pm&\cos \phi_\pm\end{matrix}\right) \ .
\label{pass1}
\end{equation}
The Pauli matrix $\sigma_3$ is inserted so that the diagonal entries of $M^D$ are always positive. The angles $\phi_\pm$ are given by
\begin{equation}
\tan 2\phi_-=2\sqrt{2}M_{W^\pm}\frac{\mu \cos \beta +M_2 \sin \beta}{\mu^2-M_2^2-2M_{W^\pm}^2
\cos 2\beta}
\label{bernard1}
\end{equation}
\begin{equation}
\tan 2\phi_+=2\sqrt{2}M_{W^\pm}\frac{\mu \sin \beta +M_2 \cos \beta}{\mu^2-M_2^2+2M_{W^\pm}^2
\cos 2\beta}
\label{bernard2}
\end{equation}
respectively. On the other hand, when $|\mu| \lesssim |M_2|$, we find that \eqref{eq:U_matrix} remains the same, as do the expressions \eqref{bernard1} and \eqref{bernard2} for the angles $\phi_{\pm}$. However, the matrix $O_{\pm}$ now becomes
\begin{equation}
O_\pm=\left(\begin{matrix}-\sin \phi_\pm&\cos \phi_\pm\\-\cos \phi_\pm&-\sin \phi_\pm\end{matrix}\right) \ .
\label{talk1}
\end{equation}
It is important to note that the $|\mu| \lesssim |M_2|$ results for the $U$, $V$ matrices can be obtained from the $|M_{2}| \lesssim |\mu|$ expressions \eqref{eq:U_matrix}, \eqref{pass1},
\eqref{bernard1} and \eqref{bernard2} simply by replacing
\begin{equation}
\phi_{\pm} \longrightarrow \phi_{\pm} + \frac{\pi}{2}
\label{pass2}
\end{equation}
in all expressions. We will use this replacement, when required, in the main numerical analysis to follow.

It is useful to note that when  $|M_{2}| \lesssim |\mu|$, it follows from \eqref{pass1} that the lightest chargino eigenstate is
\begin{equation}
{\tilde{\chi}}_{1}^{\pm}= \cos \phi_{\pm} {\tilde{W}}^{\pm} + \sin \phi_{\pm} {\tilde{H}}^{\pm} \ .
\label{ok1}
\end{equation}
Since $M^2_{W^\pm}\ll M^2_2, \mu^2$, we see from \eqref{bernard1} and \eqref{bernard2} that ${\rm tan} ~2\phi_{\pm} \ll 1$ and, hence, $|\cos \phi_{\pm}| > |\sin \phi_{\pm}|$. It follows that 
\begin{equation}
{\tilde{\chi}}_{1}^{\pm} \simeq  {\tilde{W}}^{\pm}  \ .
\label{ok2}
\end{equation}
We say that ${\tilde{\chi}}_{1}^{\pm}$ is ``predominantly'' a charged Wino and, regardless of the exact value of $\phi_{\pm}$, denote it by  ${\tilde{\chi}}^{\pm}_{{{W}}}$. On the other hand, when $|\mu| \lesssim |M_2|$ it follows from 
\eqref{talk1} that 
\begin{equation}
{\tilde{\chi}}_{1}^{\pm}= -\sin \phi_{\pm} {\tilde{W}}^{\pm} + \cos \phi_{\pm} {\tilde{H}}^{\pm} \ .
\label{ok3}
\end{equation}
Expressions \eqref{bernard1} and \eqref{bernard2} again tell us that $|\sin \phi_{\pm}| < |\cos \phi_{\pm}|$ and, hence
\begin{equation}
{\tilde{\chi}}_{1}^{\pm} \simeq  {\tilde{H}}^{\pm}  \ .
\label{ok4}
\end{equation}
That is,  ${\tilde{\chi}}_{1}^{\pm}$ is ``predominantly'' a charged Higgsino and, regardless of the exact value of $\phi_{\pm}$, we denote it by  ${\tilde{\chi}}^{\pm}_{{{H}}}$. Having made this analysis, we note that the results could have been read off directly from the leading term in the expressions for the mass eigenvalues given in \eqref{eq:Wino_mass} and \eqref{again1}. Specifically, for $|M_{2}| \lesssim |\mu|$, the leading term in \eqref{eq:Wino_mass} is $|M_{2}|$, the soft mass associated with the charged Wino-- indicating that ${\tilde{\chi}}_{1}^{\pm} \simeq  {\tilde{W}}^{\pm}$. Similarly, for $|\mu| \lesssim |M_2|$, the leading term in \eqref{again1} is $|\mu|$, the parameter associated with the charged Higgsino-- indicating that
${\tilde{\chi}}_{1}^{\pm} \simeq  {\tilde{H}}^{\pm}$, as above. As stated previously, we will refer to the mass eigenstates ${\tilde{\chi}}_{W}^{\pm}$ and
${\tilde{\chi}}^{\pm}_{{{H}}}$ simply as a Wino chargino and Higgsino chargino respectively, even though they are only ``predominantly'' the pure states of $W^{\pm}$ and $H^{\pm}$ respectively. 

%
%
%
%

Let us now {\it include the $R$-parity violation terms} in the Lagrangian. These will not significantly affect the chargino masses, but they do introduce mixing between the charginos and the standard model charged leptons. These mixings are central to our study because they allow RPV chargino decays.
In the extended bases $\psi^+=(\tilde{W}^+, \tilde{H}_u^+, {e^c}_i)$ and $\psi^-=(\tilde{W}^-, \tilde{H}^-_d, e_i)$, the mixing matrix can again be written in the form of eq. \eqref{eq:generalCharginoMatrix},
\begin{equation}
\mathcal{L}\supset -\frac{1}{2}\left(\psi^+ \> \psi^-\right)
\left(\begin{matrix}
0&{{\cal{M}}_{\tilde \chi^\pm}}^T\\
{{\cal{M}}_{\tilde \chi^\pm}}&0
\end{matrix}
\right)
\left(\begin{matrix}\psi^+\\ \psi^-\end{matrix}\right)
+h.c.
\end{equation}
now, however, where
\begin{equation}
{\cal{M}}_{\tilde \chi^\pm}=\left(
\begin{matrix}
	M_2&\frac{1}{\sqrt{2}}g_2v_u&0&0&0\\
	\frac{1}{\sqrt{2}}g_2v_d&\mu&-\frac{{v_{L1}}}{v_d}m_e&-\frac{{v_{L2}}}{v_d}m_{\mu}&-\frac{v_{L3}}{v_d}m_{\tau}\\
\frac{1}{\sqrt{2}}g_2{v_{L1}}^*&{-\epsilon_1}&m_e&0&0\\
\frac{1}{\sqrt{2}}g_2{v_{L2}}^*&{-\epsilon_2}&0&m_{\mu}&0\\
\frac{1}{\sqrt{2}}g_2{v_{L3}}^*&{-\epsilon_3}&0&0&m_{\tau}
\end{matrix}
\right)
\end{equation}
and $m_e$, $m_\mu$, and $m_\tau$ denote the Dirac masses of the standard model charged leptons.
This matrix can be expressed in a schematic form that will be useful in diagonalizing it. Let us write
\begin{equation}
{\cal{M}}_{\tilde \chi^\pm}=\left(
\begin{matrix}
{M_{\tilde \chi^\pm}}&\Gamma\\
G^T&m_{e_i}
\end{matrix}
\right)
\end{equation}
where
\begin{equation}
{M_{\tilde \chi^\pm}}=\left(
\begin{matrix}
M_2&\frac{1}{\sqrt{2}}g_2v_u \\
\frac{1}{\sqrt{2}}g_2v_d&\mu
\end{matrix}
\right),~
\Gamma=\left(\begin{matrix}0&0&0\\-\frac{{v_{L1}}}{v_d}m_e&-\frac{{v_{L2}}}{v_d}m_{\mu}&-\frac{v_{L3}}{v_d}m_{\tau} \end{matrix}\right),~
G^{T}=\left(\begin{matrix} \frac{1}{\sqrt{2}}g_2{v_{L1}}^*&{-\epsilon_1}\\ \frac{1}{\sqrt{2}}g_2{v_{L2}}^*&{-\epsilon_2}\\ \frac{1}{\sqrt{2}}g_2{v_{L3}}^*&{-\epsilon_3} \end{matrix}\right)
\end{equation}
and $m_{e_i}$ is the $3 \times 3$ matrix with diagonal entries $(m_e,m_\mu,m_\tau)$.
The $G$, $\Gamma$, and $m_{e_i}$ matrices have entries that are much smaller than the entries of the ${M_{\tilde \chi^\pm}}$ matrix and, therefore, can be used to perturbatively diagonalize the ${\cal{M}}_{\tilde \chi^\pm}$ matrix. The mass eigenstates are related to the gauge eigenstates by unitary matrices $\mathcal{V}$ and $\mathcal{U}$ defined by
\begin{equation}
\left(
\begin{matrix}
\tilde{{\chi}}_1^-\\
\tilde{{\chi}}_2^-\\
\tilde{{\chi}}_3^-\\
\tilde{{\chi}}_4^-\\
\tilde{{\chi}}_5^-\\
\end{matrix}
\right)
=\mathcal{U}\left(
\begin{matrix}
\tilde{W}^-\\
\tilde{H}^-_d\\
e_1\\
e_2\\
e_3\\
\end{matrix}
\right),
\quad \quad
\left(
\begin{matrix}
\tilde{{\chi}}_1^+\\
\tilde{{\chi}}_2^+\\
\tilde{{\chi}}_3^+\\
\tilde{{\chi}}_4^+\\
\tilde{{\chi}}_5^+\\
\end{matrix}
\right)
=\mathcal{V}\left(
\begin{matrix}
\tilde{W}^+\\
\tilde{H}^+_u\\
{e^c_1}\\
{e^c_2}\\
{e^c_3}\\
\end{matrix}
\right).
\end{equation}
%
They are chosen so that
\begin{equation}
\mathcal{U}^*\mathcal{M}_{\tilde \chi^\pm}\mathcal{V}^{-1}=\mathcal{M}_{\tilde \chi^\pm}^D=
\left(
\begin{matrix}
M_{\tilde{{\chi}}^\pm_1}&0&0&0&0\\
0&M_{\tilde{{\chi}}^\pm_2}&0&0&0\\
0&0&M_{\tilde{{\chi}}^\pm_3}&0&0\\
0&0&0&M_{\tilde{{\chi}}^\pm_4}&0\\
0&0&0&0&M_{\tilde{{\chi}}^\pm_5}\\
\end{matrix}
\right) \ ,
\end{equation}
where all eigenvalues are positive.
The first two eigenstates, $M_{\chi^\pm_{1,2}}$, are mostly charged Wino and charged Higgsino. The RPV couplings are small enough that their masses are still given by eqs. \eqref{eq:Wino_mass} and \eqref{eq:Higgsino_mass} or \eqref{again1} and \eqref{again2}, depending on the values of the parameters $M_{2}$ and $\mu$. Similarly, the masses of the leptons are 
basically unchanged; that is, $M_{\chi^\pm_{2+i}}$ are the standard model charged lepton masses, $m_{e_i}$, where $e_{1,2,3}\equiv e,\mu,\tau$. The phenomenologically important effect of the RPV mixing is the RPV decays of the charginos.
The matrices $\mathcal{U}$ and $\mathcal{V}$ can be written schematically as
\begin{equation}
\mathcal{U}=
\left(
\begin{matrix}
U&0_{2\times3}\\
0_{3\times2}&1_{3\times3}\\
\end{matrix}
\right)
\left(
\begin{matrix}
1_{2\times2}&-\xi_-\\
\xi^{\dag}_-&1_{3\times3}\\
\end{matrix}
\right)~\ ,
\quad \>
\mathcal{V}=
\left(
\begin{matrix}
V&0_{2\times3}\\
0_{3\times2}&1_{3\times3}\\
\end{matrix}
\right)
\left(
\begin{matrix}
1_{2\times2}&-\xi_+\\
\xi^{\dag}_+&1_{3\times3}\\
\end{matrix}
\right).
\end{equation}
Next, requiring that $\mathcal{U^*}\mathcal{M}_{\tilde \chi^\pm}\mathcal{V}^{-1}$ is diagonal allows one to compute the 
$\xi_+$ and $\xi_-$ matrices. To first order, they are given by
\begin{equation}
\xi_-=-\left( {M_{\tilde \chi^\pm}}^T \right)^{-1}G~, \quad \xi_+=-({M_{\tilde \chi^\pm}})^{-1}\Gamma \ .
\end{equation}
The values of all the $\mathcal{U}$ and $\mathcal{V}$ matrix elements are presented in the Appendix B.1. These matrices will be crucial in calculating the decay rates of the charginos via RPV processes. In general, the exact expressions for the five charged mass eigenstates are complicated, and will be dealt with numerically in our calculations. However, it of interest to present the complete analytic expressions, including the RPV terms, for the mass eigenstates ${\tilde{\chi}}^{\pm}_{1}$. We find that the positive eigenvector ${\tilde{\chi}}^{+}_{1}$ is
\begin{equation}
{\tilde{\chi}}^{+}_{1}=\mathcal{V}_{1\>1} {\tilde{W}}^{+}+\mathcal{V}_{1\>2} {\tilde{H}}_{u}^{+}+ \mathcal{V}_{1\>2+i} e^{c}_{i} \ ,
\label{finish1}
\end{equation}
where one sums over $i=1,2,3$. When $|M_{2}|<|\mu|$, the $\mathcal{V}$ coefficients are given by
\begin{equation}
\mathcal{V}_{1\>1}=\cos \phi_{+}~,~\mathcal{V}_{1\>2} =\sin \phi_{+}
\label{finish2}
\end{equation}
and
\begin{equation}
\mathcal{V}_{1\>2+i}=-\cos \phi_+ \frac{g_2 \tan \beta m_{e_i}}{\sqrt{2}M_2\mu}v_{L_i}+\sin \phi_+\frac{m_{e_i}}{\mu v_d}v_{L_i} \ .
\label{finish3}
\end{equation}
The result for $|\mu|<|M_{2}|$ is found by replacing $\phi_{+} \rightarrow \phi_{+} + \frac{\pi}{2}$ in these expressions, as discussed above. Similarly, the negative eigenvector ${\tilde{\chi}}^{-}_{1}$ is found to be
\begin{equation}
{\tilde{\chi}}^{-}_{1}=\mathcal{U}_{1\>1} {\tilde{W}}^{-}+\mathcal{U}_{1\>2} {\tilde{H}}_{d}^{-}+ \mathcal{U}_{1\>2+i} e_{i} \ ,
\label{finish1}
\end{equation}
where one sums over $i=1,2,3$. When $|M_{2}|<|\mu|$, the $\mathcal{U}$ coefficients are given by
\begin{equation}
\mathcal{U}_{1\>1}=\cos \phi_{-}~,~\mathcal{U}_{1\>2} =\sin \phi_{-}
\label{finish2}
\end{equation}
and
\begin{equation}
\mathcal{U}_{1\>2+i}=-\cos \phi_- \frac{g_2 v_d}{\sqrt{2}M_2\mu}\epsilon_i^*+\sin \phi_-\frac{\epsilon_i^*}{\mu} \ .
\label{finish3}
\end{equation}
The result for $|\mu|<|M_{2}|$ is found by replacing $\phi_{-} \rightarrow \phi_{-} + \frac{\pi}{2}$ in these expressions. Note that the first two terms of ${\tilde{\chi}}^{\pm}_{1}$ are independent of RPV and are identical to the expressions given in \eqref{ok1} and \eqref{ok3} for $|M_{2}|<|\mu|$ and $|\mu|<|M_{2}|$  respectively. Also, {\it note that these terms dominate over the RPV terms and, hence, are the main contributors to the mass eigenvalues. However, although they are numerically smaller, the RPV terms in ${\tilde{\chi}}^{\pm}_{1}$ play a crucial role in the $R$-parity violating decays of chargino LSPs and, hence, cannot be ignored}.

\subsection{Neutralino mass eigenstates}\label{Neutralino_masses}

\indent In the {\it absence of the RPV violating terms proportional to $\epsilon_i$ and $v_{L_i}$}, the neutral Higgsinos and gauginos of the theory mix with
the third generation right handed neutrino. In the gauge eigenstate basis $\psi^0=\left( \tilde{W}_R, \tilde{W}_0, \tilde{H}_d^0,
\tilde{H}_u^0, \tilde{B}^\prime,{\nu}^c_3 \right)$, 
\begin{equation}
\mathcal{L}\supset-\frac{1}{2}\left(\psi^0\right)^T{M}_{\tilde{{ \chi}}^0}\psi^0+c.c
\end{equation}
where 
\begin{equation}
M_{{\tilde \chi}^0}=
\left(
\begin{matrix}
M_R&0&-\frac{1}{2}g_Rv_d&\frac{1}{2}g_Rv_u&0&-\frac{1}{2}g_Rv_R\\
0&M_2&\frac{1}{2}g_2 v_d&-\frac{1}{2}g_2 v_u&0&0\\
-\frac{1}{2}g_Rv_d&\frac{1}{2}g_2 v_d&0&-\mu&0&0\\
\frac{1}{2}g_Rv_u&-\frac{1}{2}g_2 v_u&-\mu&0&0&0\\
0&0&0&0&M_{BL}&\frac{1}{{2}}g_{BL}v_R\\
-\frac{1}{{2}}g_Rv_R&0&0&0&\frac{1}{{2}}g_{BL}v_R&0\\
\end{matrix}
\right) \ .
\label{eq:neutralinoMassMatrixWithoutEpsilon}
\end{equation}
The mass eigenstates are related to the gauge states by the unitary matrix $N$ where ${\tilde{\chi }}^{0}=N \psi^{0}$. $N$ is chosen so that

\begin{equation}
N^*M_{\tilde{{ \chi}}^0}N^{\dag}= M_{\tilde{{ \chi}}^0}^{D} =
\left(
\begin{matrix}
M_{{\tilde \chi}^0_1}&0&0&0&0&0\\
0&M_{{\tilde \chi}^0_2}&0&0&0&0\\
0&0&M_{{\tilde \chi}^0_3}&0&0&0\\
0&0&0&M_{{\tilde \chi}^0_4}&0&0\\
0&0&0&0&M_{{\tilde \chi}^0_5}&0\\
0&0&0&0&0&M_{{\tilde \chi}^0_6}\\
\end{matrix}
\right) \ ,
\end{equation}
where all eigenvalues are positive.
The $B-L$ MSSM does not explicitly contain a Bino, associated with the hypercharge group $U(1)_Y$.  Instead, it contains a Blino and a Rino, the gauginos associated with $U(1)_{B-L}$ and $U(1)_{3R}$ respectively. Nevertheless, the theory does effectively contain a Bino. To see this, consider the limit $M_{W^\pm}^2,\> M_{Z^0}^2\ll M_{R}^{2}, \> M_{2}^{2},\> M_{BL}^{2}$ --- that is, when the 
EW scale is much lower than the 
soft breaking scale so that the Higgs VEV's are negligible. In this limit, the mass matrix in eq. \eqref{eq:neutralinoMassMatrixWithoutEpsilon} becomes
\begin{equation}
M_{{\tilde \chi}^0}=
\left(
\begin{matrix}
M_R&0&0&0&0&-\frac{1}{ 2}g_Rv_R\\
0&M_2&0&0&0&0\\
0&0&0&-\mu&0&0\\
0&0&-\mu&0&0&0\\
0&0&0&0&M_{BL}&\frac{1}{{2}}g_{BL}v_R\\
-\frac{1}{{2}}g_Rv_R&0&0&0&\frac{1}{{2}}g_{BL}v_R&0\\
\end{matrix}
\right)
\end{equation}
The first, fifth, and sixth columns, corresponding to the Blino, the Rino and the third generation right-handed neutrino, are now decoupled from the others and mix only with each other. In the reduced basis $\left({\nu}_3^c, \tilde W_R, \tilde B^\prime \right)$, the mixing matrix is
\begin{equation}
\left(
\begin{matrix}
0&-\cos \theta_R M_{Z_R}&\sin \theta_R M_{Z_R}\\
-\cos \theta_R M_{Z_R}&M_R&0\\
\sin \theta_R M_{Z_R}&0&M_{BL}\\
\end{matrix}
\right), \quad
\text{where} \quad
\cos \theta_R = \frac{g_R}{\sqrt{g_R^2+g_{BL}^2}} \ .
\end{equation}
The limit in which the gaugino masses are much smaller than $M_{Z_R}$ is phenomenologically relevant due to the lower bound on $M_{Z_R}$ being much higher than typical gaugino mass lower bounds. This limit is also motivated theoretically because RG running makes the gauginos masses lighter.
In this limit, the mass eigenstates can be found as an expansion in the gaugino masses. At zeroth order, they are
\begin{equation}
\tilde B=\tilde W_R \sin \theta_R+\tilde B^\prime \cos \theta_R \ ,
\end{equation}
\begin{equation}
{\nu}_{3a}^c=\frac{1}{\sqrt 2}({\nu^c}_3-\tilde W_R \cos \theta_R+\tilde B^\prime 
\sin \theta_R) \ ,
\end{equation}
\begin{equation}
{\nu}_{3b}^c=\frac{1}{\sqrt{2}}({\nu^c}_3+\tilde W_R \cos \theta_R-
\tilde B^\prime \sin \theta_R)
\end{equation}
with masses, calculated to first order, given by
\begin{equation}
M_1=\sin^2 \theta_R M_R+\cos^2 \theta_R M_{BL},\quad m_{{\nu^c}_{3a}}=M_{Z_R},
\quad m_{{\nu^c}_{3b}}=M_{Z_R} \ .
\end{equation}
respectively.
The state $\tilde B$ with mass $M_1$ is effectively a Bino.
We can rotate from the old basis, $\left( \tilde{W}_R, \tilde{W}_0, \tilde{H}_d^0,
\tilde{H}_u^0, \tilde{B}^\prime ,{\nu^c}_3 \right)$, to the new one,
$\left( \tilde{B}, \tilde{W}_0, \tilde{H}_d^0,
\tilde{H}_u^0, {\nu}_{3a}^c ,{\nu}_{3b}^c \right)$, using a rotation matrix, which at zeroth order has the form:

\begin{equation}
\left(
\begin{matrix}
\sin \theta_R&0&0&0&\cos \theta_R&0\\
0&1&0&0&0&0\\
0&0&1&0&0&0\\
0&0&0&1&0&0\\
-\frac{1}{\sqrt  2}\cos \theta_R&0&0&0&\frac{1}{\sqrt  2} \sin \theta_R&\frac{1}{\sqrt 2}\\
\frac{1}{\sqrt  2}\cos \theta_R&0&0&0&-\frac{1}{\sqrt 2}\sin \theta_R&\frac{1}{\sqrt 2}\\
\end{matrix}
\right).
\end{equation}
We then get a new neutralino mass matrix, which is in agreement with the MSSM model after B-L breaking. It is given by
\begin{equation}
M_{{\tilde \chi}^0}=
\left(
\begin{matrix}
M_1&0&-\frac{1}{\sqrt 2}g^\prime v_d&\frac{1}{\sqrt 2}g^\prime v_u&0&0\\
0&M_2&\frac{1}{\sqrt2}g_2 v_d&-\frac{1}{\sqrt 2}g_2 v_u&0&0\\
-\frac{1}{\sqrt 2}g^\prime v_d&\frac{1}{\sqrt 2}g_2 v_d&0&-\mu&0&0\\
\frac{1}{\sqrt 2}g^\prime v_u&-\frac{1}{\sqrt 2}g_2 v_u&-\mu&0&0&0\\
0&0&0&0&m_{{\nu^c}_{3a}}&0\\
0&0&0&0&0&m_{{\nu^c}_{3b}}\\
\end{matrix}
\right)
\end{equation}
Since the EW scale is generally much lower than the gaugino mass scale, the off-diagonal
terms are small,

\begin{equation}\label{eq:Bino_mass}
M_{{\tilde \chi}^0_1} \simeq |M_1|-\frac{M_{Z^0}^2\sin^2 \theta_W(M_1+\mu \sin 2\beta)}{\mu^2-M_1^2} \ ,
\end{equation}

\begin{equation}\label{eq:Wino_Neutralino_mass}
M_{{\tilde \chi}^0_2} \simeq |M_2|-\frac{M_{W^\pm}^2(M_2+\mu \sin 2\beta)}{\mu^2-M_2^2} \ ,
\end{equation}

\begin{equation}
M_{{\tilde \chi}^0_3} \simeq |\mu|+\frac{M_{Z^0}(\text{sgn}(\mu)\times 1-\sin 2\beta)(\mu+M_1\cos^2 \theta_W+M_2\sin^2 \theta_W)}
{2(\mu+M_1)(\mu+M_2)} \ ,
\end{equation}

\begin{equation}
M_{{\tilde \chi}^0_4} \simeq |\mu|+\frac{M_{Z^0}(\text{sgn}(\mu)\times 1+\sin 2\beta)(\mu-M_1\cos^2 \theta_W-M_2\sin^2 \theta_W)}
{2(\mu-M_1)(\mu-M_2)} \ ,
\end{equation}

\begin{equation}
M_{{\tilde \chi}^0_5} \simeq M_{Z_R} \ ,
\end{equation}

\begin{equation}\label{eq:neutrino_mass}
M_{{\tilde \chi}^0_6} \simeq M_{Z_R}
\end{equation}
where $\theta_W$ is the Weinberg angle.  Unlike for charginos discussed previously, our {\it labels do not imply any mass ordering}.
The exact eigenstates are more difficult to compute than in the chargino case. They are, generically, linear combinations of the six gauge neutralino states. However, as was discussed in detail for charginos, the dominant gauge neutralino in an eigenstate can be read off directly from the leading term in the associated mass eigenvalue. Using this, as well as explicit numerical computation, we find that 
\begin{equation}
{\tilde{\chi}}_{1}^{0} \simeq  {\tilde{B}}^{0}~~ ,~~{\tilde{\chi}}_{2}^{0} \simeq  {\tilde{W}}^{0}~~ ,~~{\tilde{\chi}}_{3}^{0} \simeq  {\tilde{H}}_{d}^{0} ~~,~~{\tilde{\chi}}_{4}^{0} \simeq  {\tilde{H}}_{u}^{0}~ ~,~~{\tilde{\chi}}_{5}^{0} \simeq  {{\nu}}_{3a}^{c}~~ ,~~{\tilde{\chi}}_{6}^{0} \simeq  {{\nu}}_{3b}^{c} \ .
\label{finally1}
\end{equation}
As with the charginos, we henceforth denote these mass eigenstates by ${\tilde \chi}^0_B$,~ ${\tilde \chi}^0_W$,~${\tilde \chi}^0_{H_d}$,~${\tilde \chi}^0_{H_u}$,~${\tilde \chi}^0_{\nu^{c}_{3a}}$,~${\tilde \chi}^0_{\nu^{c}_{3b}}$ respectively; and refer to them as a Bino neutralino, a Wino neutralino and so on, even though they are only ``predominantly'' the pure neutral state.


Let us now {\it add the RPV couplings $\epsilon_i$ and $v_{L_i}$}. This introduces mixing between the neutralinos and the neutral fermions of the standard model --- the neutrinos. As discussed at the beginning of Section \ref{sec:3}, mixing with the first- and second-family right-handed neutrino would lead to active-sterile neutrino oscillations. Unless and until there is more experimental evidence of such oscillations, we will continue to assume that they do not exist and, therefore, that the mixing with the first- and second-family right-handed neutrinos is negligible. Therefore, the neutrino mass matrix given below includes only mixing with the three families of left-handed neutrinos-- the seventh column --and the third-family right-handed neutrino-- the sixth column. As in the case of the charginos, the effect of adding these RPV couplings is important because it will allow RPV decays of the neutralinos. It's effect on the physical masses of the neutralinos, however, is negligible.
 The new basis is then extended to 
$\left( \tilde{W}_R, \tilde{W}_0, \tilde{H}_d^0,
\tilde{H}_u^0, \\ \tilde{B}^\prime ,{\nu}_3^c, \nu_1, \nu_2,\nu_3\right)$. The extended mass matrix is given by
\begin{equation}
\mathcal{M}_{{\tilde \chi}^0}=
\left(
\begin{matrix}
M_R&0&-\frac{1}{ 2}g_Rv_d&\frac{1}{2}g_Rv_u&0&-\frac{1}{ 2}g_Rv_R&0_{1\times 3}\\
0&M_2&\frac{1}{2}g_2 v_d&-\frac{1}{ 2}g_2 v_u&0&0&\frac{1}{ 2}g_2 v_{L_i}^*\\
-\frac{1}{ 2}g_Rv_d&\frac{1}{ 2}g_2 v_d&0&-\mu&0&0&0_{1\times 3}\\
\frac{1}{2}g_Rv_u&-\frac{1}{ 2}g_2 v_u&-\mu&0&0&0& \epsilon_i\\
0&0&0&0&M_{BL}&\frac{1}{{2}}g_{BL}v_R& -\frac{1}{ 2} g_{BL}v_{L_i}^*\\
-\frac{1}{{2}}g_Rv_R&0&0&0&\frac{1}{{2}}g_{BL}v_R&0&\frac{1}{\sqrt 2}Y_{\nu i3}v_u\\
0_{3\times 1}&\frac{1}{\sqrt 2} g_2 v_{L_j}^* &0_{3\times 1}& \epsilon_j&-\frac{1}{\sqrt 2}
g_{BL}v_{Lj}^*&\frac{1}{\sqrt 2} Y_{\nu j3}v_u&0_{3\times 3}\\
\end{matrix}
\right)
\end{equation}
This is the matrix that was introduced in eq. \eqref{eq:20}.
Just as for charginos, we can write the neutralino matrix in a schematic form that will help us diagonalize it perturbatively. As discussed in detail in Section \ref{sec:3},  $\mathcal{M}_{{\tilde \chi}^0}$ can be expressed as
\begin{equation}
\mathcal{M}_{{\tilde \chi}^0}=
\left(
\begin{matrix}
M_{{\tilde \chi}^0}&m_D\\
m_D^T&0_{3\times 3}
\end{matrix}
\right)
\end{equation}
where $M_{\chi_{0}}$ and $m^{D}$ are given in \eqref{eq:21} and \eqref{eq:22} respectively.
The mass eigenstates are related to the gauge eigenstates by the unitary matrix $\mathcal{N}$, which diagonalizes the neutralino mixing matrix
\begin{equation}
\mathcal{M}^D_{{\tilde \chi}^0}=\mathcal{N}^* \mathcal{M}_{{\tilde \chi}^0} \mathcal{N}^{\dag} \ .
\end{equation}
$\mathcal{N}$ can be written in the perturbative form
\begin{equation}
\mathcal{N}=\left(
\begin{matrix}
N&0_{3\times 3}\\
0_{3\times 3}&V^{\dag}_{PMNS}\\
\end{matrix}
\right)
\left(
\begin{matrix}
1_{6\times 6}& -\xi_0\\
\xi_0^{\dag}&1_{3\times 3}\\
\end{matrix}
\right),
\end{equation}
where $N$ is the unitary matrix introduced below eq. \eqref{eq:neutralinoMassMatrixWithoutEpsilon}. It is a $6\times 6$ matrix, analogous to the $2\times2$ matrices $U$ and $V$ in Section \ref{Chargino_masses}. However, while eq. \eqref{eq:U_matrix}, \eqref{pass1} and \eqref{talk1} provide simple analytic expressions for $U$ and $V$ in terms of the rotation angles $\phi_\pm$, it is much harder to solve for $N$ without approximations. In this paper, we will compute $N$ numerically, using the the relevant soft mass terms and couplings as input. 

The equation $\xi^0=M_{\tilde \chi^0}^{-1}m_D$ is obtained by requiring that $\mathcal{M}^D_{{\tilde \chi}^0}$ be diagonal.  We denote the mass eigenstates as $\tilde \chi^0=\mathcal{N}\psi^0$. The entries of $\mathcal{N}$ are central in calculating the neutralino decay rates and are all presented in Appendix B. However, {\it exactly as in the case of the charginos discussed above, the physical masses of the ``proper'' neutralinos are not significantly changed by introducing the RPV couplings}, so eqs. \eqref{eq:Bino_mass} -- \eqref{eq:neutrino_mass} remain valid. The states $\chi^0_{6+i}$ for $i=1,2,3$ are the three left-handed neutrinos which now receive Majorana masses. This process has been discussed in detail in Section \ref{sec:3}. As in the case of charginos, {\it it is important to note that, although small compared to the $R$-parity preserving coefficients, the RPV terms make important contributions to the RPV decays of the neutralino LSPs and, hence, cannot be ignored}.

\section{Chargino and neutralino decay channels}
\label{sec:6}

\subsection{General B-L MSSM Lagrangian}

The general $B-L$ MSSM Lagrangian, written in terms of chiral multiplet component fields, ($\phi_i, \>\psi_i$), and vector multiplet components, ($A_{\mu}^a, \>\lambda^a$), has the generic form 

\begin{multline}\label{eq:Lagrangian}
\mathcal{L}=-\partial_{\mu}\phi^{*i}\partial^{\mu}\phi_i+i\psi^{\dag i}\bar \sigma^{\mu}\partial_{\mu}\psi_i \qquad \text{ (kinetic terms)}\\
-igT^a A^a_{\mu} \phi^{*}\partial^{\mu} \phi_i+c.c+g^2(T^aA^a_{\mu})(T^bA^{b\mu})\phi_i\phi^{*i}+g\psi^{\dag}\bar \sigma^{\mu}T^aA^a_{\mu}\psi_i \qquad \text{( covariant derivative)}\\
-\sqrt{2}g(\phi^{*}T^a \psi)\lambda^a- \sqrt{2} g \lambda^{a\dag} (\psi^{\dag}T^a \phi)+g (\phi^{*} T^a \phi)D^a \qquad \text{(covariant derivative supersymmterization)}\\
+i\lambda^{\dag a}
\bar \sigma^{\mu}\partial_{\mu}\lambda^a+igf^{abc}\lambda^{\dag a}
\bar \sigma^{\mu}A_{\mu}^b\lambda^{ c}-\frac{1}{4} F^{a \mu \nu}{ F^a}_{\mu\nu}+\frac{1}{2}D^aD^a \qquad \text{(gauge field self-interactions)}\\
-\frac{1}{2}M^{ij}\psi_{i}\psi_{j}-\frac{1}{2}M^*_{ij}\psi^{\dag i}\psi^{\dag j}+M^*_{ik}M^{kj}\phi^{*i}\phi_j \qquad \text{(superpotential mass terms)} \\
-\frac{1}{2}Y^{ijk}\phi_i \psi_j \psi_k  -\frac{1}{2}Y^*_{ijk}\phi^{*i}\psi^{\dag j}\psi^{\dag k} \qquad \text{(superpotential scalar-fermion-fermion Yukawa coupling)}
\\+\frac{1}{2}M^{in}Y^*_{jkn}\phi_i\phi^{*j} \phi^{*k}+\frac{1}{2}M^{*}_{in}Y^{jkn}\phi^{*i}\phi_{j} \phi_{k}+\frac{1}{4}Y^{ijn}Y^*_{kln}\phi_i\phi_j\phi^{*k} \phi^{*l} \qquad \text{(3 and 4-scalar interactions)}.
\end{multline}

The interaction vertices that allow the charginos and neutralinos to decay into standard model particles are encoded in its complicated interactions. In order to read these RPV vertices from this Lagrangian, one needs to follow a series of steps:

\begin{itemize}

\item First, one writes this general Lagrangian in terms of the component fields of the theory. The $B-L$ MSSM matter content is give in \eqref{eq:3} and \eqref{eq:4}, whereas the gauge fields and gauginos are those associated with gauge group \eqref{eq:2}.

\item In the first step, the component fields are pure gauge states. After $B-L$ and electroweak symmetry breaking, these states mix to form massive states. In Section \ref{Chargino and Neutralino states}, we discussed how massive chargino and neutralino states are constructed. The second step then, is to write all gauge eigenstates in the Lagrangian in terms of their mass eigenstate expansion. 

\item After the second step is completed, one can identify the RPV vertices that couple a single chargino or a single neutralino to two standard model particles, typically a boson and a lepton. However, so far the theory has been written in terms of 2-component Weyl spinors, while the physical fermions are described by 4-component spinors. The final step, then, is to write the identified RPV vertices in  4-component spinor notation.

\end{itemize}

In the following sections, we will identify the RPV decay amplitudes for the charginos and neutralinos displayed in Table \ref{tab:decay_channels}, using the three steps described above. 
We find that such sparticle decays are due entirely to the RPV couplings proportional to $\epsilon_i$ and $v_{L_i}$, $i=1,2,3$ that mix the three
generations of leptons and the gauginos of the MSSM inside the chargino and neutralino mass matrices. That is, the mass eigenstate charginos and neutralinos  can decay into SM particles precisely
because they have lepton components. Only after we express the $B-L$ MSSM Lagrangian in terms of the mass eigenstates, will the decay processes in Table \ref{tab:decay_channels} become apparent. Henceforth, we use $\chi^{\pm, 0}$ when referring to chargino and neutralino 2-component Weyl fermions and $X^{\pm, 0}$ when referring to chargino and neutralino 4-component Dirac fermions. Furthermore, we use $e_i, \> i=1,2,3$ for the three families of charged leptons Weyl fermions, and $\ell_i, \>i=1,2,3$ for the three families expressed as Dirac fermions. The Dirac fermion states are defined in eq. \eqref{eq:DircaFermions}.

\begin{table}[H]
\begin{center}
\begin{tabular}{ |c|c| } 
 \hline
 Charginos & Neutralinos \\ 
 \hline
 ${\tilde X}^{\pm}\rightarrow W^{\pm}\nu_{i}$ & ${\tilde X}^0\rightarrow W^{\pm}\ell^\pm_i$\\ 
 ${\tilde X}^{\pm}\rightarrow Z^{0}\ell^\pm_i$& ${\tilde X}^{0}\rightarrow Z^0 \nu_{i}$ \\ 
  ${\tilde X}^{\pm}\rightarrow h^{0}\ell^\pm_i$& ${\tilde X}^{0}\rightarrow h^0 \nu_{i}$ \\  
 \hline
\end{tabular}
\end{center}
\caption{Chargino and Neutralino RPV decay channels, expressed in terms of 4-component spinors.}
\label{tab:decay_channels}
\end{table}

\subsection{Mass eigenstate expansion}

The chargino mass eigenstates ${\tilde \chi}^\pm=({\tilde \chi}_1^\pm,\> {\tilde \chi}_2^\pm,\> {\tilde \chi}_3^\pm,\> {\tilde \chi}_4^\pm,\> {\tilde \chi}_5^\pm)$ are related to the gauge eigenstates $\psi^+=(\tilde W^+, \> \tilde H_u^+,\> e_1^c, \>e_2^c,\> e_3^c)$ and  $\psi^-=(\tilde W^- \> \tilde H_d^-,\> e_1, \> e_2, \> e_3)$ via the unitary matrices $\mathcal{U}$ and $\mathcal{V}$ defined in Section \ref{sec:5} and given in Appendix \ref{appendix:A1} . That is,
\begin{equation}
{\tilde \chi}^-=\mathcal{U}\psi^-~, \quad \quad {\tilde \chi}^+=\mathcal{V}\psi^+ \ .
\end{equation}
There are two things worth pointing out here. First, only the mass eigenstates ${\tilde \chi}_1^\pm$ and ${\tilde \chi}_2^\pm$ are considered to be the actual charginos. They have dominant contributions from the MSSM gauginos and only small SM lepton components. Moreover, the mass eigenstates ${\tilde \chi}^\pm_3\simeq e_1, \> e_1^c$, ${\tilde \chi}^\pm_4\simeq e_2, \> e_2^c$, ${\tilde \chi}^\pm_5\simeq e_3, \> e_3^c$ are considered to be the three generations of charged leptons (to be more precise, the left-handed Weyl spinors of the negatively and positively charged leptons). Second, the $\mathcal{U}$ and $\mathcal{V}$ matrices are defined so that the chargino states ${\tilde \chi}_1^\pm$
are lighter than the chargino states ${\tilde \chi}_2^\pm$; that is, $M_{{\tilde \chi}_1^\pm} < M_{{\tilde \chi}_2^\pm}$. The state ${\tilde \chi}_1^\pm$ can be dominantly charged Wino or charged Higgsino, but it will be always be less massive than ${\tilde \chi}_2^\pm$. 

 In terms of the chargino mass eigenstates, the gauge eigenstate can be expressed as $\psi^-=\mathcal{U}^{\dag}{\tilde \chi}^-$ and  $\psi^+=\mathcal{V}^{\dag}{\tilde \chi}^+$. We then have the following mass eigenstate decomposition:

\begin{equation}\label{eq:mass_expanssion1}
e_i=\mathcal{U}^*_{1\>2+i}{\tilde \chi}^-_1+\mathcal{U}^*_{2\>2+i}{\tilde \chi}^-_2+\mathcal{U}^*_{3\>2+i}{\tilde \chi}^-_3+
\mathcal{U}^*_{4\>2+i}{\tilde \chi}^-_4+\mathcal{U}^*_{5\>2+i}{\tilde \chi}^-_5
\end{equation}

\begin{equation}\label{eq:mass_expanssion2}
 e_i^c=\mathcal{V}^*_{1\>2+i}{\tilde \chi}^+_1+\mathcal{V}^*_{2\>2+i}{\tilde \chi}^+_2+\mathcal{V}^*_{3\>2+i}{\tilde \chi}^+_3+
\mathcal{V}^*_{4\>2+i}{\tilde \chi}^+_4+\mathcal{V}^*_{5\>2+i}{\tilde \chi}^+_5
\end{equation}
Similarly, the Wino and Higgsino gauge eigenstate can be expressed as:
\begin{equation}\label{eq:mass_expanssion3}
\tilde W^-= \mathcal{U}^*_{1\>1}{\tilde \chi}^-_1+\mathcal{U}^*_{2\>1}{\tilde \chi}^-_2+\mathcal{U}^*_{3\>1}{\tilde \chi}^-_3+
\mathcal{U}^*_{4\>1}{\tilde \chi}^-_4+\mathcal{U}^*_{5\>1}{\tilde \chi}^-_5
\end{equation}

\begin{equation}\label{eq:mass_expanssion4}
\tilde W^+= \mathcal{V}^*_{1\>1}{\tilde \chi}^+_1+ \mathcal{V}^*_{2\>1}{\tilde \chi}^+_2+\mathcal{V}^*_{3\>1}{\tilde \chi}^+_3+
\mathcal{V}^*_{4\>1}{\tilde \chi}^+_4+\mathcal{V}^*_{5\>1}{\tilde \chi}^+_5
\end{equation}

\begin{equation}\label{eq:mass_expanssion5}
\tilde H^-_d=\mathcal{U}^*_{1\>2}{\tilde \chi}^-_1+ \mathcal{U}^*_{2\>2}{\tilde \chi}^-_2+\mathcal{U}^*_{3\>2}{\tilde \chi}^-_3+
\mathcal{U}^*_{4\>2}{\tilde \chi}^-_4+\mathcal{U}^*_{5\>2}{\tilde \chi}^-_5
\end{equation}

\begin{equation}\label{eq:mass_expanssion6}
\tilde H^+_u= \mathcal{V}^*_{1\>2}{\tilde \chi}^+_1+\mathcal{V}^*_{2\>2}{\tilde \chi}^+_2+\mathcal{V}^*_{3\>2}{\tilde \chi}^+_3+
\mathcal{V}^*_{4\>2}{\tilde \chi}^+_4+\mathcal{V}^*_{5\>2}{\tilde \chi}^+_5
\end{equation}

The neutralino mass eigenstates ${\tilde \chi}^0=({\tilde \chi}^0_1,\>{\tilde \chi}^0_2,\>{\tilde \chi}^0_3,\>{\tilde \chi}^0_4,\>{\tilde \chi}^0_5,\>{\tilde \chi}^0_6,\>{\tilde \chi}^0_7,\>{\tilde \chi}^0_8,\>{\tilde \chi}^0_9)$ are related to the gauge eigenstates 
$\psi^0=(\tilde W_R, \>\tilde W_0, \>\tilde H^0_d, \> \tilde H_u^0, \> \tilde B^', \> \nu^c_3, \>\nu_e,\>\nu_\mu,\>\nu_\tau )$ via the unitary matrix $\mathcal{N}$,  defined in Section \ref{sec:5} and given in  Appendix \ref{appendix:A2}. That is,
\begin{equation}
{\tilde \chi}^0=\mathcal{N}\psi^0 \ .
\end{equation}
Just as for the chargino states, it is important to remember that only the first six states ${\tilde \chi}_{1,2,3,4,5,6}$ are considered actual MSSM neutralinos, since their dominant contributions are from sparticles. The states ${\tilde \chi}_{7,8,9}$ are the three generations of left handed neutrinos, which obtain Majorana masses after the neutralino matrix diagonalization. However, the notation of the six neutralino mass eigenstates differs from that of the chargino mass eigenstates. In the case of charginos, the states are defined such that $M_{{\tilde \chi}^\pm_1}<M_{{\tilde \chi}^\pm_2}$, where both ${\tilde \chi}^\pm_1$ and ${\tilde \chi}^\pm_2$ could be dominantly charged Wino or charged Higgsino. As discussed above, usually $|M_2|<|\mu|$,  in which case ${\tilde \chi}^\pm_1$ would be dominantly charged Wino, while ${\tilde \chi}^\pm_2$ would be dominantly charged Higgsino.  In the rare case when $|\mu|<|M_2|$, $\tilde \chi_1^\pm$ would be dominantly charged Higgsino, while $\tilde \chi^\pm_2$ would be dominantly charged Wino. For the neutralino mass eigenstates, however, we always have ${\tilde \chi}_1^0$ mostly Bino,  ${\tilde \chi}_2^0$ mostly Wino,  ${\tilde \chi}_{3,4}^0$ mostly Higgsino, ${\tilde \chi}_{5,6}^0$ mostly right-handed third generation neutrino. We don't know, a priori, which state is the lightest, nor how to order them in terms of mass. Their masses are computed after we diagonalize the $6\times 6$ neutralino mixing matrix $M_{{\tilde \chi}^0}$-- neglecting RPV couplings --an operation significantly more complicated than the diagonalization of the $2\times2$ chargino mass matrix, $M_{{\tilde \chi}^\pm}$, in the absence of RPV couplings. 

In terms of the neutralino mass eigesntates, the gauge eigenstates are given by
\begin{equation}
\psi^0=\mathcal{N}^\dag {\tilde \chi}^0 \ .
\end{equation}
For the three neutrino gauge eigenstates $\nu_{e}, \nu_{\mu}, \nu_{\tau}=\nu_{{i}},~i=1,2,3$
\begin{multline}\label{eq:mass_expanssion7}
\nu_{i}=\mathcal{N}^*_{1\>6+i}
{\tilde \chi}^0_1+\mathcal{N}^*_{2\>6+i}{\tilde \chi}^0_2+
\mathcal{N}^*_{3\>6+i}{\tilde \chi}^0_3+\mathcal{N}^*_{4\>6+i}{\tilde \chi}^0_4\\+
\mathcal{N}^*_{5\>6+i}{\tilde \chi}^0_5+
\mathcal{N}^*_{6\>6+i}{\tilde \chi}^0_6+\mathcal{N}^*_{7\>6+i}{\tilde \chi}^0_7+
\mathcal{N}^*_{8\>6+i}{\tilde \chi}^0_8+\mathcal{N}^*_{9\>6+i}{\tilde \chi}^0_9 \ ,
\end{multline} 
while for the rest of the mass eigenstates, we have 
\begin{multline}\label{eq:mass_expanssion8}
\tilde W_R,\>\tilde W^0,\>\tilde H_d^0,\>\tilde H_u^0, \tilde B^', \> \tilde \nu_3^c=\mathcal{N}^*_{1\>1,2,3,4,5,6}
{\tilde \chi}^0_1+\mathcal{N}^*_{2\>1,2,3,4,5,6}{\tilde \chi}^0_2+
\mathcal{N}^*_{3\>1,2,3,4,5,6}{\tilde \chi}^0_3+\ \\+\mathcal{N}^*_{4\>1,2,3,4,5,6}{\tilde \chi}^0_4+
\mathcal{N}^*_{5\>1,2,3,4,5,6}{\tilde \chi}^0_5+
\mathcal{N}^*_{6\>1,2,3,4,5,6}{\tilde \chi}^0_6+\mathcal{N}^*_{7\>1,2,3,4,5,6}{\tilde \chi}^0_7+\\+
\mathcal{N}^*_{8\>1,2,3,4,5,6}{\tilde \chi}^0_8+\mathcal{N}^*_{9\>1,2,3,4,5,6}{\tilde \chi}^0_9 \ .
\end{multline} 

The Higgs scalar fields in the MSSM consist of two complex $SU(2)_L$ doublets; that is,
eight degrees of freedom. When electroweak symmetry is broken, 
three of them become the Goldstone bosons $G^0,\>G^{\pm}$, where  
$G^-=G^{+*}$. The rest will be Higgs scalar mass eigenstates; that is, CP-even neutral scalars $h^0$ 
and $H^0$, a CP-odd neutral scalar $\Gamma^0$ and a charged $H^+$ and a conjugate 
$H^-={H^+}^*$. They are defined by \cite{Martin:1997ns}.
\begin{equation}
\left(\begin{matrix}H_u^0\\H_d^0\end{matrix}\right)=
\left(\begin{matrix}v_u\\v_d\end{matrix}\right)+
\frac{1}{\sqrt{2}}R_{\alpha}\left(\begin{matrix}h^0\\H^0\end{matrix}\right)+
\frac{i}{\sqrt{2}}R_{\beta_0}\left(\begin{matrix}G^0\\\Gamma^0\end{matrix}\right) \ ,
\end{equation}
\begin{equation}
\left(
\begin{matrix}
H_u^+\\H_d^{-*}
\end{matrix}
\right)=
R_{\beta_\pm} \left(\begin{matrix}G^+\\H^+\end{matrix}\right) \ 
\end{equation}
where $R_{\alpha},\>R_{\beta_0}, \>R_{\beta_\pm}$ are rotation matrices.
Specifically, the matrix in front of the Standard Model Higgs Boson $h^0$ is
\begin{equation}
R_{\alpha}=\left( \begin{matrix}
\cos{\alpha}&\sin{\alpha}\\ -\sin{\alpha}&\cos{\alpha}
\end{matrix}\right),
\end{equation}
while, to lowest order, the other matrices are

\begin{equation}
R_{\beta_0}=
 R_{\beta_\pm}=\left( \begin{matrix}
 \sin \beta   & \cos \beta\\
 -\cos \beta & \sin \beta 
 \end{matrix}
 \right),
\end{equation}
where  $\tan \beta=v_u/v_d$. The mixing angle $\alpha$ is, at tree level: 
\begin{equation}\label{eq:alpha}
\frac{\tan 2 \alpha}{\tan 2 \beta}=\frac{M^2_{\Gamma^0}+M_{Z^0}^2}{M^2_{\Gamma^0}-M_{Z^0}^2}, \quad \quad
\frac{\sin 2 \alpha}{\sin 2 \beta}=-\frac{M^2_{H^0}+M^2_{h^0}}{M^2_{H^0}-M^2_{h^0}}
\end{equation}
where the masses of the Higgs eigenstates are
\begin{equation}
M^2_{\Gamma^0}=2b/\sin 2\beta=2|\mu|^2+m^2_{H_u}+m^2_{H_d} \ ,
\end{equation}
\begin{equation}
M^2_{h^0,\>H^0}=\frac{1}{2}\left(  M^2_{\Gamma^0}+M_{Z^0}^2\mp \sqrt{(M^2_{\Gamma^0}-M^2_{Z^0})^2+4M_{Z^0}^2M_{\Gamma^0}^2 
\sin^2(2\beta)} \right)
\end{equation}
and
\begin{equation}
M^2_{H^\pm}=M^2_{\Gamma^0}+M_{W^\pm}^2 \ .
\end{equation}
\subsection{Interaction vertices}

We now express Lagrangian \eqref{eq:Lagrangian}
in terms of all the matter and gauge fields in our $B-L$ MSSM theory, and then replace all gauge 
eigenstates with their mass eigenstate expansion. 
However, the full $B-L$ MSSM Lagrangian is complicated when expressed in its most general
form.  We proceed, therefore, by looking only for the terms that
can lead to chargino or neutralino decays into standard model particles. We identify the following tri-couplings:

\begin{itemize}
\item $g \psi^{i\dag} \bar \sigma^\mu T^a A_\mu^a \psi_i$, ~\quad\quad \qquad ~~from the covariant derivative of the fermionic matter fields

\item $-\sqrt{2}g(\phi^{i*}T^a \psi_i)\lambda^a$ and $-\sqrt{2}g\lambda^\dag(\psi^{i\dag}T^a\phi_i)$,~\qquad~~
from the supercovariant derivatives

\item $ig f^{abc}\lambda^{a\dag}\bar\sigma^\mu A_{\mu}^b\lambda^{c} $, ~\qquad \qquad \qquad \quad\quad \qquad \qquad \qquad  from the gauge self-interaction

\item $-\frac{1}{2}Y^{ijk}\phi_i\psi_j\psi_k$ and $-\frac{1}{2}Y^*_{ijk}\phi^{i*}\psi^{j\dag}\psi^{k\dag}$, \qquad
 from the superpotential Yukawa couplings

\end{itemize}
We now want to write these interactions in terms of the $B-L$ MSSM component fields.  The procedure is non-trivial. Hence, we split these interactions terms into two categories : 1) those responsible for the neutralino or chargino decays into a gauge boson ($Z^0$-boson or $W^\pm$-boson) and a lepton; that is 
\begin{equation}
\tilde \chi^{\pm, 0}\rightarrow Z^0, W^\pm-\text{lepton}
\end{equation}
and 2) those responsible for the decays into a Higgs boson and a lepton; that is
\begin{equation}
 {\tilde \chi}^{\pm,0}\rightarrow h^0 - \text{lepton}.
\end{equation}
 The terms with Yukawa couplings and those from the supercovariant derivatives are relevant only for 
the Higgs boson-lepton decay channel,
as we will show. 

\subsubsection{\boldmath${\tilde \chi}^{\pm,0}\rightarrow$$Z^0, W^\pm$-lepton } 

The part  of the Lagrangian responsible for the gauge boson-lepton decay channels is
\begin{equation}
\mathcal{L}_{{\tilde \chi}^{\pm,\>0}\rightarrow Z^0, W^\pm-{\rm lepton}} \supset g \psi^{i\dag} \bar \sigma^\mu T^a A_\mu^a \psi_i+ig f^{abc}\lambda^{a\dag}\bar\sigma^\mu A_{\mu}^b\lambda^{c},
\end{equation}
where this expression represents the sum over $SU(2)_L$ and $U(1)_Y$. The $i=1,2,3$ represents the three lepton families. Expressed in terms of the MSSM component fields this becomes
\begin{multline}
\mathcal{L}_{{\tilde \chi}^{\pm,\>0}\rightarrow Z^0, W^\pm-\text{lepton}} \supset g_2\left(L_i^{\dag}\bar{\sigma}^{\mu}\tau^a L_i +\tilde H_u^{\dag} \bar \sigma^{\mu
}\tau^a \tilde H_u + \tilde H_d^{\dag} \bar \sigma^{\mu
}\tau^a \tilde H_d\right)W^a_{\mu}\\
+g'\left(-\frac{1}{2}e_i^{\dag}\bar{\sigma}^{\mu}e_i+{e^c}_i^{\dag}\bar{\sigma}^
{\mu}{e^c}_i+\frac{1}{2}\tilde H_u^{\dag} \bar \sigma^{\mu
} \tilde H_u - \frac{1}{2}\tilde H_d^{\dag} \bar \sigma^{\mu
} \tilde H_d\right)B_{\mu}~~\\
+ig_2f^{abc}\tilde W^{a\dag}\bar \sigma^{\mu}\tilde{W}^{b}W_\mu^c, \quad \quad\quad ~~~
\end{multline}
where we sum over the $i$ index, $W^a_{\mu},\>a=1,2,3$ are the three vector bosons of the $SU(2)_L$ group and $B_\mu$ the vector boson of the hypercharge $U(1)_Y$ group. Here, $g_2$ and $g^{\prime}$ are the $SU(2)_L$ and $U_Y(1)$ couplings. In addition, $L_{i}$ represents the $i$-th $SU(2)_{L}$ left chiral lepton doublet defined in \eqref{eq:3}. We now make the replacements
\begin{equation}
\left(
\begin{matrix}
\gamma^0\\
Z^0
\end{matrix}
\right)=
\left(
\begin{matrix}
\cos \theta_W &  \sin \theta_W \\
-\sin \theta_W & \cos \theta_W
\end{matrix}
\right)
\left(
\begin{matrix}
B^0\\
W^0
\end{matrix}
\right) \ ,
\quad 
\quad
W^\pm=\frac{1}{\sqrt 2}(W^1 \mp i W^2),  \quad
\quad \tan \theta_W=\frac{g^{\prime}}{g_2},
\end{equation}
where $\theta_W$ is the Weinberg angle, and rearrange the previous expression to obtain
\small
\begin{multline}\label{eq:L_Z,A,W}
\mathcal{L}_{{\tilde \chi}^{\pm,\>0}
\rightarrow {Z^0, W^\pm}-{\rm lepton}}
\supset \frac{g_2}{\sqrt{2}}(J^{\mu}W^+_{\mu}+J^{\mu \dag}W_{\mu}^{-}
+J^{\mu}_HW_{\mu}^+ + J^{\mu \dag}_H W_{\mu}^{-})
+e(j^{\mu}_{{EM}}+j_{{EMH}}^{\mu})\gamma^0_{\mu}\\
+\frac{g_2}{2 \cos \theta_W}\Big(J^{\mu}_{n}
+J^{\mu}_{nH}\Big)Z^0_{\mu}
+g_2(\tilde W^{+\dag}\bar\sigma^{\mu}\tilde W^{+}-\tilde W^{-\dag}\bar \sigma^{\mu}\tilde W^{-})(\cos \theta_W Z^0_{\mu}+\sin \theta_W \gamma^0_{\mu})\\
+g_2(-\tilde W^{0\dag}\bar \sigma^{\mu}\tilde W^{+}+\tilde W^{-\dag}\bar\sigma^{\mu}\tilde W^{0})W^+_{\mu}
+g_2(\tilde W^{0\dag}\bar \sigma^{\mu}\tilde W^{-}-\tilde W^{+\dag}\bar\ \sigma^{\mu}\tilde W^{0})W^-_{\mu}.
\end{multline}
\normalsize
$J^{\mu}$, $J^\mu_{n}$ and $j^{\mu}_{{EM}}$ are the usual weak charged, neutral and electromagnetic currents from the standard model theory of EW breaking, while 
$J^{\mu}_H$, $J^{\mu}_{nH}$ and $j^{\mu}_{EMH}$ are the equivalent currents of the 
Higgsino fermionic fields. Also note that $e$ is the electromagnetic coupling $
e=\frac{g_2 g^{\prime}}{\sqrt{g_2^2+g^{\prime 2}}}. $
In 2-component Weyl notation these currents are:
\begin{itemize}

\item Weak charged currents, coupling to $W^\pm$ bosons

\begin{equation}
J^{\mu}=\nu_{i}^{\dag}\bar{\sigma}^{\mu}e_i \ , \quad \quad  
J^{\mu}_H=\tilde H_u^{+^{\dag}}\bar \sigma^{\mu} \tilde H_u^0
+\tilde H_d^{0^{\dag}}\bar \sigma^{\mu}\tilde H_d^-
\end{equation}

\item Electromagnetic currents, coupling to the photon $\gamma^0$

\begin{equation}
j^{\mu}_{{EM}}=+{e_i^c}^{\dag}\bar \sigma^{\mu}{e_i^c}-e_i^{\dag}\bar \sigma^{\mu}e_i  \ ,
\quad \quad
j^{\mu}_{EMH}=\tilde H_u^{+^{\dag}}\bar \sigma^{\mu}\tilde H_u^+
-\tilde H_d^{-^{\dag}}\bar \sigma^{\mu}\tilde H_d^-
\end{equation}

\item Neutral currents, coupling to the $Z^0$ boson

\begin{equation}
J_{n}^{\mu}=\nu_{i}^{\dag}\bar \sigma^{\mu}\nu_{i}-(1-2\sin^2\theta_W)e_i^{\dag}\bar{\sigma}^{\mu}e_i-2\sin^2\theta_W{e^c_i}^{\dag}\bar{\sigma}^{\mu}{e^c_i} \ ,
\end{equation}
\begin{multline}
J_{nH}^{\mu}=(1-2\sin^2\theta_W)\tilde H_u^{+^{\dag}}\bar \sigma^{\mu}\tilde H_u^+
-(1+2\sin^2\theta_W)\tilde H_u^{0^{\dag}}\bar \sigma^{\mu}\tilde H_u^0\\
+(1+2\sin^2\theta_W)\tilde H_d^{0^{\dag}}\bar \sigma^{\mu}\tilde H_d^0
-(1-2\sin^2\theta_W)\tilde H_d^{-^{\dag}}\bar \sigma^{\mu}\tilde H_d^-
\end{multline}

\end{itemize}
where in $J^{\mu}$, $j^{\mu}_{{EM}}$ and $J_{n}^{\mu}$ we sum over $i=1,2,3$.
Plugging these currents into eq. \eqref{eq:L_Z,A,W}, and arranging the couplings in terms of $W^\pm$, $Z^0$ and $\gamma^0$ respectively, we get
\small
\begin{multline}\label{eq:633}
\mathcal{L}_{{\tilde \chi}^{\pm,\>0}\rightarrow Z^0, W^\pm-\text{lepton}}\supset 
\frac{g_2}{\sqrt 2}W_\mu^- \Big[e_i^{\dag}\bar \sigma^\mu \nu_{i}+\tilde H_u^{0\dag}\bar
\sigma^\mu \tilde H^+_u+ \tilde H_d^{-\dag}\bar
\sigma^\mu \tilde H^0_d +\sqrt{2}(\tilde W^{0\dag}\bar \sigma^\mu \tilde W^{+}\\-
\tilde W^{-\dag}\bar\sigma^\mu \tilde W^{0}) \Big]
+\frac{g_2}{\sqrt 2}W_\mu^+\left[ \nu_{i}^{\dag} \bar \sigma^\mu e_i+\tilde H_u^{+\dag}\bar \sigma^\mu \tilde H^0_u+ \tilde H_d^{0\dag} \bar \sigma^\mu \tilde H^-_d
 +\sqrt 2(\tilde W^{+\dag}\bar \sigma^\mu \tilde W^{0}-
\tilde W^{0\dag}\bar \sigma^\mu \tilde W^{-}) \right]\\
+g_2c_W Z^0_{\mu}\Big[\tilde W^{+\dag}\bar \sigma^{\mu}\tilde W^{+}-\tilde W^{-\dag}\bar \sigma^{\mu}\tilde W^{-}  \Big]
+\frac{g_2}{2c_W}Z^0_\mu \Big[\nu_{i}^{\dag}\bar \sigma^{\mu}\nu_{i}-(1-2s_W^2)e_i^{\dag}\bar{\sigma}^{\mu}e_i-2 s_W^2{e_i^c}^{\dag}\bar{\sigma}^{\mu}{e_i^c}\\
+(1-2s_W^2)\tilde H_u^{+^{\dag}}\bar \sigma^{\mu}\tilde H_u^+
-(1+2s_W^2)\tilde H_u^{0^{\dag}}\bar \sigma^{\mu}\tilde H_u^0
+(1+2s_W^2)\tilde H_d^{0^{\dag}}\bar \sigma^{\mu}\tilde H_d^0
-(1-2s_W^2)\tilde H_d^{-^{\dag}}\bar \sigma^{\mu}\tilde H_d^-
\Big]\\
+g_2s_W \gamma^0_{\mu}\Big[\tilde W^{+\dag}\bar \sigma^{\mu}\tilde W^{+}-\tilde W^{-\dag} \bar \sigma^{\mu}\tilde W^{-}  \Big]
+e\gamma^0_\mu\Big[  e_i^{c\dag}\bar \sigma^\mu {e_i^c}-e_i^{\dag} \bar\sigma^\mu e_i
+\tilde H_u^{+^{\dag}}\bar \sigma^{\mu}\tilde H_u^+
-\tilde H_d^{-^{\dag}}\bar \sigma^{\mu}\tilde H_d^- \Big],
\end{multline}
\normalsize
where we have used the notation $s_W=\sin \theta_W$, $c_W=\cos \theta_W$ and summed over $i=1,2,3$.

Finally, we are in a position to expand all gauge eigenstates in terms of the mass eigenstates, as in equations \eqref{eq:mass_expanssion1}-\eqref{eq:mass_expanssion6} and \eqref{eq:mass_expanssion7}-\eqref{eq:mass_expanssion8}. After this procedure, the Lagrangian (\ref{eq:633}) is expressed in terms of the mass eigenstates $\tilde \chi_1^\pm$, $\tilde \chi_n^0$, $e_i$, $\nu_i$ for $i=1,2,3$, and their hermitian conjugates. Notice that we keep only $\tilde \chi^\pm_1$ to simplify our results, since $\tilde \chi^\pm_1$ are always lighter than $\tilde \chi^\pm_2$ and, hence, have better prospects to be detected. Their charged Wino and charged Higgsino content are determined from the rotation matrices $U$ and $V$ in eq. 
\eqref{eq:U_matrix}. At the same time, we study the vertices of general neutralino $\tilde \chi^0_n$ states for $n=1,2,3,4,5,6$, since we have no a priori mass ordering for these states. We remind the reader that $n=1$ means a mostly Bino neutralino ${\tilde \chi}^0_1={\tilde \chi}^0_B$,   $n=2$ means a mostly Wino neutralino ${\tilde \chi}^0_2={\tilde \chi}^0_W$,  $n=3,4$ means a mostly Higgsino neutralino ${\tilde \chi}^0_{3,4}={\tilde \chi}^0_H$ and $n=5,6$ means a mostly third generation right-handed neutrino neutralino ${\tilde \chi}^0_{5,6}={\tilde \chi}^0_{\nu_3^c}$. 

Until now, we have used 2-component Weyl spinor notation for all of our matter fields. Since we are interested in the decays of physical particles, we will henceforth introduce and use 4-component spinor notation for the initial and final states of the interacting particles. The 4-component spinors are defined in terms of the 2-component Weyl spinors as
\begin{equation}\label{eq:DircaFermions}
\ell_i^-=\left(\begin{matrix}e_i\\ {e_i^c}^\dag\end{matrix}\right),~
\ell_i^+=\left(\begin{matrix}{e_i^c}\\ e_i^\dag\end{matrix}\right),~
\nu_i=\left(\begin{matrix}{\nu_i}\\ \nu_i^\dag\end{matrix}\right),~
{\tilde X}_1^-=\left(\begin{matrix}{\tilde \chi}^-_1\\{\tilde \chi}^{+\dag}_1\end{matrix}\right),~
{\tilde X}_1^+=\left(\begin{matrix}{\tilde \chi}^+_1\\{\tilde \chi}^{-\dag}_1\end{matrix}\right),~
{\tilde X}_n^0=\left(\begin{matrix}{\tilde \chi}_n^0\\{\tilde \chi}^{0\dag}_n\end{matrix}\right).\quad
\end{equation}
In our model, $\ell^\pm_i,\>X^\pm_1$ are Dirac fermions, while $\nu_i,\>X^0_n$ are Majorana fermions. Note that, for simplicity, we use the same symbol, $\nu_{i}$, for both a Weyl and Majorana neutrino. The Lagrangian (\ref{eq:633}) then becomes 
\small
\begin{multline*}\label{eq:gauge_amplitudes}
\mathcal{L}_{{\tilde \chi}^{\pm,\>0}\rightarrow Z^0, W^\pm-\text{lepton}}\supset \\g_2Z^0_{\mu}\bar {{\tilde X}}^-_1\gamma^\mu\Bigg[\Bigg(
-\frac{1}{c_W}
\left(\frac{1}{2}-s_W^2\right)\mathcal{U}_{2+j\>1}\mathcal{U}_{2+j\>2+i}^*
-\frac{1}{c_W}\left(\frac{1}{2}-s_W^2 \right)\mathcal{U}_{1\>2}\mathcal{U}^*_{2+i\>2}-c_W\mathcal{U}^*_{2+i\>1}\mathcal{U}_{1\>1}
\Bigg)P_L\\+
\Bigg(\frac{1}{c_W} s_W^2\mathcal{V}_{2+i\>2+j}\mathcal{V}_{1\>2+j}^*-\frac{1}{c_W}\left(\frac{1}{2}-s_W^2\right)\mathcal{V}_{2+i\>2}\mathcal{V}_{1\>2}^*
-c_W\mathcal{V}^*_{1\>1}\mathcal{V}_{2+i\>1}
\Bigg)P_R
 \Bigg]\ell_i^-\\
-g_2Z^0_{\mu}\bar {{\tilde X}}^+_1\gamma^\mu\Bigg[
\left(-\frac{1}{c_W}\left(\frac{1}{2}-s_W^2\right)\mathcal{U}^*_{1\>2+j}\mathcal{U}_{2+i\>2+j}
-\frac{1}{c_W}\left(\frac{1}{2}-s_W^2 \right)\mathcal{U}^*_{1\>2}\mathcal{U}_{2+i\>2}-c_W\mathcal{U}_{2+i\>1}\mathcal{U}^*_{1\>1}
\right)P_R\\+
\left(\frac{1}{c_W} s_W^2\mathcal{V}^*_{2+i\>2+j}\mathcal{V}_{1\>2+j}-\frac{1}{c_W}\left(\frac{1}{2}-s_W^2\right)\mathcal{V}^*_{2+i\>2}\mathcal{V}_{1\>2}
-c_W\mathcal{V}_{1\>1}\mathcal{V}^*_{2+i\>1}
\right)P_L
 \Bigg]\ell_i^+\\
+ \frac{g_2}{\sqrt 2}W_\mu^- \bar {{\tilde X}}^-_1 \gamma^\mu \Big[(\mathcal{U}_{1\>2}\mathcal{N}^*_{6+i\>3}+\mathcal{U}_{1\>2+j}\mathcal{N}^*_{6+j\>6+i}-\sqrt{2}\mathcal{N}^*_{6+i\>2}\mathcal{U}_{1\>1})P_L
\hfill-(\mathcal{N}_{6+i\>4}\mathcal{V}^*_{1\>2}+\sqrt{2}\mathcal{V}^*_{1\>1}\mathcal{N}_{6+i\>2})P_R \Big]\nu_i\\
-\frac{g_2}{\sqrt 2}W_\mu^+\bar {{\tilde X}}^+_1 \gamma^\mu\Big[
(\mathcal{U}^*_{1\>2}\mathcal{N}_{6+i\>3}+\mathcal{U}^*_{1\>2+j}\mathcal{N}_{6+j\>6+i}-\sqrt{2}\mathcal{N}_{6+i\>2}\mathcal{U}^*_{1\>1})P_R
\hfill-(\mathcal{N}^*_{6+i\>4}\mathcal{V}_{1\>2}+\sqrt{2}\mathcal{V}_{1\>1}\mathcal{N}^*_{6+i\>2})P_L
 \Big] { \nu}_{i}\\
 +\frac{g_2}{\sqrt 2}W_\mu^-\bar {{\tilde X}}^0_n \gamma^\mu
\Big[
(\mathcal{N}_{n\>4}\mathcal{V}^*_{2+i\>2}+\sqrt{2}\mathcal{V}^*_{2+i\>1}\mathcal{N}_{n\>2})P_L+(-\mathcal{U}_{2+i\>2+j}\mathcal{N}^*_{n\>6+j}-\mathcal{U}_{2+i\>2}\mathcal{N}^*_{n\>3}+\sqrt{2}\mathcal{N}^*_{n\>2}\mathcal{U}_{2+i\>1})P_R
\Big]\ell_i^+\\
-\frac{g_2}{\sqrt 2}W_\mu^+\bar {{\tilde X}}^0_n \gamma^\mu
\Big[
(\mathcal{N}^*_{n\>4}\mathcal{V}_{2+i\>2}+\sqrt{2}\mathcal{V}_{2+i\>1}\mathcal{N}^*_{n\>2})P_R+(-\mathcal{U}^*_{2+i\>2+j}\mathcal{N}_{n\>6+j}-\mathcal{U}^*_{2+i\>2}\mathcal{N}_{n\>3}+\sqrt{2}\mathcal{N}_{n\>2}\mathcal{U}^*_{2+i\>1})P_L
\Big]\ell_i^-\\
+{g_2}Z^0_\mu \bar {{\tilde X}}^0_n \gamma^{\mu}\Big[
\Big(\frac{1}{2c_W}\mathcal{N}_{n\>6+j}\mathcal{N}^*_{6+j\>6+i}-\frac{1}{c_W}\left(\frac{1}{2}+s_W^2\right)\mathcal{N}_{n\>4}\mathcal{N}^*_{6+i\>4} \Big)P_L
- \frac{1}{c_W}\left(\frac{1}{2}+s_W^2\right) \mathcal{N}_{n\>3}\mathcal{N}^*_{6+i\>3}P_R
\Big]\nu_i\\
\end{multline*}
\begin{multline}
-{g_2}Z^0_\mu \bar {{\tilde X}}^0_n \gamma^{\mu}\Big[
\Big(\frac{1}{2c_W}\mathcal{N}^*_{n\>6+j}\mathcal{N}_{6+j\>6+i}-\frac{1}{c_W}\left(\frac{1}{2}+s_W^2\right)\mathcal{N}^*_{n\>4}\mathcal{N}_{6+i\>4} \Big)P_R
- \frac{1}{c_W}\left(\frac{1}{2}+s_W^2\right) \mathcal{N}^*_{n\>3}\mathcal{N}_{6+i\>3}P_L
\Big]\nu_i\\
-e\gamma^0_{\mu} \bar {{\tilde X}}^-_1\gamma^\mu\Bigg[\big(
\mathcal{U}_{2+j\>1}\mathcal{U}_{2+i\>2+j}^*
+\mathcal{U}_{1\>2}\mathcal{U}^*_{2+i\>2}+\mathcal{U}^*_{2+i\>1}\mathcal{U}_{1\>1}
\big)P_L+
\big(\mathcal{V}_{2+i\>2+j}\mathcal{V}_{1\>2+j}^*+\mathcal{V}_{2+i\>2}\mathcal{V}_{1\>2}^*+\mathcal{V}^*_{1\>1}\mathcal{V}_{2+i\>1}
\big)P_R
\Bigg]\ell_i^-\\
+e\gamma^0_{\mu}\bar {{\tilde X}}^+_1\gamma^\mu\Bigg[\big(
\mathcal{U}^*_{2+j\>1}\mathcal{U}_{2+i\>2+j}
+\mathcal{U}^*_{1\>2}\mathcal{U}_{2+i\>2}+\mathcal{U}_{2+i\>1}\mathcal{U}^*_{1\>1}
\big)P_R +
\big(\mathcal{V}^*_{2+i\>2+j}\mathcal{V}_{1\>2+j}+\mathcal{V}^*_{2+i\>2}\mathcal{V}_{1\>2}
+\mathcal{V}_{1\>1}\mathcal{V}^*_{2+i\>1}
\big)P_L
\Bigg]\ell_i^+
\end{multline}
\normalsize
where we sum over all neutralino states $n=1,2,3,4,5,6$, all lepton families $i=1,2,3$ and $j=1,2,3$. $P_L$ and $P_R$ are the projection operators $\frac{1-\gamma^5}{2}$ and $\frac{1+\gamma^5}{2}$ respectively. 

It is important to note, however, that the last terms in this expression--that is, those proportional to the photon $\gamma_{0}$--exactly {\it cance}l. This occurs 
because the unitary matrices $\mathcal{U}$ and $\mathcal{V}$ satisfy the identities
\begin{equation}
\mathcal{U}^*_{2+j\>1}\mathcal{U}_{2+i\>2+j}
+\mathcal{U}^*_{1\>2}\mathcal{U}_{2+i\>2}+\mathcal{U}_{2+i\>1}\mathcal{U}^*_{1\>1}=0
\end{equation}
and
\begin{equation}
\mathcal{V}^*_{2+i\>2+j}\mathcal{V}_{1\>2+j}+\mathcal{V}^*_{2+i\>2}\mathcal{V}_{1\>2}
+\mathcal{V}_{1\>1}\mathcal{V}^*_{2+i\>1}=0 \ .
\end{equation}
It follows that the amplitude for the decay channel $\tilde X^\pm_1\rightarrow \gamma_0\ell^\pm$ vanishes. Therefore,  charginos cannot decay into a photon and a lepton

\subsection{\boldmath${\tilde \chi}^{\pm, 0}\rightarrow$$h^0$-lepton }

The part  of the Lagrangian responsible for the Higgs boson-lepton decay channels is

\begin{equation}\label{eq:636}
\mathcal{L}_{{\tilde \chi}^{\pm,\>0}\rightarrow h^0- \text{lepton}} \supset -\sqrt{2}g(\phi^{i*}T^a \psi_i)\lambda^a
-\sqrt{2}g\lambda^{a\dag}(\psi^{i\dag}T^a\phi_i)\\-\frac{1}{2}Y^{ijk}\phi_i\psi_j\psi_k-\frac{1}{2}Y^*_{ijk}\phi^{i*}\psi^{j\dag}\psi^{k\dag},
\end{equation}
where this expression represents the sum over $SU(2)_L$ and $U(1)_Y$. The index $i=1,2,3$ sums over the three lepton families.
The terms with Yukawa couplings arising from the $B-L$ MSSM superpotential
%
%
enter the Lagrangian as 
\begin{equation}\label{eq:Higgs1}
 -Y_{e_i}(H_d^0 e_i {e_i^c}+{H_d^0}^*{e_i^c}^{\dag}e_i^{\dag})
+Y_{\nu_i}(H_u^0 \nu_{i}{\nu^c}_{i}+{H_u^0}^*{\nu^c}_{i}^{\dag}\nu_{i}^{\dag}).
\end{equation}
Notice that we have kept only the terms with neutral Higgs scalar components, since only those have a Higgs boson mass eigenstate component $h^0$. Other terms responsible for this decay channel arise from the supercovariant derivatives of the Higgs fields of the type
\begin{equation}
-\sqrt{2}g(\phi_i^*T^a\psi_i)\lambda^a+h.c.
\end{equation}
in Lagrangian (\ref{eq:636}).
For the $B-L$ MSSM, these produce the terms
\begin{multline}\label{eq:Higgs2}
-\frac{1}{\sqrt{2}}g_2(H_u^{+^*}\tilde H_u^0)\tilde W^+-\frac{1}{\sqrt{2}}
g_2(H_u^{0^*}\tilde H_u^+)\tilde W^-
 -\frac{1}{\sqrt{2}}g_2(H_d^{0^*}\tilde H_d^-)\tilde W^+ -\frac{1}{\sqrt{2}}g_2(H_d^{-^*}
\tilde H_d^0)\tilde W^-\\
-\frac{1}{\sqrt{2}}g_2(H_u^{+^*}\tilde H_u^+)\tilde W^0+\frac{1}{\sqrt{2}}
g_2(H_u^{0^*}\tilde H_u^0)\tilde W^0
 +\frac{1}{\sqrt{2}}g_2(H_d^{-^*}\tilde H_d^-)\tilde W^0 -\frac{1}{\sqrt{2}}g_2(H_d^{0^*}
\tilde H_d^0)\tilde W^0\\
-\frac{1}{\sqrt{2}}g'(H_u^{+^*}\tilde H_u^+)\tilde B-\frac{1}{\sqrt{2}}
g'(H_u^{0^*}\tilde H_u^0)\tilde B
 +\frac{1}{\sqrt{2}}g'(H_d^{-^*}\tilde H_d^-)\tilde B +\frac{1}{\sqrt{2}}g'(H_d^{0^*}
\tilde H_d^0)\tilde B+h.c.
\end{multline}
 It follows that the part of the Lagrangian responsible for the ${\tilde \chi}^{\pm,\>0}\rightarrow h^0- \text{lepton}$ decays is the sum of the equations \eqref{eq:Higgs1} and \eqref{eq:Higgs2}. Note that these expressions are written in terms of the gauge eigenstates. As in the previous section, we expand the gauge eigenstates in terms of the mass eigenstates $\tilde \chi_1^\pm$, $\tilde \chi_n^0$, $e_i$, $\nu_i$ for $n=1,2,3,4,5,6$, $i=1,2,3$ and their hermitian conjugates. Once this step is completed, we group the terms into 4-component spinors to get
\small
\begin{multline}\label{eq:Higgs_amplitudes}
\mathcal{L}_{{\tilde \chi}^{\pm,\>0}\rightarrow h^0-\text{lepton}}\supset 
-\frac{1}{\sqrt{2}}Y_{e_i}\sin \alpha h^0\bar{ \tilde X}_1^-\Big[\mathcal{V}^*_{1\>2+j}\mathcal{U}^*_{2+i\>2+j}P_L+\mathcal{V}_{2+i\>2+j}\mathcal{U}_{1\>2+j}P_R\Big]\ell^-_j\\
 +\frac{1}{\sqrt{2}}Y_{e_i}\sin \alpha h^0\bar{ \tilde X}_1^+\Big[  \mathcal{V}^*_{2+i\>2+j}\mathcal{U}^*_{1\>2+j}P_L 
+\mathcal{V}_{1\>2+j}\mathcal{U}_{2+i\>2+j}P_R\Big]
\ell^+_j\\
+\frac{g_2}{{2}}
h^0 \bar {\tilde{{X}}}_1^- \Big[(-\cos \alpha \mathcal{V}^*_{1\>2}\mathcal{U}^*_{2+i\>1}-\sin \alpha
\mathcal{U}^*_{2+i\>2}\mathcal{V}_{1\>1}^*)P_L
+(-\cos\alpha \mathcal{V}_{2+i\>2}\mathcal{U}_{1\>1} -\sin \alpha  \mathcal{U}_{1\>2}\mathcal{V}_{2+i\>1})P_R
\Big] \ell_i^-\\
+\frac{g_2}{{2}}
h^0 \bar {\tilde{{ X}}}_1^+ \Big[ (\cos\alpha \mathcal{V}^*_{2+i\>2}\mathcal{U}^*_{1\>1} +\sin \alpha  \mathcal{U}^*_{1\>2}\mathcal{V}^*_{2+i\>1})P_L
+(\cos \alpha \mathcal{V}_{1\>2}\mathcal{U}_{2+i\>1}+\sin \alpha
\mathcal{U}_{2+i\>2}\mathcal{V}_{1\>1})P_R
\Big] \ell_i^+\\
+\frac{g_2}{{2}}\bar {\tilde{{ X}}}^0_n h^0\Big[\Big(
\cos \alpha (\mathcal{N}^*_{n\>4}\mathcal{N}^*_{6+i\>2}+\mathcal{N}^*_{6+i\>4}\mathcal{N}_{n\>2}^*)+\sin \alpha (\mathcal{N}^*_{n\>3}\mathcal{N}^*_{6+i\>2}+\mathcal{N}^*_{6+i\>3}\mathcal{N}_{n\>2}^*)\Big)P_L\\-
\Big(
\cos \alpha (\mathcal{N}_{n\>4}\mathcal{N}_{6+i\>2}+\mathcal{N}_{6+i\>4}\mathcal{N}_{n\>2})+\sin \alpha (\mathcal{N}_{n\>3}\mathcal{N}_{6+i\>2}+\mathcal{N}_{6+i\>3}\mathcal{N}_{n\>2})\Big)P_R
\Big]\nu_{i}\\
-\frac{g'}{{2}}\bar {\tilde{{ X}}}^0_n h^0\Big[\Big(
\cos\alpha \left(\sin \theta_R(\mathcal{N}^*_{n\>4}\mathcal{N}^*_{6+i\>1}+\mathcal{N}^*_{6+i\>4}\mathcal{N}^*_{n\>1})+\cos \theta_R(\mathcal{N}^*_{n\>4}\mathcal{N}^*_{6+i\>5}+\mathcal{N}^*_{6+i\>4}\mathcal{N}^*_{n\>5})   \right)\\
+\sin \alpha \left( \sin \theta_R(\mathcal{N}^*_{n\>3}\mathcal{N}^*_{6+i\>1}+\mathcal{N}^*_{6+i\>3}\mathcal{N}^*_{n\>1})+\cos \theta_R(\mathcal{N}^*_{n\>3}\mathcal{N}^*_{6+i\>5}+\mathcal{N}^*_{6+i\>3}\mathcal{N}^*_{n\>5}) \right)\Big)P_L\\
-\Big(\cos\alpha \left(\sin \theta_R(\mathcal{N}_{4\>n}\mathcal{N}_{1\>6+i}+\mathcal{N}_{4\>6+i}\mathcal{N}_{n\>1})+\cos \theta_R(\mathcal{N}_{n\>4}\mathcal{N}_{6+i\>5}+\mathcal{N}_{6+i\>4}\mathcal{N}_{n\>5})   \right)\\
+\sin \alpha \left( \sin \theta_R(\mathcal{N}_{n\>3}\mathcal{N}_{6+i\>1}+\mathcal{N}_{6+i\>3}\mathcal{N}_{n\>1})+\cos \theta_R(\mathcal{N}_{n\>3}\mathcal{N}_{6+i\>5}+\mathcal{N}_{6+i\>3}\mathcal{N}_{n\>5}) \right)\Big)P_R
 \Big]\nu_{i}\\
+\frac{1}{\sqrt 2}Y_{\nu3i}\cos\alpha {\tilde{{X}}}^0_n h^0
\Big[\Big(-\mathcal{N}^*_{n\>6+i}\mathcal{N}^*_{6+j\>6}
+\mathcal{N}^*_{6+j\>6+i}
\mathcal{N}^*_{n\>6}\Big)P_L
+\Big(-\mathcal{N}_{n\>6+i}\mathcal{N}_{6+j\>6}
+\mathcal{N}_{6+j\>6+i}
\mathcal{N}_{n\>6}\Big)P_R
\Big]\nu_{j}
\end{multline}
\normalsize
where the angle $\alpha$ is defined in equation (\ref{eq:alpha}) and we sum over all lepton families $i,j=1,2,3$.

One now has the information required to compute the amplitude for each of the processes listed in Table \ref{tab:decay_channels}. To do this, we need the exact expression for the vertex coefficient associated with each such process. These can be read off from the Lagrangians in Eqs. \eqref{eq:gauge_amplitudes} and \eqref{eq:Higgs_amplitudes}.
For example, consider the the decay channel $\tilde X_1^{-}\rightarrow Z^0\ell^{-}_i$. Then it follows from \eqref{eq:gauge_amplitudes} that the vertex coupling is
\begin{equation}
g_{\tilde X^-_1\rightarrow Z^0\ell^-_i}
={G_L}_{\tilde X^-_1\rightarrow Z^0\ell^-_i}P_L+ {G_R}_{\tilde X^-_1\rightarrow Z^0\ell^-_i} P_R \ ,
\end{equation}
where 
\begin{equation}
{G_L}_{\tilde X^-_1\rightarrow Z^0\ell^-_i}=g_{2}\gamma^{\mu}\left(\frac{1}{c_W}
\left(\frac{1}{2}-s_W^2\right)\mathcal{U}_{2+j\>1}\mathcal{U}_{2+j\>2+i}^*
-\frac{1}{c_W}\left(\frac{1}{2}-s_W^2 \right)\mathcal{U}_{1\>2}\mathcal{U}^*_{2+i\>2}+2c_W\mathcal{U}^*_{2+i\>1}\mathcal{U}_{1\>1}
\right) 
\label{hanger1}
\end{equation}
and
\begin{equation}
{G_R}_{\tilde X^-_1\rightarrow Z^0\ell^-_i}=g_{2} \gamma^{\mu}\left(\frac{1}{c_W} s_W^2\mathcal{V}_{2+i\>2+j}\mathcal{V}_{1\>2+j}^*-\frac{1}{c_W}\left(\frac{1}{2}-s_W^2\right)\mathcal{V}_{2+i\>2}\mathcal{V}_{1\>2}^*
+2c_W\mathcal{V}^*_{1\>1}\mathcal{V}_{2+i\>1}
\right) \ .
\label{hanger2}
\end{equation}
 However, their form and derivations are somewhat cumbersome. With this in mind,we provide a series of diagrams to pictorially express the origin of the interaction terms in the Lagrangian.


\subsection{Chargino decay diagrams}

\subsubsection{\boldmath${\tilde X}_1^{+}\rightarrow W^{+}\nu_{{i}}$}

\begin{figure}[H]
 \begin{minipage}{0.48\textwidth}
     \centering
   \begin{subfigure}[b]{0.49\linewidth}
   \centering
       \includegraphics[width=0.77\textwidth]{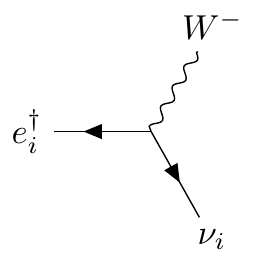}
\caption*{(a1)\\ $ \frac{g_2}{\sqrt{2}} \bar \sigma^\mu$}
       \label{fig:table2}
   \end{subfigure} \hspace{0.01\linewidth}\\
   \centering
   \begin{subfigure}[b]{0.49\linewidth}
   \centering
       \includegraphics[width=0.77\textwidth]{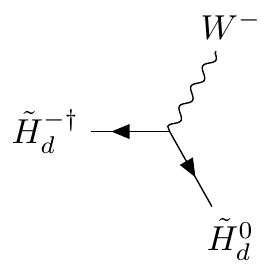}
\caption*{(a2)\\ $\frac{g_2}{\sqrt{2}} \bar \sigma^{\mu}$}
       \label{fig:table2}
\end{subfigure}
   \centering
     \begin{subfigure}[b]{0.49\textwidth}
   \centering
\includegraphics[width=0.77\textwidth]{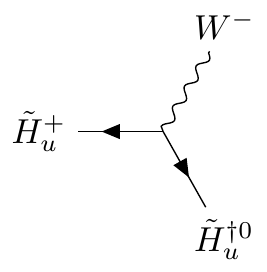}
\caption*{(a3) \\$\frac{g_2}{\sqrt{2}} \bar \sigma^{\mu}$}
\end{subfigure}\\
   \centering
   \begin{subfigure}[b]{0.49\textwidth}
   \centering
\includegraphics[width=0.77\textwidth]{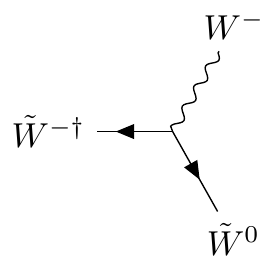}
\caption*{(a4) \\$ -g_2 \bar \sigma^{\mu}$}
\end{subfigure}
   \centering
   \begin{subfigure}[b]{0.49\textwidth}
   \centering
   \includegraphics[width=0.77\textwidth]{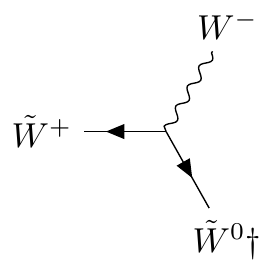}    
\caption*{(a5) \\$ g_2 \bar \sigma^{\mu}$}
\end{subfigure}\subcaption{ }\label{fig:X11}
\end{minipage}\hfill
\rulesep
 \begin{minipage}{0.48\textwidth}
     \centering
   \begin{subfigure}[b]{0.49\linewidth}
   \centering
    \includegraphics[width=0.77\textwidth]{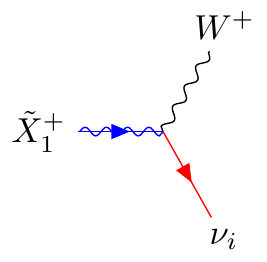} 
       \caption*{\centerline{(b1)}\\$+\frac{g_2}{\sqrt{2}}\textcolor{red}{\mathcal{U}_{1\>2+j}}
\textcolor{blue}{\mathcal{N}^*_{6+j\>6+i}} \gamma^{\mu}P_L$}
       \label{fig:table2}
   \end{subfigure}\\
   \begin{subfigure}[b]{0.49\linewidth}
   \centering
      \includegraphics[width=0.77\textwidth]{651b.pdf} 
\caption*{\centerline{(b2)}\\$\frac{g_2}{\sqrt{2}}\textcolor{blue}{\mathcal{U}_{1\>2}}
\textcolor{red}{\mathcal{N}^*_{6+i\>3}} \gamma^{\mu}P_L$}
       \label{fig:table2}
\end{subfigure}
   \begin{subfigure}[b]{0.49\textwidth}
   \centering
       \includegraphics[width=0.77\textwidth]{651b.pdf} 
\caption*{\centerline{(b3)}\\$-\frac{g_2}{\sqrt{2}}\textcolor{blue}{\mathcal{V}^*_{1\>2}}
\textcolor{red}{\mathcal{N}_{6+i\>4}} \gamma^{\mu}P_R$}
       \label{fig:X1}
\end{subfigure}\\
   \begin{subfigure}[b]{0.49\textwidth}
   \centering
        \includegraphics[width=0.77\textwidth]{651b.pdf} 
\caption*{\centerline{(b4)}\\$-g_2\textcolor{blue}{\mathcal{U}_{1\>1}}
\textcolor{red}{\mathcal{N}^*_{6+i\>2}} \gamma^{\mu}P_L$}
       \label{fig:X1}
\end{subfigure}
   \begin{subfigure}[b]{0.49\textwidth}
   \centering
       \includegraphics[width=0.77\textwidth]{651b.pdf} 
\caption*{\centerline{(b5)}\\ $-g_2\textcolor{blue}{\mathcal{V}^*_{1\>1}}
\textcolor{red}{\mathcal{N}_{6+i\>2}} \gamma^{\mu}P_R$}
       \label{fig:X1}
\end{subfigure}\\
\subcaption{  }\label{fig:X12}
\end{minipage}
\caption{a) We show the vertices as they 
appear in the MSSM Lagrangian in terms of the gauge eigenstates, expressed as 2-component Weyl fermions. Vertices (a1), (a2)  and (a3)
arise from the covariant derivatives of the lepton and Higgsino matter fields, respectively.
Vertices (a4) and (a5) come from the covariant derivative of the non-Abelian gaugino fields. b) We 
express the interactions in terms of the mass eigenstates relevant for the  chargino decay into SM 
particles, expressed as 4-component Dirac fermions. We assume ${\tilde X}_1$ is the lightest chargino and is either dominantly charged Higgsino or charged Wino. 
The "red" matrix elements are proportional to $\epsilon_i/M_{soft}$,  while "blue" matrix elements are of order unity with small RPV corrections $1-\epsilon_i/M_{soft}$. At first order, the decay amplitudes are proportional to  $(1-\epsilon_i/M_{soft})\times \epsilon_i/M_{soft} \simeq \epsilon_i/M_{soft}$.}\label{fig:X1}
\end{figure}

The vertices associated with positively charged chargino decays are shown in Figures \ref{fig:X11} and \ref{fig:X12}. The vertices associated with the negatively charged chargino decays are the hermitian conjugates of those. 
The diagrams in Figure \ref{fig:X11} are expressed in terms of gauge eigenstates written as 2-component Weyl spinors. The diagrams in Figure \ref{fig:X12} are the Feynman diagrams of the same vertices, in terms of 4-component mass eigenstates.

\subsubsection{\boldmath${\tilde X}_1^{+} \rightarrow Z^0 \ell_i^+$}

\begin{figure}[H]
\begin{minipage}{0.48\textwidth}
     \centering
   \begin{subfigure}[b]{0.49\textwidth}
   \centering
     \includegraphics[width=0.76\textwidth]{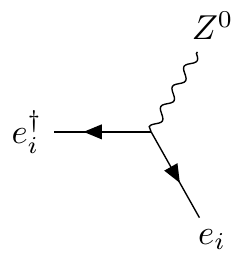} 
\caption*{\centerline{(a1)}\\ \centerline{$ -\frac{g_2}{c_W}\left(\frac{1}{2}-s_W^2\right) \bar \sigma^\mu$}}
       \label{fig:table2}
   \end{subfigure}
   \begin{subfigure}[b]{0.49\textwidth}
   \centering
        \includegraphics[width=0.76\textwidth]{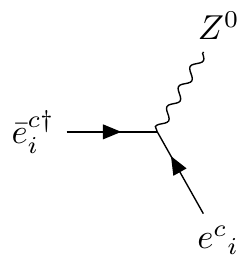} 
\caption*{\centerline{(a2)}\\ \centerline{$-\frac{g_2}{c_W}s_W^2 \bar \sigma^\mu$}}
       \label{fig:table2}
   \end{subfigure}\ \\
   \begin{subfigure}[b]{0.49\textwidth}
   \centering
       \includegraphics[width=0.76\textwidth]{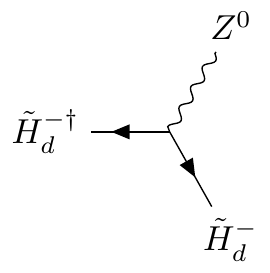} 
\caption*{\centerline{(a3)}\\ \centerline{$-\frac{g_2}{c_W}\left(\frac{1}{2}-s_W^2\right) \bar \sigma^{\mu}$}}
       \label{fig:table2}
       \bigskip
\end{subfigure}
   \begin{subfigure}[b]{0.49\textwidth}
   \centering
      \includegraphics[width=0.76\textwidth]{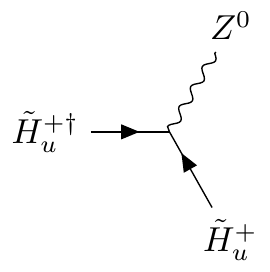} 
\caption*{\centerline{(a4)}\\ \centerline{$\frac{g_2}{c_W}\left(\frac{1}{2}-s_W^2\right)\bar \sigma^{\mu}$}}
       \label{fig:table2}
\bigskip
 \end{subfigure}\\
   \begin{subfigure}[b]{0.49\textwidth}
   \centering
      \includegraphics[width=0.76\textwidth]{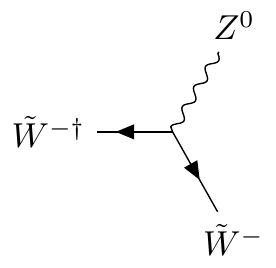}
\caption*{\centerline{(a5)}\\ \centerline{$-g_2c_W\bar \sigma^{\mu}$}}
       \label{fig:table2} 
\end{subfigure}
   \begin{subfigure}[b]{0.49\textwidth}
   \centering
      \includegraphics[width=0.76\textwidth]{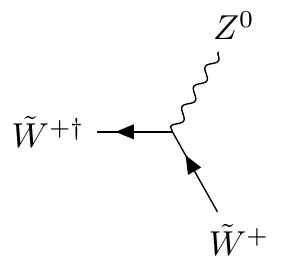}
 \caption*{\centerline{(a6)}\\ \centerline{$g_2c_W\bar  \sigma^{\mu}$}}
       \label{fig:table2}
\end{subfigure}
    \subcaption{ }\label{fig:X21}
\end{minipage}\hfill
\rulesep
\begin{minipage}{0.48\textwidth}
     \centering
   \begin{subfigure}[b]{0.49\textwidth}
   \centering
 \includegraphics[width=0.76\textwidth]{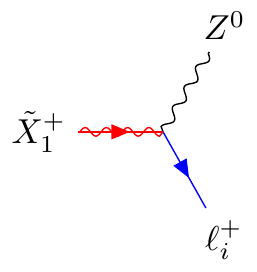}
      \caption*{\centerline{(b1)}\\ \centerline{$ -\frac{g_2}{c_W}\left(\frac{1}{2}-s_W^2\right) \textcolor{red}{\mathcal{U}_{1\>2+i}}\gamma^{\mu}P_L$}}
       \label{fig:table2}
   \end{subfigure}\
   \begin{subfigure}[b]{0.49\textwidth}
   \centering
 \includegraphics[width=0.76\textwidth]{652b.pdf}
       \caption*{\centerline{(b2)}\\ \centerline{$ \frac{g_2}{c_W}s_W^2 \textcolor{red}{\mathcal{V}^*_{1\>2+i}} \gamma^{\mu}P_R$}}
       \label{fig:table2}
   \end{subfigure}\\
   \begin{subfigure}[b]{0.49\textwidth}
   \centering
  \includegraphics[width=0.76\textwidth]{652b.pdf}
\caption*{\centerline{(b3)}\\ $-\frac{g_2}{c_W}\left(\frac{1}{2}-s_W^2\right) \times \\ \times \textcolor{blue}
{\mathcal{U}_{ 1\>2}}
\textcolor{red}{\mathcal{U}^*_{2+i\>2}} \gamma^{\mu}P_L$}
       \label{fig:table2}
\end{subfigure}
   \centering
   \begin{subfigure}[b]{0.49\textwidth}
   \centering
       \includegraphics[width=0.76\textwidth]{652b.pdf}
\caption*{\centerline{(b4)}\\ $-\frac{g_2}{c_W}\left(\frac{1}{2}-s_W^2\right) \times \\ \times \textcolor{blue}
{\mathcal{V}^*_{ 1\>2}}
\textcolor{red}{\mathcal{V}_{2+i\>2}} \gamma^{\mu}P_R$}
       \label{fig:table2}
 \end{subfigure}\\
   \begin{subfigure}[b]{0.49\textwidth}
   \centering
        \includegraphics[width=0.76\textwidth]{652b.pdf}
\caption*{\centerline{(b5)} \\\centerline{$-g_2c_W\textcolor{blue}{\mathcal{U}_{ 1\>1}}
\textcolor{red}{\mathcal{U}^*_{2+i\>1}} \gamma^{\mu}P_L$}}
       \label{fig:table2}
\end{subfigure}
   \begin{subfigure}[b]{0.49\textwidth}
   \centering
      \includegraphics[width=0.76\textwidth]{652b.pdf}
       \caption*{\centerline{(b6)}\\ \centerline{ $-g_2c_W\textcolor{blue}{\mathcal{V}^*_{ 1\>1}} \textcolor{red}{\mathcal{V}_{2+i\>1}} \gamma^{\mu}P_R$}}
       \label{fig:table2}
\end{subfigure}\\  
    \subcaption{}\label{fig:X22}
\end{minipage}
 \caption{a) We show the vertices as they 
appear in the MSSM Lagrangian in terms of the gauge eigenstates, expressed as 2-component Weyl fermions. Vertices (a1), (a2)  and (a3)
arise from the covariant derivatives of the lepton and Higgsino matter fields, respectively.
Vertices (a4) and (a5) come from the covariant derivative of the non-Abelian gaugino fields. b) We 
express the interactions in terms of the mass eigenstates relevant for the  chargino decay into SM 
particles, expressed as 4-component Dirac fermions. We assume ${\tilde X}_1$ is the lightest chargino and is either dominantly charged Higgsino or charged Wino. 
The "red" matrix elements have small values and are proportional to $\epsilon_i/M_{soft}$,  while "blue" matrix elements are of order unity with small RPV corrections of the form $1-\epsilon_i/M_{soft}$. At first order, the decay amplitudes are proportional to  $(1-\epsilon_i/M_{soft})\times \epsilon_i/M_{soft} \simeq \epsilon_i/M_{soft}$.}\label{fig:X2}
\end{figure}

The vertices associated with positively charged chargino decays are shown in Figures \ref{fig:X21} and \ref{fig:X22}. The vertices associated with the negatively charged chargino decays are the  hermitian conjugates of those. 
The diagrams in Figure \ref{fig:X21} are expressed in terms of gauge eigenstates written as 2-component Weyl spinors. The diagrams in Figure \ref{fig:X22} are the Feynman diagrams of the same vertices, in terms of 4-component mass eigenstates.

\subsubsection{\boldmath${\tilde X}_1^{+}\rightarrow h^0 \ell_i^+$}

\begin{figure}[H]
   \centering
   \begin{minipage}{0.48\textwidth}
   \begin{subfigure}[b]{0.49\textwidth}
   \centering
     \includegraphics[width=0.75\textwidth]{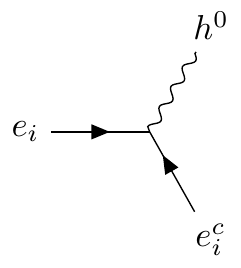}
\caption*{\centerline{(a1)}\\\centerline{ $ -\frac{1}{\sqrt 2}Y_{e_i} \sin \alpha$}}
       \label{fig:table2}
   \end{subfigure}
   \begin{subfigure}[b]{0.49\textwidth}
   \centering
      \includegraphics[width=0.75\textwidth]{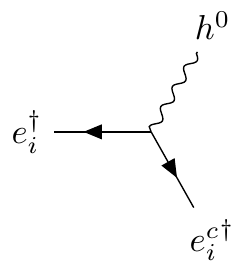}
\caption*{\centerline{(a2)}\\ \centerline{ $ -\frac{1}{\sqrt 2}Y_{e_i}\sin \alpha $}}
       \label{fig:table2}
   \end{subfigure}\\
   \begin{subfigure}[b]{0.49\textwidth}
   \centering
      \includegraphics[width=0.75\textwidth]{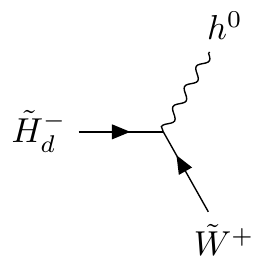}
\caption*{\centerline{(a3)}\\  \centerline{ $\frac{1}{ 2}g_2\sin \alpha$}}
       \label{fig:table2}
\end{subfigure}
   \begin{subfigure}[b]{0.49\textwidth}
   \centering
       \includegraphics[width=0.75\textwidth]{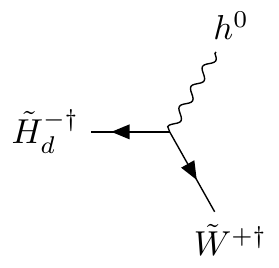}
\caption*{\centerline{(a4)}\\ \centerline{$\frac{1}{ 2}g_2\sin \alpha$}}
       \label{fig:table2}
 \end{subfigure}\\
   \begin{subfigure}[b]{0.49\textwidth}
   \centering
        \includegraphics[width=0.75\textwidth]{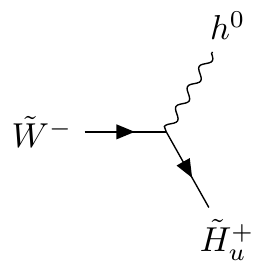}
\caption*{\centerline{(a5)}\\ \centerline{ $-\frac{1}{2}g_2\cos \alpha$}}
\bigskip
       \label{fig:table2}
\end{subfigure}
   \begin{subfigure}[b]{0.49\textwidth}
   \centering
      \includegraphics[width=0.75\textwidth]{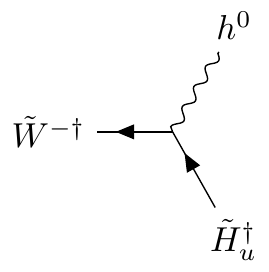}
\caption*{\centerline{(a6)}\\ \centerline{ $-\frac{1}{2}g_2\cos \alpha$}}
\bigskip
       \label{fig:table2}
 \end{subfigure}
    \subcaption{}\label{fig:X41}
\end{minipage}\hfill
\rulesep\begin{minipage}{0.48\textwidth}
\centering
   \begin{subfigure}[b]{0.49\textwidth}
   \centering
 \includegraphics[width=0.75\textwidth]{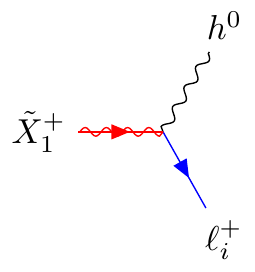}
       \caption*{\centerline{(b1)}\\ $ -\frac{1}{\sqrt 2}Y_{e_i}\sin \alpha 
\textcolor{red}{\mathcal{V}^*_{1\> 2+i}}P_L$}
       \label{fig:table2}
   \end{subfigure}
   \begin{subfigure}[b]{0.49\textwidth}
   \centering
       \includegraphics[width=0.75\textwidth]{654b.pdf}
       \caption*{\centerline{(b2)}\\ $-\frac{1}{\sqrt 2}Y_e\sin \alpha  \textcolor{red}{\mathcal{U}_{1\>2+i}}P_R$}
       \label{fig:table2}
   \end{subfigure}\\
   \begin{subfigure}[b]{0.49\textwidth}
   \centering
       \includegraphics[width=0.75\textwidth]{654b.pdf}
\caption*{\centerline{(b3)}\\ $-\frac{1}{ 2}g_2\sin \alpha \textcolor{blue}{\mathcal{V}^*_{ 1\>1}}
\textcolor{red}{\mathcal{U}^*_{2+i\>2}} P_L$}
       \label{fig:table2}
\end{subfigure}
   \begin{subfigure}[b]{0.49\textwidth}
   \centering
   \includegraphics[width=0.75\textwidth]{654b.pdf}
\caption*{\centerline{(b4) }\\ $-\frac{1}{ 2}g_2 \sin \alpha \textcolor{blue}{\mathcal{U}_{ 1\>2}}
\textcolor{red}{\mathcal{V}_{2+i\>1}} P_R$}
       \label{fig:table2}
 \end{subfigure}\\
   \begin{subfigure}[b]{0.49\textwidth}
   \centering
      \includegraphics[width=0.75\textwidth]{654b.pdf}
\caption*{\centerline{(b5)}\\ $-\frac{1}{2}g_2\cos \alpha \times \\ \times \textcolor{blue}{\mathcal{V}^*_{ 1\>2}}
\textcolor{red}{\mathcal{U}^*_{2+i\>1}} P_L$}
       \label{fig:table2}
\end{subfigure}
   \begin{subfigure}[b]{0.49\textwidth}
   \centering
       \includegraphics[width=0.75\textwidth]{654b.pdf}
\caption*{\centerline{(b6)} $-\frac{1}{ 2}g_2 \cos \alpha \times \\ \times \textcolor{blue}{\mathcal{U}_{ 1\>1}}
\textcolor{red}{\mathcal{V}_{2+i\>2}} P_R$}
       \label{fig:table2}
 \end{subfigure}\\
    \subcaption{}\label{fig:X42}
\end{minipage}
\caption{a) We show the vertices as they 
appear in the MSSM Lagrangian in terms of the gauge eigenstates, expressed as 2-component Weyl fermions. Vertices (a1), (a2)  and (a3)
arise from the covariant derivatives of the lepton and Higgsino matter fields, respectively.
Vertices (a4) and (a5) come from the covariant derivative of the non-Abelian gaugino fields. b) We 
express the interactions in terms of the mass eigenstates relevant for the  chargino decay into SM 
particles, expressed as 4-component Dirac fermions. We assume ${\tilde X}_1$ is the lightest chargino and is either dominantly charged Higgsino or charged Wino. 
The "red" matrix elements have small values and are proportional to $\epsilon_i/M_{soft}$,  while "blue" matrix elements are of order unity with small RPV corrections of the form $1-\epsilon_i/M_{soft}$. At first order, the decay amplitudes are proportional to  $(1-\epsilon_i/M_{soft})\times \epsilon_i/M_{soft} \simeq \epsilon_i/M_{soft}$.}
\label{fig:X4}
\end{figure}

The vertices associated with positively charged chargino decays are shown in Figures \ref{fig:X41} and \ref{fig:X42}. The vertices associated with the negatively charged chargino decays are the hermitian conjugates of those. 
The diagrams in Figure \ref{fig:X41} are expressed in terms of gauge eigenstates written as 2-component Weyl spinors. The diagrams in Figure \ref{fig:X42} are the Feynman diagrams of the same vertices, written in terms of 4-component mass eigenstates.

\subsection{Neutralino decay diagrams}

\subsubsection{\boldmath${\tilde X}^0_n\rightarrow Z^0\nu_{i}$}

\begin{figure}[H]
\begin{minipage}{0.48\textwidth}
\centering
   \begin{subfigure}[b]{0.49\textwidth}
       \includegraphics[width=0.76\textwidth]{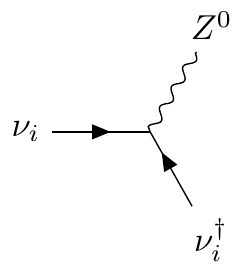} 
\caption*{\centerline{(a1)}\\ \centerline{ $ \frac{g_2}{2c_W}$}}
\bigskip
       \label{fig:table2}
   \end{subfigure}\\
   \begin{subfigure}[b]{0.49\textwidth}
       \includegraphics[width=0.76\textwidth]{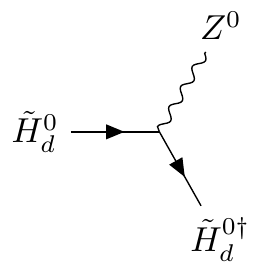} 
\caption*{\centerline{(a2)}\\ \centerline{ $\frac{g_2}{c_W}\left(\frac{1}{2}+s_W^2\right)  \sigma^{\mu}$}}
\bigskip
\bigskip
       \label{fig:table2}
\end{subfigure}
   \begin{subfigure}[b]{0.49\textwidth}
       \includegraphics[width=0.76\textwidth]{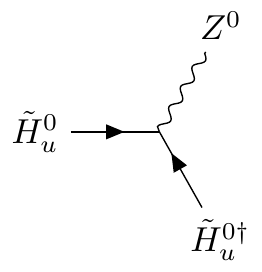} 
\caption*{\centerline{(a3)} \\ \centerline{ $-\frac{g_2}{c_W}\left(\frac{1}{2}+s_W^2\right) \sigma^{\mu}$}}
\bigskip
\bigskip
       \label{fig:table2}
 \end{subfigure}\\
    \subcaption{}\label{fig:X51}
\end{minipage}\hfill
\rulesep
\begin{minipage}{0.48\textwidth}
\centering
   \begin{subfigure}[b]{0.49\textwidth}
        \includegraphics[width=0.76\textwidth]{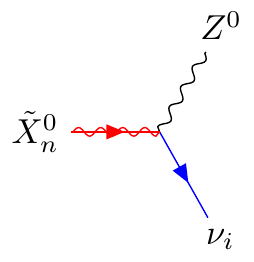} 
       \caption*{\centerline{(b1)} $\frac{g_2}{2c_W}[\textcolor{red}{\mathcal{N}_{n\>6+j}}\textcolor{blue}{\mathcal{N}^*_{6+j\>6+i}}P_L-\textcolor{red}{\mathcal{N}^*_{n\>6+j}}\textcolor{blue}{\mathcal{N}_{6+i\>6+j}}P_R]\gamma^{\mu}$}
       \label{fig:table2}
   \end{subfigure}\\
   \begin{subfigure}[b]{0.49\textwidth}
       \includegraphics[width=0.76\textwidth]{655b.pdf} 
\caption*{\centerline{(b2)}\\ $\frac{g_2}{c_W}\left(\frac{1}{2}+s_W^2\right) \times \\ \times 
[\textcolor{blue}{\mathcal{N}_{n\>3}^*}
\textcolor{red}{\mathcal{N}_{6+i\>3}}P_L\\-\textcolor{blue}{\mathcal{N}_{n\>3}}
\textcolor{red}{\mathcal{N}_{6+i\>3}^*}P_R ] \gamma^{\mu}$}
       \label{fig:table2}
\end{subfigure}
   \begin{subfigure}[b]{0.49\textwidth}
   \centering
        \includegraphics[width=0.76\textwidth]{655b.pdf} 
\caption*{\centerline{(b3) } \\$-\frac{g_2}{c_W}\left(\frac{1}{2}+s_W^2\right) \times \\ \times 
[\textcolor{blue}{\mathcal{N}_{n\>4}}
\textcolor{red}{\mathcal{N}^*_{6+i\>4}}P_L\\
-\textcolor{blue}{\mathcal{N}^*_{n\>4}}
\textcolor{red}{\mathcal{N}_{6+i\>4}}^*P_R] \gamma^{\mu}$}
       \label{fig:table2}
 \end{subfigure}\\
    \subcaption{}\label{fig:X52}
\end{minipage}
 \caption{a) We show the MSSM vertices, in terms of the gauge eigenstates, expressed as 2-component Weyl fermions.  (a1), (a2) and (a3) 
come from the covariant derivatives of the lepton and Higgsino matter fields, respectively.
Unlike for chargino, the lightest neutralino can be any state ${\tilde \chi}_n^0$ with $n=1$ Bino dominant,
$n=2$ for neutral Wino dominant, $n=3,4$ for neutral Higgsino dominant, and $n=5,6$ for right handed 
sterile neutrino dominant. b) We 
express the interactions in terms of the mass eigenstates relevant for neutralino decay into SM 
particles, expressed as 4-component fermions. 
The "red" matrix elements have small values and are proportional to $\epsilon_i/M_{soft}$  while with "blue" matrix elements that have small RPV corrections $1-\epsilon_i/M_{soft}$. At first order, the decay amplitudes are proportional to  $(1-\epsilon_i/M_{soft})\times \epsilon_i/M_{soft} \simeq \epsilon_i/M_{soft}$.
}\label{fig:X5}
\end{figure}

The vertices associated with neutralino decays are shown in Figures \ref{fig:X51} and \ref{fig:X52}. 
The diagrams in Figure \ref{fig:X51} are expressed in terms gauge eigenstates written as 2-component Weyl spinors. The diagrams in Figure \ref{fig:X52} are the same vertices in terms of mass eigenstates, expressed as 4-component spinors.

\subsubsection{\boldmath${\tilde X}_n^0 \rightarrow W^\pm \ell_i^\mp$}

\begin{figure}[H]
\centering
\begin{minipage}{0.48\textwidth}
\centering
   \begin{subfigure}[b]{0.49\textwidth}
        \includegraphics[width=0.76\textwidth]{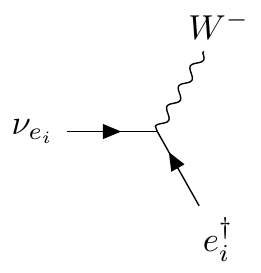} 
\caption*{\centerline{(a1)}\\ \centerline{ $ \frac{g_2}{\sqrt{2}} \bar \sigma^{\mu}$}}
       \label{fig:table2}
   \end{subfigure}\\
   \begin{subfigure}[b]{0.49\textwidth}
        \includegraphics[width=0.76\textwidth]{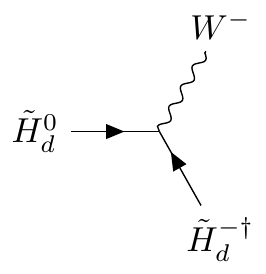} 
\caption*{\centerline{(a2)}\\ \centerline{ $\frac{g_2}{\sqrt{2}} \bar \sigma^{\mu}$}}
       \label{fig:table2}
\end{subfigure}
   \begin{subfigure}[b]{0.49\textwidth}
 \includegraphics[width=0.76\textwidth]{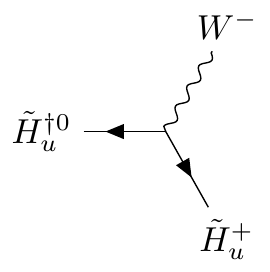} 
\caption*{\centerline{(a3)} \\ \centerline{ $\frac{g_2}{\sqrt{2}} \bar \sigma^{\mu}$}}
       \label{fig:table2}
 \end{subfigure}\\
   \begin{subfigure}[b]{0.49\textwidth}
   \centering
       \includegraphics[width=0.76\textwidth]{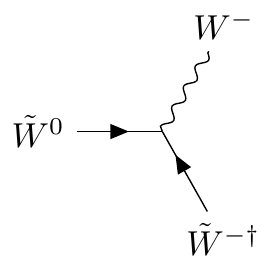} 
\caption*{\centerline{(a4)}\\ \centerline{ $-g_2 \bar \sigma^{\mu}$}}
       \label{fig:table2} 
\end{subfigure}
   \begin{subfigure}[b]{0.49\textwidth}
   \centering
    \includegraphics[width=0.76\textwidth]{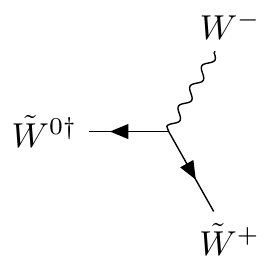}
      \caption*{\centerline{(a5)} \\\centerline{ $g_2 \bar \sigma^{\mu}$}}
       \label{fig:table2}
\end{subfigure}
    \subcaption{}\label{fig:X61}
\end{minipage}\hfill
\rulesep
\begin{minipage}{0.48\textwidth}
\centering
   \begin{subfigure}[b]{0.49\textwidth}
      \includegraphics[width=0.76\textwidth]{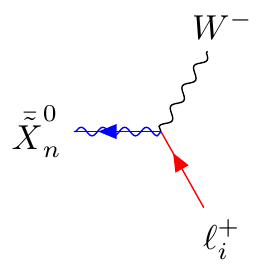}
       \caption*{\centerline{(b1)}\\ \centerline{ $-\frac{g_2}{\sqrt{2}}\textcolor{red}{\mathcal{N}^*_{n\>6+i}}
\textcolor{blue} \gamma^{\mu}P_R$}}
       \label{fig:table2}
   \end{subfigure}\\
   \begin{subfigure}[b]{0.49\textwidth}
           \includegraphics[width=0.76\textwidth]{656b.pdf}
\caption*{\centerline{(b2)}\\ $-\frac{g_2}{\sqrt{2}}\textcolor{blue}{\mathcal{N}^*_{n\>3}}
\textcolor{red}{\mathcal{U}_{2+i\>2}} \gamma^{\mu}P_R$}
       \label{fig:table2}
\end{subfigure}
   \begin{subfigure}[b]{0.49\textwidth}
   \centering
          \includegraphics[width=0.76\textwidth]{656b.pdf}
\caption*{\centerline{(b3)}\\ $\frac{g_2}{\sqrt{2}}\textcolor{blue}{\mathcal{N}_{n\>4}}
\textcolor{red}{\mathcal{V}^*_{2+i\>2}} \gamma^{\mu}P_L$}
       \label{fig:table2}
 \end{subfigure}\\
   \begin{subfigure}[b]{0.49\textwidth}
   \centering
           \includegraphics[width=0.76\textwidth]{656b.pdf}
\caption*{\centerline{(b4)} \\ $+g_2\textcolor{blue}{\mathcal{N}^*_{n\>2}}
                     \textcolor{red}{\mathcal{U}_{2+i\>1}} \gamma^{\mu}P_R$}
       \label{fig:table2}
\end{subfigure}
   \begin{subfigure}[b]{0.49\textwidth}
   \centering
            \includegraphics[width=0.76\textwidth]{656b.pdf} \caption*{\centerline{(b5)} \\ $g_2\textcolor{blue}{\mathcal{N}_{n\>2}}
                     \textcolor{red}{\mathcal{V}^*_{2+i\>1}} \gamma^{\mu}P_L$}
       \label{fig:table2}
\end{subfigure}
    \subcaption{}\label{fig:X62}
\end{minipage}
\caption{ a) We show the MSSM vertices, in terms of the gauge eigenstates, expressed as 2-component Weyl fermions.  (a1), (a2) and (a3) 
come from the covariant derivatives of the lepton and Higgsino matter fields, respectively. (a4) and (a5) come from the covariant derivatives of the non-Abelian gauge fileds.
Unlike for chargino, the lightest neutralino can be any state ${\tilde \chi}_n^0$ with $n=1$ Bino dominant,
$n=2$ for neutral Wino dominant, $n=3,4$ for neutral Higgsino dominant, and $n=5,6$ for right handed 
sterile neutrino dominant. b) We 
express the interactions in terms of the mass eigenstates relevant for the neutralino decay into SM 
particles, expressed as 4-component fermions. 
The "red" matrix elements have small values and are proportional to $\epsilon_i/M_{soft}$  while with "blue" matrix elements that have small RPV corrections $1-\epsilon_i/M_{soft}$. At first order, the decay amplitudes are proportional to  $(1-\epsilon_i/M_{soft})\times \epsilon_i/M_{soft} \simeq \epsilon_i/M_{soft}$.
}\label{fig:X6}
\end{figure}

The vertices associated with neutralino decays are shown in Figures \ref{fig:X61} and \ref{fig:X62}. 
The diagrams in Figure \ref{fig:X61} are expressed in terms gauge eigenstates written as 2-component Weyl spinors. The diagrams in Figure \ref{fig:X62} are the same vertices in terms of mass eigenstates, expressed as 4-component spinors.

\subsubsection{\boldmath${\tilde X}_n^0\rightarrow h^0\nu_{L_i}$}

\begin{figure}[H]
   \centering
\begin{minipage}{0.48\textwidth}
\centering
   \begin{subfigure}[b]{0.49\textwidth}
      \includegraphics[width=0.76\textwidth]{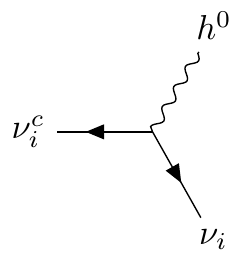}

\caption*{\centerline{(a1)}\\ \centerline{$ \frac{1}{\sqrt 2}Y_{\nu_{j}}\cos \alpha $}}
\bigskip
       \label{fig:table2}
   \end{subfigure}\\
   \begin{subfigure}[b]{0.49\textwidth}
       \includegraphics[width=0.76\textwidth]{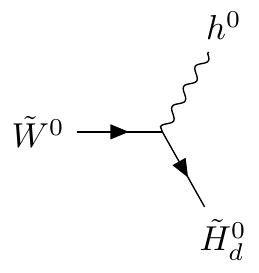}
\caption*{\centerline{(a2)}\\ \centerline{ $\frac{1}{2}g_2\sin \alpha$}}
\bigskip
\bigskip
       \label{fig:table2}
\end{subfigure}
   \begin{subfigure}[b]{0.49\textwidth}
       \includegraphics[width=0.76\textwidth]{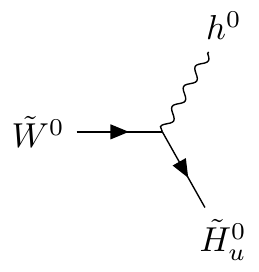}
\caption*{\centerline{(a3)} \\ \centerline{ $\frac{1}{2}g_2\cos \alpha$}}
\bigskip
\bigskip
       \label{fig:table2}
 \end{subfigure}\\
   \begin{subfigure}[b]{0.49\textwidth}
      \includegraphics[width=0.76\textwidth]{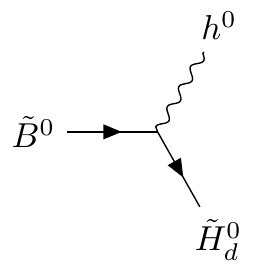}
\caption*{\centerline{(a4)} \\ \centerline{ $-\frac{1}{2}g^{\prime}\sin \alpha$}}
       \label{fig:table2}
\end{subfigure}
   \begin{subfigure}[b]{0.49\textwidth}
      \includegraphics[width=0.76\textwidth]{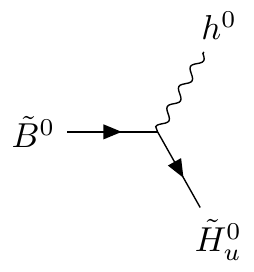}
\caption*{\centerline{(a5)} \\ \centerline{ $-\frac{1}{2}g^{\prime}\cos \alpha$}}
       \label{fig:table2}
 \end{subfigure}
\bigskip
\bigskip
\bigskip
    \subcaption{}\label{fig:X71}
\end{minipage}\hfill
\rulesep
\begin{minipage}{0.48\textwidth}
\centering
   \begin{subfigure}[b]{0.49\textwidth}
\centering
        \includegraphics[width=0.76\textwidth]{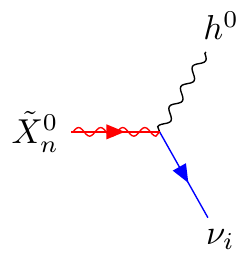}
       \caption*{\centerline{(b1)}\\ $\frac{1}{\sqrt 2}Y_{\nu_j}\cos \alpha \Big[ \textcolor{red}
{\mathcal{N}^*_{n\>6}} \textcolor{blue}{\mathcal{N}^*_{6+j\>6+i}}
- \textcolor{red}
{\mathcal{N}^*_{n\>6+i}} \textcolor{blue}{\mathcal{N}^*_{6+j\>6}}\Big] P_L$}
       \label{fig:table2}
   \end{subfigure}\\
   \begin{subfigure}[b]{0.49\textwidth}
         \includegraphics[width=0.76\textwidth]{657b.pdf}
\caption*{\centerline{(b2)} \\ $+\frac{1}{ 2}g_2\sin \alpha \times \\ \times (\textcolor{blue}{\mathcal{N}^*_{n\>2}}
\textcolor{red}{\mathcal{N}^*_{6+i\>3}}\\
+\textcolor{blue}{\mathcal{N}^*_{n\>3}}
\textcolor{red}{\mathcal{N}^*_{6+i\>2}})P_L$}
       \label{fig:table2}
\end{subfigure}
   \begin{subfigure}[b]{0.49\textwidth}
   \centering
         \includegraphics[width=0.76\textwidth]{657b.pdf}
\caption*{\centerline{(b3)} \\ $\frac{1}{ 2}g_2 \cos \alpha   \times \\ \times (-\textcolor{blue}{\mathcal{N}^*_{n\>2}}
\textcolor{red}{\mathcal{N}^*_{6+i\>4}  }\\
+\textcolor{blue}{\mathcal{N}^*_{n\>4}}
\textcolor{red}{\mathcal{N}^*_{6+i\>2}})P_L
$}
       \label{fig:table2}
 \end{subfigure}
   \begin{subfigure}[b]{0.49\textwidth}
         \includegraphics[width=0.76\textwidth]{657b.pdf}
\caption*{\centerline{(b4)} $-\frac{1}{ 2}g^{\prime}\sin \alpha \times \\ \times [\sin \theta_R(\textcolor{blue}{\mathcal{N}^*_{n\>1}}
\textcolor{red}{\mathcal{N}^*_{6+i\>3}}\\
+\textcolor{blue}{\mathcal{N}^*_{n\>3}}
\textcolor{red}{\mathcal{N}^*_{6+i\>1}}
)\\+\cos \theta_R (\textcolor{blue}{\mathcal{N}^*_{n\>5}}
\textcolor{red}{\mathcal{N}^*_{6+i\>3}}\\
+\textcolor{blue}{\mathcal{N}^*_{n\>3}}
\textcolor{red}{\mathcal{N}^*_{6+i\>5}})]P_L$}
       \label{fig:table2}
\end{subfigure}
   \begin{subfigure}[b]{0.49\textwidth}
   \centering
         \includegraphics[width=0.76\textwidth]{657b.pdf}
\caption*{\centerline{(b5)} $-\frac{1}{ 2}g^{\prime}\cos \alpha   \times \\ \times [\sin \theta_R(\textcolor{blue}{\mathcal{N}^*_{n\>1}}
\textcolor{red}{\mathcal{N}^*_{6+i\>4}}\\
+\textcolor{blue}{\mathcal{N}^*_{n\>4}}
\textcolor{red}{\mathcal{N}^*_{6+i\>1}}
)\\+\cos \theta_R (\textcolor{blue}{\mathcal{N}^*_{n\>5}}
\textcolor{red}{\mathcal{N}^*_{6+i\>4}}\\
+\textcolor{blue}{\mathcal{N}^*_{n\>4}}
\textcolor{red}{\mathcal{N}^*_{6+i\>5}})]P_L$}
       \label{fig:table2}
 \end{subfigure}\\
    \subcaption{}\label{fig:X72}
\end{minipage}
\caption{ a) We show the MSSM vertices in terms of the gauge eigenstates, expressed as 2-component Weyl fermions. (a1)
comes from the Yukawa couplings of the leptons to the Higgs
field in the superpotential, whereas
(a2), (a3), (a4), (a5) come from the supercovariant derivative terms. b) We 
express these interactions in terms of the mass eigenstates relevant for the neutralino decay into SM 
particles, expressed as 4-component fermions. Unlike for chargino, the lightest neutralino can be any state ${\tilde \chi}_n^0$ with $n=1$ Bino dominant,
$n=2$ for neutral Wino dominant, $n=3,4$ for neutral Higgsino dominant, and $n=5,6$ for right handed 
sterile neutrino dominant. The "red" matrix elements have small values and are proportional to $\epsilon_i/M_{soft}$,  while "blue" matrix elements are of order unity with small RPV corrections of the form $1-\epsilon_i/M_{soft}$. At first order, the decay amplitudes are proportional to  $(1-\epsilon_i/M_{soft})\times \epsilon_i/M_{soft} \simeq \epsilon_i/M_{soft}$.}\label{fig:X7}
\end{figure}

The vertices associated with neutralino decays are shown in Figures \ref{fig:X71} and \ref{fig:X72}. 
The diagrams in Figure \ref{fig:X71} are expressed in terms gauge eigenstates written as 2-component Weyl spinors. The diagrams in Figure \ref{fig:X72} are the same vertices in terms of mass eigenstates, expressed as 4-component spinors.

\section{Decay rates}
\label{sec:7}

In the previous discussion, we presented all relevant RPV decay channels of charginos $\tilde \chi^\pm_1$ and neutralinos $\tilde \chi^0_n,\>n=1,2,3,4,5,6$ into standard model particles and gave the associated Lagrangian interactions. In this section, we will use these results to calculate the decay rates associated with each such process. The calculations are carried out using the dominant linear terms in the RPV couplings $\epsilon_i$ and $v_{L_i}$ only, since higher order terms are highly suppressed. The analysis is completely general, regardless of whether or not the charginos or the neutralinos are the LSPs. However, only the lightest sparticles decay exclusively via RPV processes into SM particles. Furthermore, they have best prospects for detection at the LHC. Therefore, in subsequent publications, we will use these results to compute the RPV branching ratios of chargino and neutralinos LSPs and NLSPs.

\subsection{Wino/Higgsino Chargino}

We are ultimately interested in LSP and NSLP decays via RPV channels. Thus, we calculate the decay rates of the $\tilde \chi^\pm_1$ charginos only, which are defined to be lighter than the $\tilde \chi^\pm_2$ charginos. As discussed in detail at the end of subsection 5.1, a chargino mass eigenstate is a superposition of a charged Wino, a charged Higgsino and an RPV sum over left-chiral and right-chiral charged leptons gauge eigenstates. 
In the present analysis, we will consider the decays of a {\it generic} chargino with arbitrary mixed charged Wino, charged Higgsino and RPV charged lepton content. More specialized decays involving predominantly Wino chargino or Higgsino chargino mass eigenstates, will be considered in future publications.\\

\begin{itemize}

\item ${\tilde X}_1^\pm\rightarrow W^\pm \nu_{i}$

The terms in the Lagrangian associated with these decay channels have been calculated in eq. \eqref{eq:gauge_amplitudes} and illustrated in Figure \ref{fig:X1}. Summing all the vertices, one finds
\begin{multline}
g_{{\tilde X}^+_1\rightarrow W^-\nu_i}
=\gamma^\mu {G_L}_{{\tilde X}^+_1\rightarrow W^+ \nu_i} P_L+\gamma^\mu {G_R}_{{\tilde X}^+_1\rightarrow W^+ \nu_i} P_R  \label{F1} \\
~~=\frac{g_2}{\sqrt 2} \gamma^\mu \Big[(\mathcal{U}_{1\>2+j}\mathcal{N}^*_{6+j\>6+i}+\mathcal{U}_{1\>2}\mathcal{N}^*_{6+i\>3}-\sqrt{2}\mathcal{N}^*_{6+i\>2}\mathcal{U}_{1\>1})P_L\\
\hfill+(-\mathcal{N}_{6+i\>4}\mathcal{V}^*_{1\>2}-\sqrt{2}\mathcal{V}^*_{1\>1}\mathcal{N}_{6+i\>2})P_R \Big] \qquad
\end{multline}
and
\begin{multline}
g_{{\tilde X}^-_1\rightarrow W^+\nu_i}  =\gamma^\mu {G_L}_{{\tilde X}^-_1\rightarrow W^- \nu_i} P_L+\gamma^\mu {G_R}_{{\tilde X}^-_1\rightarrow W^- \nu_i} P_R \\
~~~~~=-\frac{g_2}{\sqrt 2} \gamma^\mu\Big[
(\mathcal{U}^*_{1\>2+i}\mathcal{N}_{6+j\>6+i}+\mathcal{U}^*_{1\>2}\mathcal{N}_{6+i\>3}-\sqrt{2}\mathcal{N}_{6+i\>2}\mathcal{U}^*_{1\>1})P_R\\
\hfill+(-\mathcal{N}^*_{6+i\>4}\mathcal{V}_{1\>2}-\sqrt{2}\mathcal{V}_{1\>1}\mathcal{N}^*_{6+i\>2})P_L
 \Big] \qquad 
\end{multline}
Next, using the expressions for the matrix elements of $\mathcal{U}$, $\mathcal{V}$ and $\mathcal{N}$ given in Appendices B.1 and B.2, we get
\begin{multline}\label{eq:72}
 {G_L}_{{\tilde X}^+_1\rightarrow W^+ \nu_i}= -{G^*_R}_{{\tilde X}^-_1\rightarrow W^- \nu_i} =\frac{g_2}{\sqrt 2}\Bigg[
\left( -\cos \phi_- \frac{g_2 v_d}{\sqrt{2}M_2\mu}\epsilon_j^*+\sin \phi_-\frac{\epsilon_j^*}{\mu}\right)\\
-\sin \phi_- \frac{1}{16d_{\tilde \chi^0}}\left( M_{\tilde \gamma}v_R^2v_u(v_d\epsilon_j-\mu v_{L_j}^*)-4M_2\mu(M_{\tilde Y}v_R^2+g_R^2M_{BL}v_u^2)\epsilon_j \right)\\
+\sqrt{2}\cos \phi_- \frac{g_2\mu}{8d_{\tilde \chi_0}}\left( 2g_R^2M_{BL}v_dv_u^2\epsilon_j+M_{\tilde Y}v_R^2(v_d\epsilon_j+\mu v_{L_j}^*) \right)\Bigg]\left[V_{\text{PMNS}}\right]_{ji}
\end{multline}
and
\begin{multline}\label{eq:73}
 {G_R}_{{\tilde X}^+_1\rightarrow W^+ \nu_i}= -{G^*_L}_{{\tilde X}^-_1\rightarrow W^- \nu_i} = \\
\frac{g_2}{\sqrt 2}\Bigg[\sin \phi_+ \frac{1}{16d_{{\tilde \chi}^0}}
[ M_{\tilde \gamma} v_R^2v_u (v_d\epsilon_j^*+\mu v_{L_j})
-4g_R^2\mu M_2M_{BL}v_d v_u\epsilon_j^*]  
 \\-\sqrt{2} \cos \phi_+ 
\frac{g_2\mu}{8d_{{\tilde \chi}^0}}
[2g_R^2M_{BL}v_dv_u^2\epsilon_j+M_{\tilde Y}v_R^2(v_d\epsilon_j^*+\mu v_{L_j})]
\Bigg] \left[V_{\text{PMNS}}^\dag \right]_{ij} \ .
 \end{multline}
 
 The decay width $\Gamma$ is proportional to the square of the amplitude of this process. Note that we account for the longitudinal degrees of freedom of the resultant $W^\pm$ gauge bosons (Goldostone equivalence theorem) in calculating this decay width. This results in an amplification of this channel, which becomes more significant as the mass of the decaying chargino increases. The result is
\begin{equation}\label{eq:Chargino_Decay1}
\Gamma_{{\tilde X}_1^\pm\rightarrow W^\pm \nu_{i}}=\frac{\left(|{G_L}|^2_{{\tilde X}^\pm_1\rightarrow W^\pm \nu_i}+|{G_R}|^2_{{\tilde X}^\pm_1\rightarrow W^\pm \nu_i} 
\right)}{64\pi}
\frac{M_{{\tilde \chi}_1^\pm}^3}{M_{W^\pm}^2}\left(1-\frac{M_{W^\pm}^2}{M_{{\tilde \chi}_1^\pm}^2}\right)^2\left(1+2\frac{M_{W^\pm}^2}{M_{{\tilde \chi}_1^\pm}^2}\right).
\end{equation}
Note that $|{G_L}|^2_{{\tilde X}^\pm_1\rightarrow W^\pm \nu_i} $ and $|{G_R}|^2_{{\tilde X}^\pm_1\rightarrow W^\pm \nu_i} $ are proportional to the RPV couplings $\epsilon_i$ and $v_{L_i}$ at first order. Therefore, the decay of the chargino into the SM particles $W^\pm$ boson and neutrino would vanish in the absence of RPV.\\

\item ${\tilde X_1}^\pm\rightarrow Z^0 \ell_{i}^\pm $

The terms in the Lagrangian associated with these decay channels have been calculated in eq. \eqref{eq:gauge_amplitudes} and illustrated in Figure \ref{fig:X2}. Summing all the vertices, one finds
\begin{multline}
g_{{\tilde X}^+_1\rightarrow Z^0 \ell^+_i}=\gamma^\mu {G_L}_{{\tilde X}^+_1\rightarrow Z^0 \ell^+_i} P_L+\gamma^\mu {G_R}_{{\tilde X}^+_1\rightarrow Z^0 \ell^+_i} P_R\\
=g_2\gamma^\mu\Bigg[
\left(\frac{1}{c_W}
\left(-\frac{1}{2}-s_W^2\right)\mathcal{U}_{2+j\>1}\mathcal{U}_{2+i\>2+j}^*
-\frac{1}{c_W}\left(\frac{1}{2}-s_W^2 \right)\mathcal{U}_{1\>2}\mathcal{U}^*_{2+i\>2}-c_W\mathcal{U}^*_{2+i\>1}\mathcal{U}_{1\>1}
\right)P_L\\+
\left(\frac{1}{c_W} s_W^2\mathcal{V}_{2+j\>2+i}\mathcal{V}_{1\>2+i}^*-\frac{1}{c_W}\left(\frac{1}{2}-s_W^2\right)\mathcal{V}_{2+i\>2}\mathcal{V}_{1\>2}^*
-c_W\mathcal{V}^*_{1\>1}\mathcal{V}_{2+i\>1}
\right)P_R\Bigg]
\end{multline}
and
\begin{multline}
g_{{\tilde X}^-_1\rightarrow Z^0 \ell_i^-}=\gamma^\mu {G_L}_{{\tilde X}^-_1\rightarrow \gamma^0 \ell^-_i} P_L+Z^\mu {G_R}_{{\tilde X}^-_1\rightarrow Z^0 \ell^-_i} P_R\\
=-g_2\gamma^\mu\Bigg[
\left(-\left(\frac{1}{2}-s_W^2\right)\mathcal{U}^*_{1\>2+j}\mathcal{U}_{2+i\>2+j}
-\frac{1}{c_W}\left(\frac{1}{2}-s_W^2 \right)\mathcal{U}^*_{1\>2}\mathcal{U}_{2+i\>2}-c_W\mathcal{U}_{2+i\>1}\mathcal{U}^*_{1\>1}
\right)P_R\\+
\left(\frac{1}{c_W} s_W^2\mathcal{V}^*_{2+j\>2+i}\mathcal{V}_{1\>2+i}-\frac{1}{c_W}\left(\frac{1}{2}-s_W^2\right)\mathcal{V}^*_{2+i\>2}\mathcal{V}_{1\>2}
-c_W\mathcal{V}_{1\>1}\mathcal{V}^*_{2+i\>1}
\right)P_L
 \Bigg] \ .
\end{multline}
Using the expressions for the matrix elements of $\mathcal{U}$ and $\mathcal{V}$ from the Appendix B.1, we get
\begin{multline}
{G_L}_{{\tilde X}^+_1\rightarrow Z^0 \ell^+_i}=-{G_R}^*_{{\tilde X}^-_1\rightarrow Z^0 \ell^-_i}=-g_2c_W\Big(\frac{g_2}{\sqrt{2}M_2\mu}(v_d\epsilon_i+\mu v_{L_i}^*)\Big)
\cos \phi_- +\\ +\frac{g_2}{c_W}
\left(\frac{1}{2}-s_W^2\right)
\Big(-\cos \phi_- \frac{g_2 v_d}{\sqrt{2}M_2\mu}\epsilon_i+\sin \phi_-\frac{\epsilon_i}{\mu}\Big)-\frac{g_2}{c_W}\left(\frac{1}{2}-s_W^2\right)
\Big(\frac{\epsilon_i}{\mu} \Big)\sin \phi_-
\end{multline}
and
\begin{multline}
{G_R}_{{\tilde X}^+_1\rightarrow Z^0 \ell^+_i}=-{G_L}^*_{{\tilde X}^-_1\rightarrow Z^0 \ell^-_i}=-g_2c_W\cos \phi_+ \Big(-\frac{1}{\sqrt{2}M_2\mu}g_2\tan \beta m_{e_i}v_{L_i} \Big)-\\+
\frac{g_2}{c_W}s_W^2\Big( -\cos \phi_+ \frac{g_2 \tan \beta m_{e_i}}{\sqrt{2}M_2\mu}v_{L_i}+\sin \phi_+\frac{m_{e_i}}{\mu v_d}v_{L_i}\Big)
-\frac{g_2}{c_W}\left(\frac{1}{2}-s_W^2\right)\sin \phi_+\Big(\frac{m_{e_i}}{ v_d \mu}v_{L_i} \Big),
\end{multline}
where there is no sum over the $i$ in $v_{L_i}m_{e_i}$.

The decay width $\Gamma$ is proportional to the square of the amplitude of this process. We note that we have accounted for the longitudinal degrees of freedom of the resultant $Z^0$ gauge bosons (Goldstone equivalence theorem) in calculating this decay width. This results in an amplification of this channel which becomes more significant as the mass of the decaying chargino increases. We find that
\begin{equation}\label{eq:Chargino_Decay2}
\Gamma_{{\tilde X}_1^\pm\rightarrow Z^0 \ell_i^\pm}=\frac{\Big( |{G_L}|_{{\tilde X}^\pm_1\rightarrow Z^0 \ell^\pm_i}^2+|{G_R}|_{{\tilde X}^\pm_1\rightarrow Z^0 \ell^\pm_i}^2\Big)}{64\pi}
\frac{M_{{\tilde \chi}_1^\pm}^3}{M_{Z^0}^2}\left(1-\frac{M_{Z^0}^2}{M_{{\tilde \chi}_1}^2}\right)^2
\left(1+2\frac{M_{Z^0}^2}{M_{{\tilde \chi}_1^\pm}^2}\right).
\end{equation}
Note that both $G_{\tilde X^\pm\rightarrow Z^0 \ell_i^\pm}$ coefficients are proportional to the RPV couplings $\epsilon_i$ and $v_{L_i}$ at first order. Therefore, there would be no decay of the chargino into the SM $Z^0$ boson and charged leptons if the RPV effects were non-existent.\\

\item ${\tilde X}^\pm_1\rightarrow h^0 \ell_i^\pm$\\
The terms in the Lagrangian associated with these decay channels have been calculated in eq. \eqref{eq:Higgs_amplitudes} and illustrated in Figure \ref{fig:X4}. Summing all the vertices, one finds
\begin{multline}
g_{{\tilde X}^+_1\rightarrow h^0 \ell_i^+}={G_L}_{{\tilde X}^-_1\rightarrow h^0 \ell_i^-} P_L+ {G_R}_{{\tilde X}^-_1\rightarrow h^0 \ell_i^-} P_R\\
=-\frac{1}{\sqrt{2}}Y_{e_i}\sin \alpha \Big[\mathcal{V}^*_{1\>2+j}\mathcal{U}^*_{2+i\>2+j}P_L+\mathcal{V}_{2+i\>2+j}\mathcal{U}_{1\>2+j}P_R\Big] \qquad ~~\\
+\frac{g_2}{{2}}\Big[(-\cos \alpha \mathcal{V}^*_{1\>2}\mathcal{U}^*_{2+i\>1}-\sin \alpha
\mathcal{U}^*_{2+i\>2}\mathcal{V}_{1\>1}^*)P_L
+(-\cos\alpha \mathcal{V}_{2+i\>2}\mathcal{U}_{1\>1} -\sin \alpha  \mathcal{U}_{1\>2}\mathcal{V}_{2+i\>1})P_R
\Big]\\
\end{multline}
and
\begin{multline}
g_{{\tilde X}^-_1\rightarrow h^0 \ell_i^-}={G_L}_{{\tilde X}^+_1\rightarrow h^0 \ell_i^+} P_L+ {G_R}_{{\tilde X}^+_1\rightarrow h^0 \ell_i^+} P_R\\
=\frac{1}{\sqrt{2}}Y_{e_i}\sin \alpha \Big[  \mathcal{V}^*_{2+j\>2+i}\mathcal{U}^*_{1\>2+j}P_L 
+\mathcal{V}_{1\>2+i}\mathcal{U}_{2+j\>2+i}P_R\Big] \qquad ~~~~~~\\+ \frac{g_2}{{2}}\Big[ (\cos\alpha \mathcal{V}^*_{2+i\>2}\mathcal{U}^*_{1\>1} +\sin \alpha  \mathcal{U}^*_{1\>2}\mathcal{V}^*_{2+i\>1})P_L
+(\cos \alpha \mathcal{V}_{1\>2}\mathcal{U}_{2+i\>1}+\sin \alpha
\mathcal{U}_{2+i\>2}\mathcal{V}_{1\>1})P_R 
\Big] \ .
\end{multline}
Next, using the expressions for the matrix elements of $\mathcal{U}$ and $\mathcal{V}$ from the Appendix B.1, we get

\begin{multline}
{G_L}_{{\tilde X}^+_1\rightarrow h^0 \ell_i^-}=-{G_R}^*_{{\tilde X}^+_1\rightarrow h^0 \ell_i^+} =-\frac{1}{\sqrt 2}Y_{e_i}\sin \alpha 
\Big(-\cos \phi_+ \frac{g_2 \tan \beta m_{e_i}}{\sqrt{2}M_2\mu}v_{L_i}^*+\sin \phi_+\frac{m_{e_i}}{\mu v_d}v_{L_i}^* \Big)+\\-
\frac{1}{ 2}g_2 \sin \alpha \cos \phi_+ \Big( \frac{\epsilon_i}{\mu} \Big)-
\frac{1}{2}g_2\cos \alpha \sin \phi_+\Big( \frac{g_2}{\sqrt{2}M_2\mu}(v_d\epsilon_i+\mu v_{L_i}^*) \Big)
\end{multline}
and
\begin{multline}
{G_R}_{{\tilde X}^-_1\rightarrow h^0 \ell_i^-}=-{G_L}^*_{{\tilde X}^+_1\rightarrow h^0 \ell_i^+}=-\frac{1}{\sqrt 2}Y_{e_i}\sin \alpha 
\Big(-\cos \phi_- \frac{g_2v_d}{\sqrt{2}M_2\mu}\epsilon_i^*+\sin \phi_-\frac{\epsilon_i^*}{\mu}\Big)\\
+\frac{1}{2}g_2 \sin \alpha \sin \phi_-\Big(-\cos \phi_+\frac{1}{\sqrt{2}M_2\mu}g_2\tan \beta m_{e_i}v_{L_i} -\sin \phi_+ \frac{m_{e_i}}{\mu v_d}v_{L_i}\Big)\\
-\frac{1}{2}g_2\cos \alpha \cos \phi_-\Big( \frac{m_{e_i}}{ v_d \mu}v_{L_i} \Big),
\end{multline}

where we do not sum over the $i$ in either of these expressions. The decay width $\Gamma$ is proportional to the square of the amplitude of this process, and is found to be

\begin{equation}\label{eq:Chargino_Decay4}
\Gamma_{{\tilde X}_1^\pm\rightarrow h^0 \ell_i^\pm}=\frac{\Big(|{G_L}|_{{\tilde X}^\pm_1\rightarrow h^0 \ell_i^\pm}^2+|{G_R}|_{{\tilde X}^\pm_1\rightarrow h^0 \ell_i^\pm}^2\Big)}{64\pi}
M_{{\tilde X}_1^\pm}\left(1-\frac{M_{h^0}^2}{M_{{\tilde X}_1^\pm}^2}\right)^2.
\end{equation}

Again, note that $G_{\tilde X^\pm\rightarrow h^0 \ell_i^\pm}$ are proportional to the RPV couplings $\epsilon_i$ and $v_{L_i}$ at first order. Hence, there would be no decay of the chargino into the SM Higgs boson and charged leptons if there would be no RPV effects.

\end{itemize}

\subsection{Neutralinos}

Recall that the index $n$ indicates the neutralino species as follows:
\begin{equation}
{\tilde X}_1^0={\tilde X}_B^0,\quad  {\tilde X}_2^0={\tilde X}_W^0, \quad {\tilde X}_3^0={\tilde X}_{H_d}^0,
\quad {\tilde X}_4^0={\tilde X}_{H_u}^0, \quad {\tilde X}_5^0={\tilde X}_{\nu_{3a}}^0, \quad {\tilde X}_6^0={\tilde X}_{\nu_{3b}}^0.
\end{equation}
For a general neutralino state $n$, we found expressions for the parameters of the decay to two standard model particles. 

\begin{itemize}

\item{${\tilde X}^0_n\rightarrow Z^0 \nu_{i}$}

To begin with, it follows from
eq. \eqref{eq:gauge_amplitudes} and the vertices in Figure \ref{fig:X5} that
\begin{multline}
g_{{\tilde X}^0_n\rightarrow Z^0 \nu_{i}}=\gamma^\mu {G_L}_{{\tilde X}^0_n\rightarrow Z^0 \nu_{i}}+\gamma^{\mu}{G_R}_{{\tilde X}^0_n\rightarrow Z^0 \nu_{i}} \\
={g_2}\gamma^{\mu}\Big[
\Big(\frac{1}{2c_W}\mathcal{N}_{n\>6+j}\mathcal{N}^*_{6+j\>6+i}-\frac{1}{c_W}\left(\frac{1}{2}+s_W^2\right)\mathcal{N}_{n\>4}\mathcal{N}^*_{6+i\>4} \Big)P_L\\
-\Big( \frac{1}{c_W}\left(\frac{1}{2}+s_W^2\right) \mathcal{N}_{n\>3}\mathcal{N}^*_{6+i\>3}\Big)P_R
\Big]\\
-{g_2}\gamma^{\mu}\Big[
\Big(\frac{1}{2c_W}\mathcal{N}^*_{n\>6+j}\mathcal{N}_{6+j\>6+i}-\frac{1}{c_W}\left(\frac{1}{2}+s_W^2\right)\mathcal{N}^*_{n\>4}\mathcal{N}_{6+i\>4} \Big)P_R\\
-\Big( \frac{1}{c_W}\left(\frac{1}{2}+s_W^2\right) \mathcal{N}^*_{n\>3}\mathcal{N}_{6+i\>3}\Big)P_L
\Big] \,
\end{multline}
where we can read 
\begin{multline}
 {G_L}_{{\tilde X}^0_n\rightarrow Z^0 \nu_{i}}=
g_2\Big(\frac{1}{2c_W}\mathcal{N}_{n\>6+j}\mathcal{N}^*_{6+j\>6+i}-\frac{1}{c_W}\left(\frac{1}{2}+s_W^2\right)\mathcal{N}_{n\>4}\mathcal{N}^*_{6+i\>4} \Big)\\
+g_2\Big( \frac{1}{c_W}\left(\frac{1}{2}+s_W^2\right) \mathcal{N}^*_{n\>3}\mathcal{N}_{6+i\>3}\Big)
\end{multline}
and
\begin{multline}
{G_R}_{{\tilde X}^0_n\rightarrow Z^0 \nu_{i}}=g_2\Big(- \frac{1}{c_W}\left(\frac{1}{2}+s_W^2\right) \mathcal{N}_{n\>3}\mathcal{N}^*_{6+i\>3}\Big)\\
-{g_2}
\Big(\frac{1}{2c_W}\mathcal{N}^*_{n\>6+j}\mathcal{N}_{6+j\>6+i}-\frac{1}{c_W}\left(\frac{1}{2}+s_W^2\right)\mathcal{N}^*_{n\>4}\mathcal{N}_{6+i\>4} \Big).
\end{multline}

Using these results, one can compute the associated decay rate. It is found to be
\begin{equation}
\Gamma_{{\tilde X}^0_n\rightarrow Z^0\nu_{i}}=
\frac{\Big(|{G_L}|_{{\tilde X}^0_n\rightarrow Z^0\nu_{i}}^2
+|{G_R}|_{{\tilde X}^0_n\rightarrow Z^0\nu_{i}}^2 \Big)
}{64\pi}
\frac{M_{{\tilde \chi}_n^0}^3}{M_{Z^0}^2}\left(1-\frac{M_{Z^0}^2}{M_{{\tilde \chi}_n^0}^2}\right)^2
\left(1+2\frac{M_{Z^0}^2}{M_{{\tilde \chi}^0_n}^2}\right),
\end{equation}

\item{${\tilde X}^0_n\rightarrow W^\pm \ell_i^\mp$}

Similarly, it follows from eq. \eqref{eq:gauge_amplitudes} and the vertices in Figure \ref{fig:X6} that
\begin{multline}
g_{{\tilde X}^0_n\rightarrow W^- \ell_i^+}=\gamma^\mu {G_L}_{{\tilde X}^0_n\rightarrow W^- \ell_i^+} P_L +\gamma^\mu {G_R}_{{\tilde X}^0_n\rightarrow W^- \ell_i^+} P_R \\
=\frac{g_2}{\sqrt 2} \gamma^\mu
\Big[
(\mathcal{N}_{n\>4}\mathcal{V}^*_{2+i\>2}+\sqrt{2}\mathcal{V}^*_{2+i\>1}\mathcal{N}_{n\>2})P_L\\+(-\mathcal{U}_{2+i\>2+j}\mathcal{N}^*_{n\>6+j}-\mathcal{U}_{2+i\>2}\mathcal{N}^*_{n\>3}+\sqrt{2}\mathcal{N}^*_{n\>2}\mathcal{U}_{2+i\>1})P_R
\Big]
\end{multline}
and its conjugate
\begin{multline}
g_{{\tilde X}^0_n\rightarrow W^+ \ell_i^-}=\gamma^\mu {G_L}_{{\tilde X}^0_n\rightarrow W^+ \ell_i^-} P_L +\gamma^\mu {G_R}_{{\tilde X}^0_n\rightarrow W^+ \ell_i^-} P_R\\
=-\frac{g_2}{\sqrt 2} \gamma^\mu
\Big[
(\mathcal{N}^*_{n\>4}\mathcal{V}_{2+i\>2}+\sqrt{2}\mathcal{V}_{2+i\>1}\mathcal{N}^*_{n\>2})P_R\\+(-\mathcal{U}^*_{2+i\>2+j}\mathcal{N}_{n\>6+j}-\mathcal{U}^*_{2+i\>2}\mathcal{N}_{n\>3}+\sqrt{2}\mathcal{N}_{n\>2}\mathcal{U}^*_{2+i\>1})P_L 
\Big] \ ,
\end{multline}
where we can read 
\begin{equation}
 {G_L}_{{\tilde X}^0_n\rightarrow W^- \ell_i^+}=-{G_R}_{{\tilde X}^0_n\rightarrow W^+ \ell_i^-}=\frac{g_2}{\sqrt{2}}\Big[\mathcal{N}_{n\>4}\mathcal{V}^*_{2+i\>2}-2\sqrt{2}\mathcal{V}^*_{2+i\>1}\mathcal{N}_{n\>2}\Big]
\end{equation}
and
\begin{equation}
{G_R}_{{\tilde X}^0_n\rightarrow W^- \ell_i^+}=-{G_L}_{{\tilde X}^0_n\rightarrow W^+ \ell_i^-}=\frac{g_2}{\sqrt{2}}\Big[-\mathcal{U}_{2+i\>2+j}\mathcal{N}^*_{n\>6+j}-\mathcal{U}_{2+i\>2}\mathcal{N}^*_{n\>3}+2\sqrt{2}\mathcal{N}^*_{n\>2}\mathcal{U}_{2+i\>1}\Big].
\end{equation}
Using these results, one can compute the decay rate
\begin{equation}
\Gamma_{{\tilde X}^0_n\rightarrow W^\mp \ell_i^\pm}=\frac{\Big(|{G_L}|_{{\tilde X}^0_n\rightarrow W^\pm \ell_i^\mp}^2+|{G_R}|_{{\tilde X}^0_n\rightarrow W^\pm \ell_i^\mp}^2\Big)}{64\pi}
\frac{M_{{\tilde \chi}_1^\pm}^3}{M_{W^\pm}^2}\left(1-\frac{M_{W^\pm}^2}{M_{{\tilde \chi}_n^0}^2}\right)^2
\left(1+2\frac{M_{W^\pm}^2}{M_{{\tilde \chi}_n^0}^2}\right),
\end{equation}

\item{${\tilde X}^0_n\rightarrow h^0 \nu_{i}$}

Finally, from eq. \eqref{eq:gauge_amplitudes}  and the vertices in Figure \ref{fig:X6} we find
\begin{multline}
g_{{\tilde X}^0_n\rightarrow h^0 \nu_{i}} ={G_L}_{{\tilde X}^0_n\rightarrow h^0 \nu_{i} } P_L +{G_R}_{{\tilde X}^0_n\rightarrow h^0 \nu_{i} } P_R \\
\qquad =+\frac{g_2}{{2}}\Big[\Big(
\cos \alpha (\mathcal{N}^*_{n\>4}\mathcal{N}^*_{6+i\>2}+\mathcal{N}^*_{6+i\>4}\mathcal{N}_{n\>2}^*)+\sin \alpha (\mathcal{N}^*_{n\>3}\mathcal{N}^*_{6+i\>2}+\mathcal{N}^*_{6+i\>3}\mathcal{N}_{n\>2}^*)\Big)P_L\\-
\Big(
\cos \alpha (\mathcal{N}_{n\>4}\mathcal{N}_{6+i\>2}+\mathcal{N}_{6+i\>4}\mathcal{N}_{n\>2})+\sin \alpha (\mathcal{N}_{n\>3}\mathcal{N}_{6+i\>2}+\mathcal{N}_{6+i\>3}\mathcal{N}_{n\>2})\Big)P_R
\Big]\\
-\frac{g'}{{2}}\Big[\Big(
\cos\alpha \left(\sin \theta_R(\mathcal{N}^*_{n\>4}\mathcal{N}^*_{6+i\>1}+\mathcal{N}^*_{6+i\>4}\mathcal{N}^*_{n\>1})+\cos \theta_R(\mathcal{N}^*_{n\>4}\mathcal{N}^*_{6+i\>5}+\mathcal{N}^*_{6+i\>4}\mathcal{N}^*_{n\>5})   \right)\\
+\sin \alpha \left( \sin \theta_R(\mathcal{N}^*_{n\>3}\mathcal{N}^*_{6+i\>1}+\mathcal{N}^*_{6+i\>3}\mathcal{N}^*_{n\>1})+\cos \theta_R(\mathcal{N}^*_{n\>3}\mathcal{N}^*_{6+i\>5}+\mathcal{N}^*_{6+i\>3}\mathcal{N}^*_{n\>5}) \right)\Big)P_L\\
-\Big(\cos\alpha \left(\sin \theta_R(\mathcal{N}_{n\>4}\mathcal{N}_{1\>6+i}+\mathcal{N}_{6+i\>4}\mathcal{N}_{n\>1})+\cos \theta_R(\mathcal{N}_{n\>4}\mathcal{N}_{6+i\>5}+\mathcal{N}_{6+i\>4}\mathcal{N}_{n\>5})   \right)\\
+\sin \alpha \left( \sin \theta_R(\mathcal{N}_{n\>3}\mathcal{N}_{6+i\>1}+\mathcal{N}_{6+i\>3}\mathcal{N}_{n\>1})+\cos \theta_R(\mathcal{N}_{n\>3}\mathcal{N}_{6+i\>5}+\mathcal{N}_{6+i\>3}\mathcal{N}_{n\>5}) \right)\Big)P_R
 \Big]\\
+\frac{1}{\sqrt 2}Y_{\nu i3}\cos\alpha 
\Big[\Big(\mathcal{N}^*_{n\>6+j}\mathcal{N}^*_{6+i\>6}
+\mathcal{N}^*_{6+i\>6+j}
\mathcal{N}^*_{n\>6}\Big)P_L
+\Big(\mathcal{N}_{n\>6+j}\mathcal{N}_{6+i\>6}
+\mathcal{N}_{6+i\>6+j}
\mathcal{N}_{n\>6}\Big)P_R
\Big],
\end{multline}
where
\begin{multline}
{G_L}_{{\tilde X}^0_n\rightarrow h^0 \nu_{i} }=\frac{g_2}{{2}}\Big(
\cos \alpha (\mathcal{N}^*_{n\>4}\mathcal{N}^*_{6+i\>2}+\mathcal{N}^*_{6+i\>4}\mathcal{N}_{n\>2}^*)+\sin \alpha (\mathcal{N}^*_{n\>3}\mathcal{N}^*_{6+i\>2}+\mathcal{N}^*_{6+i\>3}\mathcal{N}_{n\>2}^*)\Big)\\
-\frac{g'}{{2}}\Big(
\cos\alpha \left(\sin \theta_R(\mathcal{N}^*_{n\>4}\mathcal{N}^*_{6+i\>1}+\mathcal{N}^*_{6+i\>4}\mathcal{N}^*_{n\>1})+\cos \theta_R(\mathcal{N}^*_{n\>4}\mathcal{N}^*_{6+i\>5}+\mathcal{N}^*_{6+i\>4}\mathcal{N}^*_{n\>5})   \right)\\
+\sin \alpha \left( \sin \theta_R(\mathcal{N}^*_{n\>3}\mathcal{N}^*_{6+i\>1}+\mathcal{N}^*_{6+i\>3}\mathcal{N}^*_{n\>1})+\cos \theta_R(\mathcal{N}^*_{n\>3}\mathcal{N}^*_{6+i\>5}+\mathcal{N}^*_{6+i\>3}\mathcal{N}^*_{n\>5}) \right)\Big)\\
+\frac{1}{\sqrt 2}Y_{\nu i3}\cos\alpha \Big(\mathcal{N}^*_{n\>6+j}\mathcal{N}^*_{6+i\>6}
+\mathcal{N}^*_{6+i\>6+j}
\mathcal{N}^*_{n\>6}\Big)
\end{multline}
and
\begin{multline}
{G_R}_{{\tilde X}^0_n\rightarrow h^0 \nu_{i} }=\frac{g_2}{{2}}\Big(
\cos \alpha (\mathcal{N}_{n\>4}\mathcal{N}_{6+i\>2}+\mathcal{N}_{6+i\>4}\mathcal{N}_{n\>2})+\sin \alpha (\mathcal{N}_{n\>3}\mathcal{N}_{6+i\>2}+\mathcal{N}_{6+i\>3}\mathcal{N}_{n\>2})\Big)\\
+\frac{g'}{{2}}\Big(\cos\alpha \left(\sin \theta_R(\mathcal{N}_{n\>4}\mathcal{N}_{1\>6+i}+\mathcal{N}_{6+i\>4}\mathcal{N}_{n\>1})+\cos \theta_R(\mathcal{N}_{n\>4}\mathcal{N}_{6+i\>5}+\mathcal{N}_{6+i\>4}\mathcal{N}_{n\>5})   \right)\\
+\sin \alpha \left( \sin \theta_R(\mathcal{N}_{n\>3}\mathcal{N}_{6+i\>1}+\mathcal{N}_{6+i\>3}\mathcal{N}_{n\>1})+\cos \theta_R(\mathcal{N}_{n\>3}\mathcal{N}_{6+i\>5}+\mathcal{N}_{6+i\>3}\mathcal{N}_{n\>5}) \right)\Big)\\
+\Big(\mathcal{N}_{n\>6+j}\mathcal{N}_{6+i\>6}
+\frac{1}{\sqrt 2}Y_{\nu i3}\cos\alpha \Big(\mathcal{N}_{6+i\>6+j}
\mathcal{N}_{n\>6}\Big)
\end{multline}
The decay rate is given by
\begin{equation}
\Gamma_{{\tilde X}^0_n\rightarrow h^0\nu_{i}}=\frac{\Big(|{G_L}|_{{\tilde X}^0_n\rightarrow h^0\nu_{i}}^2+|{G_R}|_{{\tilde X}^0_n\rightarrow h^0\nu_{i}}^2\Big)}{64\pi}
M_{{\tilde \chi}_n^0}\left(1-\frac{M_{h^0}^2}{M_{{\tilde \chi}_n^0}^2}\right)^2.
\end{equation}

\end{itemize}

Note that in the above neutralino expressions, we sum over $j=1,2,3$. Using the matrix elements of $\mathcal{U}$, $\mathcal{V}$ and $\mathcal{N}$ given in Appendices B.1 and B.2, one can can calculate the values of the decay rates numerically. Just as for charginos, it can be shown that the decay amplitudes are proportional to the RPV couplings $\epsilon_i$ and $v_{L_i}$ at first order. Hence, RPV is directly responsible for the neutralino decays into SM particles.

\section*{Acknowledgments}
The authors would like to thank Evelyn Thomson, Elliot Lipeles, Jeff Dandoy, Christopher Mauger, Nuno Barros, Leigh Schaefer, and Christian Herwig for helpful suggestions. Ovrut and Purves would also like to acknowledge many informative conversations with Zachary Marshall and Sogee Spinner concerning RPV decays of a stop LSP. Burt Ovrut and Sebastian Dumitru are supported in part by DOE No. DE-SC0007901 and SAS Account 020-0188-2-010202-6603-0338.

\begin{appendix}

\section {Notation}

In this Appendix, we present for clarity all the notation used throughout the paper.
\subsection{Gauge Eigenstates}

\begin{itemize}
\item{Bosons}

\underline{{\it vector gauge bosons}}

~~~ $SU(2)_L-\quad\> W^1_\mu\>, W^2_\mu\>, W^3_\mu$, $\quad $   
coupling parameter $g_2$

~~~ $U(1)_{B-L}-\quad \> B^'_\mu \> $, $\quad $   
coupling parameter $g_{BL}$

~~~ $U(1)_{3R}-\quad \> {W_R}_\mu \> $, $\quad $   
coupling parameter $g_R$

~~~ $U(1)_{Y}-\quad \> {B}_\mu \> $, $\quad $   
coupling parameter $g'$

~~~ $U(1)_{EM}-\quad \> {\gamma}^0_\mu \> $, $\quad $   
coupling parameter $e$

~~~B-L Breaking:    $U(1)_{3R}\otimes U(1)_{B-L}\rightarrow U(1)_Y,\quad$    massive boson ${Z_R}_\mu$,$\>\>$ coupling $g_{Z_R}$

~~~EW Breaking:    $SU(2)_L\otimes U(1)_Y \rightarrow U(1)_{EM},\quad$    massive bosons $Z^0_\mu
,\>W^\pm_\mu\quad$

\underline{{\it Higgs scalars}}

~~~$H_u^0\>, H_u^+\>, H_d^0\>, H_d^-\quad $

\item{Weyl Spinors}

\underline{ {\it gauginos}}

~~~ $SU(2)_L-\> \tilde W^0\>, \tilde W^\pm$,\quad
$U(1)_{B-L}- \>\tilde B^' , \quad$ $U(1)_{3R}-\> {\tilde W_R} \> $,\quad
$U(1)_{Y}-\> \tilde {B},\quad$ $U(1)_{EM}- \> \tilde {\gamma}^0 \> $

\underline{{\it Higgsinos}}

~~~$\tilde H_u^0\>, \tilde H_u^+\>, \tilde H_d^0\>, \tilde H_d^-$

\underline{{\it leptons}}

 ~~~left chiral\quad  $e_i,\>\nu_i, \>\> i=1,2,3 \quad \text{where} \quad e_1=e,\>e_2=\mu,\>e_3=\tau$
          
 ~~~right chiral\quad  $e^c_i,\>\nu^c_i, \>\> i=1,2,3 \quad \text{where} \quad e^c_1=e^c,\>e_2^c=\mu^c,\>e_3^c=\tau^c$

\underline{{\it sleptons}}

 ~~~left chiral\quad  $\tilde e_i,\>\tilde \nu_i, \>\> i=1,2,3 \quad \text{where} \quad \tilde e_1=\tilde e,\>\tilde e_2=\tilde \mu,\>\tilde e_3=\tilde \tau$
          
 ~~~right chiral\quad  $\tilde e^c_i,\>\tilde \nu^c_i, \>\> i=1,2,3 \quad \text{where} \quad \tilde e^c_1=\tilde e^c,\>\tilde e_2^c=\tilde \mu^c,\>\tilde e_3^c=\tilde \tau^c$

\end{itemize}

\subsection{Mass Eigenstates}

\begin{itemize}

\item{Weyl Spinors}

\underline{{\it leptons}}\\
 \quad  $e_i,\>\nu_i, \>\> i=1,2,3 \quad \text{where} \quad e_1=e,\>e_2=\mu,\>e_3=\tau$

\underline{{\it charginos and neutralinos}}\\
  \quad $\tilde \chi^\pm_1, \quad \tilde \chi^\pm_2, \quad \tilde \chi_n^0,\quad n=1,2,3,4,5,6$

\item{4-component Spinors}

\underline{{\it  leptons}}\\
~~~$\ell_i^-=\left(\begin{matrix}e_i\\ {e_i^c}^\dag\end{matrix}\right),\quad
\ell_i^+=\left(\begin{matrix}{e_i^c}\\ e_i^\dag\end{matrix}\right), \quad
\nu_i=\left(\begin{matrix}\nu_i\\ {\nu_i}^\dag\end{matrix}\right)\quad i=1,2,3$

\underline{{\it charginos and neutralinos}}

$\tilde X^-_1=\left(\begin{matrix}\tilde \chi^-_1\\ \tilde {\chi}^{+\dag}_1\end{matrix}\right),\quad
\tilde X^+_1=\left(\begin{matrix}\tilde \chi^+_1\\ \tilde {\chi}^{-\dag}_1\end{matrix}\right), \quad
\tilde X^0_n=\left(\begin{matrix}\tilde \chi^0_n\\ \tilde {\chi}^{0\dag}_n\end{matrix}\right)$

\end{itemize}

\subsection{VEV's}

\begin{itemize}
\item {sneutrino VEV's \\
\quad $\left<\tilde \nu^c_{3}\right> \equiv \frac{1}{\sqrt 2} {v_R} \quad 
\epsilon_i=\frac{1}{2}Y_{\nu i3}v_R \quad \left<\tilde \nu_{i}\right> \equiv \frac{1}{\sqrt 2} {v_L}_i, \quad i=1,2,3$}

\item {Higgs VEV's \\
$\left< H_u^0\right> \equiv \frac{1}{\sqrt 2}v_u, \ \ \left< H_d^0\right> \equiv \frac{1}{\sqrt 2}v_d, \quad \tan \beta=v_u/v_d$}

\end{itemize}

\section{ Mass Matrix elements}

\subsection{Chargino mass matrix}\label{appendix:A1}

The matrices $\mathcal{U}$ and $\mathcal{V}$ can be written schematically as

\begin{equation}
\mathcal{U}=
\left(
\begin{matrix}
U&0_{2\times3}\\
0_{3\times2}&1_{3\times3}\\
\end{matrix}
\right)
\left(
\begin{matrix}
1_{2\times2}&-\xi_-\\
\xi^{\dag}_-&1_{3\times3}\\
\end{matrix}
\right) \ ,
\quad \> 
\mathcal{V}=
\left(
\begin{matrix}
V&0_{2\times3}\\
0_{3\times2}&1_{3\times3}\\
\end{matrix}
\right)
\left(
\begin{matrix}
1_{2\times2}&-\xi_+\\
\xi^{\dag}_+&1_{3\times3}\\
\end{matrix}
\right)
\end{equation}
Assuming that the lighter chargino is ${\tilde \chi}_1^\pm$, and since we are interested in its decays, it follows that we will need the elements $\mathcal{U}_{1\>2+i}$ and $\mathcal{V}_{1\>2+i}$ and their conjugates when replacing 
a lepton state with the lightest chargino mass eigenstate. It follows from the above that

\begin{equation}
\mathcal{U}_{1\>2+i}=-\cos \phi_- \frac{g_2 v_d}{\sqrt{2}M_2\mu}\epsilon_i^*+\sin \phi_-\frac{\epsilon_i^*}{\mu} \ ,
\end{equation}
\begin{equation}
\mathcal{V}_{1\>2+i}=-\cos \phi_+ \frac{g_2 \tan \beta m_{e_i}}{\sqrt{2}M_2\mu}v_{L_i}+\sin \phi_+\frac{m_{e_i}}{\mu v_d}v_{L_i} \ .
\end{equation}

When replacing a charged Wino gaugino with the lightest chargino mass eigenstate,
 we need the elements $\mathcal{U}_{1\>1}$ and $\mathcal{V}_{1\>1}$ given by  
\begin{equation}
\mathcal{U}_{1\>1}=\cos \phi_-  \ ,\quad \quad
\mathcal{V}_{1\>1}=\cos \phi_+
\end{equation}
and their conjugates. Similarly for replacing a charged Higgsino, one needs $\mathcal{U}_{1\>2}$ and $\mathcal{V}_{1\>2}$, which we find to be
\begin{equation}
\mathcal{U}_{1\>2}=\sin \phi_-  \ , \quad \quad
\mathcal{V}_{1\>2}=\sin \phi_+
\end{equation}
and their conjugates.

We also need the elements $\mathcal{U}_{2+i\>1}$ and $\mathcal{V}_{2+i\>1}$ and their complex conjugates when 
replacing a  charged Wino state with a charged lepton, where
\begin{equation}
\mathcal{U}_{2+i\>1}=\frac{g_2}{\sqrt{2}M_2\mu}(v_d\epsilon^*_i+\mu v_{L_i}) \ ,
\quad \quad
\mathcal{V}_{2+i\>1}=-\frac{1}{\sqrt{2}M_2\mu}g_2\tan \beta m_{e_i}v_{L_i} \ .
\end{equation}
The elements $\mathcal{U}_{2+i\>2}$ and $\mathcal{V}_{2+i\>2}$ and their complex conjugates when required when
replacing a charged Higgsino state with a charged lepton,
\begin{equation}
\mathcal{U}_{2+i\>2}=\frac{\epsilon_i^*}{\mu} \ , \quad \quad
\mathcal{V}_{2+i\>2}=\frac{m_{e_i}}{ v_d \mu}v_{L_i} \ .
\end{equation}

The angles $\phi_\pm$ are defined in Section 5.1. They express the charged Wino and charged Higgsino content of the chargino mass eigenstates, in the absence of the RPV couplings $\epsilon_i$ and $v_{L_i}$
\begin{equation}
\tilde \chi^\pm_1=\cos \phi_\pm {\tilde{W}}^\pm+\sin \phi_\pm {\tilde{H}}^\pm
\end{equation}
and
\begin{equation}
\tilde \chi^\pm_2=-\sin \phi_\pm {\tilde{W}}^\pm+\cos \phi_\pm {\tilde{H}}^\pm.
\end{equation}

Hence, for $\phi^\pm=0$, we have purely Wino chargino states $\tilde \chi^\pm_1$ and purely Higgsino chargino
states $\tilde \chi^\pm_2$. Conversely, for $\phi^\pm=\pi/2$, we have purely Higgsino chargino states $\tilde \chi^\pm_1$ and purely Wino chargino
states $\tilde \chi^\pm_2$.

\subsection{Neutralino mass matrix}\label{appendix:A2}

The $\mathcal{N}$ matrix can be written schematically as
\begin{equation}
\mathcal{N}=\left(
\begin{matrix}
N&0_{3\times 3}\\
0_{3\times 3}&V^{\dag}_{PMNS}\\
\end{matrix}
\right)
\left(
\begin{matrix}
1_{6\times 6}& -\xi_0\\
\xi_0^{\dag}&1_{3\times 3}\\
\end{matrix}
\right) \ .
\end{equation}
The rows of $\xi_0$ are the gaugino gauge eigenstates, whereas the columns correspond to the
neutrino gauge eigenstates. These are explicitly labeled and presented below. They are
\begin{equation}
\xi_{0_{\tilde W_R\nu_{L_i}}}=\frac{g_R\mu}{8d_{{\tilde \chi}^0}}
[2M_{BL}v_u(g_2^2v_dv_u-2M_2\mu)\epsilon_i-g_{BL}^2M_2v_R^2(v_d\epsilon_i+\mu v_{L_i}^*)] \ ,
\end{equation}
\begin{equation}
\xi_{0_{\tilde W_2\nu_{L_i}}}=\frac{g_2\mu}{8d_{{\tilde \chi}^0}}
[2g_R^2M_{BL}v_dv_u^2\epsilon_i+M_{\tilde Y}v_R^2(v_d\epsilon_i+\mu v_{L_i}^*)] \ ,
\end{equation}
\begin{equation}
\xi_{0_{\tilde H^0_d\nu_{L_i}}}=\frac{1}{16d_{{\tilde \chi}^0}}
[ M_{\tilde \gamma} v_R^2v_u (v_d\epsilon_i-\mu v_{L_i}^*)
-4M_2\mu(M_{\tilde Y}v_R^2+g^2_RM_{BL}v_u^2)\epsilon_i] \ ,
\end{equation}
\begin{equation}
~~\qquad \xi_{0_{\tilde H^0_u\nu_{L_i}}}=\frac{1}{16d_{{\tilde \chi}^0}}
[ M_{\tilde \gamma} v_R^2v_u (v_d\epsilon_i+\mu v_{L_i}^*)
-4g_R^2\mu M_2M_{BL}v_d v_u\epsilon_i] \ ,
\end{equation}
\begin{multline}
\qquad\qquad \qquad \xi_{0_{\tilde B^'\nu_{L_i}}}=-\frac{1}{8d_{{\tilde \chi}^0}}
[ g_{BL}g_R^2 M_2\mu v_R^2 (v_d\epsilon_i+\mu v_{L_i}^*)\\
+2g_{BL}\mu v_u((  g_R^2M_2+g_2^2 M_R  )v_dv_u-2M_RM_2\mu)\epsilon_i] \ , \qquad\qquad \qquad
\end{multline}
\begin{multline}
\qquad\qquad \qquad\xi_{0_{\tilde \nu_3^c\nu_{L_i}}}=\frac{\mu}{8v_Rd_{{\tilde \chi}^0}}[(
M_{\tilde \gamma}v_R^2v_dv_u-2g_{BL}^2M_RM_2\mu v_R^2)v_{L_i}^*\\
+2M_{BL}(M_2(g_R^2v_R^2v_d-4M_R\mu v_u)+2(g_R^2M_2+g_2^2M_R)v_dv_u^2)\epsilon_i] \qquad
\end{multline}
where
\begin{equation}
d_{{\tilde \chi}^0}=\frac{1}{4}M_2M_1\mu^2v_R^2-\frac{1}{8}M_{\tilde \gamma}\mu v_R^2v_dv_u \ ,
\end{equation}
\begin{equation}
M_Y=g_R^2M_{BL}+g^2_{BL}M_R \ ,
\end{equation}
\begin{equation}
M_{\gamma} =g^2_{BL}g^2_RM+g_2^2g^2_RM_{BL}+g_2^2g_{BL}^2M_R \ .
\end{equation}
We can now express the matrix elements of 
\begin{equation}\label{eq:A19}
\mathcal{N}=\left(
\begin{matrix}
N&0_{3\times 3}\\
0_{3\times 3}&V^{\dag}_{PMNS}\\
\end{matrix}
\right)
\left(
\begin{matrix}
1_{6\times 6}& -\xi_0\\
\xi_0^{\dag}&1_{3\times 3}\\
\end{matrix}
\right)=\left(
\begin{matrix}
N&-N\xi_0\\
V_{PMNS}^\dag\xi_0^\dag&V_{PMNS}^\dag
\end{matrix}
\right)
\end{equation}
$N$ is the matrix that diagonalizes the neutralino mass matrix in the absence of RPV couplings to the three families of left-handed neutrinos. If the soft masses in the neutralino mass matrix, eq. \eqref{eq:A19}, are much larger than the Higgs VEV's $v_u$ and $v_d$, then, at zeroth order, we have
\begin{equation}
N=\left(
\begin{matrix}
\sin \theta_R&0&0&0&\cos \theta_R&0\\
0&1&0&0&0&0\\
0&0&0&1&0&0\\
0&0&1&0&0&0\\
-\frac{1}{\sqrt 2}\cos \theta_R&0&0&0&\frac{1}{\sqrt 2}\sin \theta_R&\frac{1}{\sqrt 2}\\
\frac{1}{\sqrt 2}\cos \theta_R&0&0&0&-\frac{1}{\sqrt 2}\sin \theta_R&\frac{1}{\sqrt 2}\\
\end{matrix}
\right) \ .
\end{equation}
However, in the regimes with small chargino and neutralino masses that we analyze, this approximation is no longer valid. The elements of $N$ will, in general,  have complicated expressions and we choose to evaluate them numerically. We use as input the numerical values of the neutralino mass matrix. We expect, however, based on the zeroth order form of $N$, that $N_{11}$, $N_{22}$, $N_{34}$, $N_{43}$, $N_{15}$, $N_{51}$, $N_{61}$,
$N_{55}$, $N_{65}$, $N_{56}$ and $N_{66}$ are of order $ \mathcal{O}(1)$, while the remaining matrix elements are of order $\mathcal{O}({M_{EW}/M_{soft}})<<1$.

Elements form the top-right block $N\xi_0$ have the form 
\begin{multline}
\mathcal{N}_{n\>6+i}=-N_{n\>1}\xi_{0_{\tilde W_R\nu_{L_i}}}-N_{n\>2}\xi_{0_{\tilde W_2\nu_{L_i}}}-N_{n\>3}\xi_{0_{\tilde H_d^0\nu_{L_i}}}-N_{n\>4}\xi_{0_{\tilde H_u^0\nu_{L_i}}}-N_{n\>5}\xi_{0_{\tilde B^'\nu_{L_i}}}-N_{n\>6}\xi_{0_{\tilde \nu_3^c\nu_{L_i}}}\\
\simeq N_{n\>1}\left[\frac{2g_RM_{BL}v_u}{M_1v_R^2}\epsilon_i+\frac{g_Rg_{BL}^2}{2M_1}v_{L_i}^*\right]
-N_{n\>2}\left[\frac{g_2v_d}{2M_2\mu}\epsilon_i+\frac{g_2}{2M_2}v_{L_i}^*\right]\\
+N_{n\>3}\left[\frac{\epsilon_i}{16\mu}\right]-N_{n\>4}\left[\frac{\tilde M_\gamma v_u}{4M_2M_1\mu^2}(v_d\epsilon_i+\mu v_{L_i})-\frac{g_R^2M_{BL}v_uv_d}{M_1v_R^2\mu}\epsilon_i\right]\\
-N_{n\>5}\left[ \frac{g_{BL}g_R^2}{2M_1\mu}(v_d\epsilon_i+\mu v_{L_i}^*) -\frac{2g_{BL}M_Rv_u}{M_1v_R^2}\epsilon_i \right] \qquad \\
-N_{n\>6}\left[ \frac{-g_{BL}^2M_R}{v_RM_1}v_{L_i}^*+\frac{ M_{BL}}{v_R^3M_2M_1}\mu(g_R^2M_2v_R^2v_d-4M_R\mu v_u)\epsilon_i \right] \qquad \qquad
\end{multline}
for $n=1,2,3,4,5,6$ and $i=1,2,3$. Elements of the bottom left block $V_{PMNS}^\dag\xi_0^\dag$ are computed in a similar fashion. One can then determine $\mathcal{N}_{6+i\>n}$ as
\begin{equation}
\mathcal{N}_{6+i\>n}={[V_{PMNS}^\dag]}_{6+i\>6+j}{[\xi_0^\dag]}_{6+j\>n}
\end{equation}
\end{appendix}

\end{document}